\documentclass[preprint]{aastex}
\usepackage{natbib}
\usepackage{float}
\usepackage{morefloats}
\usepackage{amsmath}
\usepackage{graphicx}
\usepackage{verbatim}
\usepackage{color}
\usepackage{ulem}
\usepackage{appendix}

   \usepackage{graphicx}
   \usepackage{epstopdf}
\newcommand{\strikethrough}[1]                   {}

\newcommand{\func}[2]                   { #1 \left[#2\right] }
\newcommand{\ddt}[1]                    {\frac{\partial }{\partial t} #1}
\newcommand{\divergence}[1]             { \mathbf{\nabla} \cdot {#1}}
\newcommand{\cylindricaldivergence}[3]  { 	\frac{1}{R} \frac{\partial}{\partial R} R #1 + 
						\frac{1}{R} \frac{\partial}{\partial \phi} #2 + 
						\frac{\partial}{\partial z} #3 }
\newcommand{\seperatedivergence}[4]  { 	\frac{1}{R} \frac{\partial}{\partial R} R #1 + \frac{\partial}{\partial R} #2 + 
						\frac{1}{R} \frac{\partial}{\partial \phi} #3 + 
						\frac{\partial}{\partial z} #4 }

\newcommand{\gradient}[1]               { \mathbf{\nabla} {#1}}

\definecolor{red}{rgb}{1.0,0.0,0}

\begin{document}

		\title{A Numerical Method for Studying Super-Eddington Mass Transfer in Double White Dwarf Binaries}
		\author{Dominic C. Marcello}	
		\email{dmarcello@phys.lsu.edu}
		\affil{Louisiana State University and Agricultural \& Mechanical College}
		\affil{Department of Physics and Astronomy}
		\affil{202 Nicholson Hall, Baton Rouge, Louisiana}
		\and
		\author{Joel E. Tohline}
		\email{tohline@phys.lsu.edu}
		\affil{Louisiana State University and Agricultural \& Mechanical College}
		\affil{Department of Physics and Astronomy}
		\affil{202 Nicholson Hall, Baton Rouge, Louisiana}
		\begin{abstract}
	
	We present a numerical method for the study of double white dwarf (DWD) binary systems at the onset of super-Eddington mass transfer.
 	We incorporate the physics of ideal inviscid hydrodynamical flow, Newtonian self-gravity, and radiation transport on a 
	three-dimensional uniformly rotating cylindrical Eulerian grid. Care has been taken to conserve the key physical quantities such as 
	angular momentum and energy. Our new method conserves total energy to a higher degree of accuracy than other codes that are presently being used to model mass-transfer in DWD systems.
 We present the results of 
	verification tests and we simulate the first $20+$ orbits of a binary system of mass ratio $q=0.7$  at the onset of 
	dynamically unstable direct impact mass transfer. The mass transfer rate quickly exceeds the critical Eddington limit by many orders of magnitude, and thus we are unable to model a trans-Eddington phase.  It appears that radiation pressure does not significantly effect the accretion flow in the highly super-Eddington regime. An optically thick common envelope forms around the binary within a few orbits. Although this envelope quickly exceeds the spatial domain of the computational grid, the fraction of the common envelope that exceeds zero gravitational binding energy is extremely small, suggesting that radiation-driven mass loss is insignificant in this regime. It remains to be seen whether simulations that capture the trans-Eddington phase of such flows will lead to the same conclusion or show that substantial material gets expelled.

Keywords: binaries: close -- gravitation -- hydrodynamics -- methods: numerical -- radiative transfer -- white dwarfs

		\end{abstract}

	\section{Introduction }

	Theoretical evidence suggests there are approximately $3 \times 10^8$ close double white dwarf (DWD) binary systems in the Galaxy, with birth rates of 
	$5 \times 10^{-2} / \mathrm{yr}$  (\cite{NYPV2001}). These systems are thought to be the progenitors of a wide array of astronomical phenomena. Due to their short 
	orbital periods, they emit significant gravitational radiation which may form a low frequency background noise limiting the sensitivity of detectors such as the 
	Laser Interferometer Space Antenna (LISA) (\cite{HBW1990}, \cite{NYP2001}, \cite{FP2002}).  Loss of  angular momentum due to this gravitational radiation will cause 
	a significant fraction of these systems to undergo Roche lobe overflow within a Hubble time. If the mass transfer causes the accretor's mass to exceed the Chandrasekhar mass limit, nuclear detonation and a Supernovae Type Ia is a possible result (\cite{W1984}, \cite{IT1984}, \cite{LR2003}, \cite{S2010}), although it is also possible such a system could avoid nuclear detonation and collapse to form a more compact object (\cite{NI1985}, \cite{SN1985}, \cite{ML1990}, \cite{HN2000}). Less massive DWD's may merge to form hydrogen poor objects such as R Coronae Borealis variable stars, extreme helium stars , or sub-dwarf B and sub-dwarf O stars (\cite{W1984}, \cite{SJ2000}, \cite{HPMMI2002}, \cite{CGHFA2007}). Systems that survive the initial onset of mass transfer likely become AM Canum Vanaticorum (AM CVn) systems (\cite{P1967}, \cite{FFW1972}, \cite{NPVY2001}). 

When a DWD mass transfer event is dynamically unstable, the mass-transfer rate can quickly grow to exceed the Eddington limit (\cite{W1984}, \cite{I1988}).  If the result is roughly equivalent to the Eddington limit in the context of spherical accretion, mass transfer onto the accretor will cease at the Eddington limit, and the remaining mass lost from the donor may be driven from the system. Analytic work  suggests that if a significant mass fraction is unbound from the system, the dynamics of the system may be altered in favor of survival. However, if the mass is retained within a common envelope which extends beyond the orbit of the binary, dissipative effects could cause the orbital separation to shrink, resulting in eventual merger (\cite{HW1999}, \cite{GPF2007}). The geometry of accretion in a close DWD, however, is far from spherical. If the radiation from accretion is able to escape interaction with the accreting mass before it becomes captured by the accretor, the radiation may have little effect and mass transfer onto the accretor may proceed at super-Eddington rates. If mass loss occurs as the system approaches the Eddington limit, however, it is possible the system may never enter the super-Eddington regime. To account for the effects of radiative transport in the complex geometry of DWD accretion, potentially in the presence of a common envelope, requires numerical simulations that can couple radiation to hydrodynamic flows.

In recent years there has been much progress in the study of DWD's using computational fluid dynamic techniques. The two dominant numerical paradigms for this purpose are the smoothed particle hydrodynamics (SPH) codes (e.g. \cite{BCPB1990}, \cite{RS1995}) and the Eulerian grid based codes (e.g. \cite{MTF2002}, \cite{DMTF2006}, \cite{MFTD2007}). In both cases, the laws of fluid hydrodynamics and Newtonian gravity are applied in a three-dimensional space. Some of the more recent SPH codes also employ detailed equations of state and/or nuclear reaction networks (e.g. \cite{SCM1997}, \cite{GGI2004}, \cite{YPR2007},  \cite{DRB2009}), and recently our group has produced an Eulerian code that incorporates a cold white dwarf equation of state (\cite{E2010}).  None of the aforementioned codes simulate radiative transport. \cite{GDRR2010} combined results from an SPH simulation with the FLASH (\cite{FORTZLMRTT2000}) code to simulate DWD mass transfer. The FLASH code models radiative transport, as well as nuclear physics, however, it was used to model only the accretion stream and accretor. The boundary conditions for the FLASH code portion of the simulation were set based on the results of an SPH simulation of the complete binary.  \cite{P2009} suggested that the DJEHUTY code (\cite{B2003}), which incorporates radiation transport, could be modified to simulate the common envelope phase of DWD mass transfer. At the time of this writing, we are unaware of any three-dimensional simulations of DWD mass transfer which simulate an entire DWD self-consistently and incorporate radiation transport.

Below we describe the capabilities of our most recent Eulerian computer code that has been designed to simulate mass transferring DWD's. In addition to improvements over our previous codes in the treatment of the fluid transport equations, we incorporate the flux-limited diffusion (FLD) approximation to the radiation transport equation. The FLD approximation has been applied for use in other astrophysical contexts by other codes, such as FLASH and ZeusMP2 code of \cite{HNF2006}. The radiation hydrodynamics portion of our code is adopted from these previous approaches to FLD, however, its application in the context of interacting DWD's is original. Although the accuracy of the FLD approximation has limitations, we believe it is suitable as a first step in the numerical study of DWD's undergoing super-Eddington mass transfer. We will first describe the numerical method in detail, then describe verification tests to which the code has been subjected. Finally, we present results from a pair of runs simulating a DWD during the initial phase of mass transfer.

\pagebreak

\section{The Model Equation Set}
\label{eqs_section}	

As with our previous codes, our new code models the DWD system as a self-gravitating inviscid fluid, governed by the equations of Newtonian gravity coupled to the classical hydrodynamic equations for density, momentum, and gas energy transport. Our earliest codes   assumed adiabatic flow and hence did not properly account for the conversion of kinetic energy into thermal energy at shock fronts (i.e. \cite{MTF2002}). This code, as well as another of our recent codes (\cite{E2010}), properly accounts for entropy generation at shock fronts. This is crucial for the proper modeling of super-Eddington accretion flows, as the generation of heat at the accretion stream's point of impact on the accretor is responsible for the conversion of accretion luminosity into radiative luminosity. Here we model the gas temperature and pressure based on the ideal gas equation of state. The most important additions to the new code are radiation energy transport and the coupling of  radiation energy to the momentum and gas energy. These physical processes are the bare minimum required to simulate super-Eddington accretion. They can account for: (1) the conversion of kinetic energy into thermal energy at the accretion stream's point of impact and thermal energy into radiation energy ; (2) the transport of radiation energy through space; and (3) the interaction of the resulting radiation energy flux on the momentum of the accretion stream. 

\subsection{Governing Equations }
\label{governor}
Fundamentally we adopt the same basic set of dynamical governing equations as presented by \cite{HNF2006} but with the magnetic field set to zero. Specifically, by
taking the Eulerian form of equations (1) through (4) in \cite{HNF2006}, adding rotational terms and removing contributions from the magnetic field, our adopted governing equations are:
\begin{eqnarray}
\label{continuity}
\ddt{ \rho}             + \divergence{\rho \mathbf{u}} & = & 0; \\
\label{momenta}
\ddt{ \rho \mathbf{u} } + \divergence{\left( \rho \mathbf{u} \mathbf{u} + p \mathbf{I} \right) }  & = & 
-\rho \gradient{\Phi} + \frac{\chi}{c} \mathbf{F} - 2 \Omega \times \rho \mathbf{u} - \rho \Omega \times \left(\Omega \times \mathbf{r}\right); \\
\label{internal_gas_energy}
\ddt{ e }             + \divergence{ e \mathbf{u} } + p \divergence{\mathbf{u}} & = &
 - 4 \pi \kappa_p B_p + c \kappa_E E_R;\\
\label{radiation_energy}
\ddt{ E_R }             + \divergence{ E_R \mathbf{u} } + \divergence{\mathbf{F}}  + \mathbf{P} : \gradient{\mathbf{u}} & = &
4 \pi \kappa_p B_p - c \kappa_E E_R.
\end{eqnarray}
The quantity $\mathbf{I}$ is the identity second rank tensor. The fluid velocity, $\mathbf{u}$, is defined in the \textit{rotating} frame.
The internal gas energy density is $e$.  The radiation energy density is $E_R$. 
The gravitational potential, $\Phi$, is determined from Poisson's equation, 
\begin{equation}
\label{poisson}
\nabla^2 \Phi = 4 \pi G \rho,
\end{equation}
where $G$ is the gravitational constant.
The gas pressure, $p$, is given by, 
\begin{equation}
\label{pressure}
p = \left(\gamma - 1\right) e,
\end{equation}
where $\gamma$ is the ratio of specific heats. 
The frequency integrated Planck function, $B_p$, is
\begin{equation}
B_p = \frac{\sigma}{\pi} T^4,
\end{equation}
where $\sigma$ is the Stefan-Boltzmann constant and $T$ is the gas temperature. We compute $T$ by using equation (\ref{pressure}) and the ideal gas equation,
\begin{equation}
p = \frac{\mathcal{R}}{\mu} \rho T,
\label{idealgas}
\end{equation}
where $\mu$ is the mean molecular weight and $\mathcal{R}$ is the gas constant. 

The quantities $\chi$, $\kappa_p$, and $\kappa_E$ are, respectively, the flux mean opacity, Planck mean opacity, and energy mean opacity.  Their general definitions are provided by \cite{HNF2006}. In Appendix \ref{opac_app} we describe how we treat the opacities in the code. The radiative flux, $\mathbf{F}$, is given by the FLD approximation
\begin{equation}
\label{fld}
\mathbf{F} = -\frac{c \Lambda_E}{\chi} \gradient{E_R},
\end{equation}
where $c$ is the speed of light and $\Lambda_E$ is the flux limiter. We use the flux limiter of \cite{LP1981}:
\begin{equation}
\label{limiter}
\Lambda_E := \frac{1}{\Theta} \left( \mathrm{coth} \Theta - \frac{1}{\Theta} \right),
\end{equation}
where 
\begin{equation}
\Theta := \frac{\left|\gradient{E_R}\right|}{\chi E_R}.
\end{equation}
The symmetric radiative stress tensor, $\mathbf{P}$, is given by 
\begin{equation}
\mathbf{P} = \mathbf{f_{\mathrm{Edd}}} E_R.
\end{equation}
Using the Eddington factor,
\begin{equation}
f_{\mathrm{Edd}} := \Lambda_E + \left( \Lambda_E \Theta \right)^2,
\end{equation}
the Eddington tensor, $\mathbf{f_{\mathrm{Edd}}}$, is defined as
\begin{equation}
\mathbf{f_{\mathrm{Edd}}} := \frac{1}{2}\left(1-f_{\mathrm{Edd}}\right) \mathbf{I} + \frac{1}{2}\left(3 f_{\mathrm{Edd}}-1\right) \mathbf{\hat{n}}\mathbf{\hat{n}},
\end{equation}
where the vector normal to the flow of radiation is
\begin{equation}
\mathbf{\hat{n}} := - \frac{\gradient{E_R}}{\left|\gradient{E_R}\right|}.
\end{equation} 
The propagation speed of the radiation energy density under this simple diffusion approximation is not bounded by the speed of light. However, the use of the flux limiter, $\Lambda_E$, ensures that this propagation speed never exceeds $\left|\mathbf{u}\right| + c$.  Although physically we expect that the propagation speed never exceeds $c$, the elimination of higher order terms in the FLD approximation can, in principle, result in super-luminal radiation transport. Our adopted flux limiter satisfies
\begin{equation}
\lim_{\Theta \rightarrow 0} \Lambda_E = \frac{1}{3}
\end{equation}
and
\begin{equation}
\lim_{\Theta \rightarrow \infty} \Lambda_E = \frac{1}{\Theta}.
\end{equation}
Hence, when $\Theta \rightarrow 0$, we recover the diffusion limit, 
\begin{equation}
\mathbf{F} \rightarrow -\frac{1}{3} \frac{c}{\chi} \gradient{E_R} 
\end{equation}
and
\begin{equation}
\mathbf{P} \rightarrow \frac{1}{3} E \mathbf{I}.
\end{equation}
In this limit, the radiation intensity is isotropic.
In the ``streaming" limit, where $\Theta \rightarrow \infty$, 
\begin{equation}
\mathbf{F} \rightarrow c E_R \mathbf{n}
\end{equation}
and
\begin{equation}
\mathbf{P} \rightarrow E_R \mathbf{n} \mathbf{n}.
\end{equation}
In this limit the radiation intensity is modeled as a single ray of light with a delta function angular distribution. 

At shock discontinuities, kinetic energy is converted into internal gas energy so updating $e$ via equation (\ref{internal_gas_energy}) will produce lower internal energies than physically expected. For this reason we generally prefer to update $e$ by first evolving the total gas energy density, 
\begin{equation}
E_G := e + \frac{1}{2} \rho u^2, 
\end{equation}
then subtracting $\frac{1}{2} \rho u^2$ from $E_G$ to obtain $e$. Note that $E_G$ is defined in the rotating frame. The equation governing the time-evolution of $E_G$ is obtained by dotting $\mathbf{u}$ into equation (\ref{momenta}), realizing that,
 \begin{equation}
\label{convert}
\mathbf{u} \cdot \left( \ddt{\rho \mathbf{u}} + \divergence{ \rho \mathbf{u} \mathbf{u} } \right)  \ = \
\frac{1}{2}\left(\ddt{\rho u^2} + \divergence{\rho u^2 \mathbf{u} } \right),
\end{equation}
and adding it to equation (\ref{internal_gas_energy}). Specifically, we obtain
\begin{equation}
\label{total_gas_energy}
\ddt{ E_G }             + \divergence{ \left(E_G + p\right) \mathbf{u} } \ = \
-\rho \mathbf{u} \cdot \gradient{\Phi} + \mathbf{u} \cdot \frac{\chi}{c} \mathbf{F} - 4 \pi \kappa_p B_p + c \kappa_E E_R - \rho \mathbf{u} \cdot \Omega \times \left(\Omega \times \mathbf{r}\right).
\end{equation}
Equations (\ref{total_gas_energy}) and (\ref{internal_gas_energy}) should both correctly describe the time-evolution of $e$ in regions of space where $\mathbf{u}$ is continuous but, in the vicinity of shocks, only equation (\ref{total_gas_energy}) provides the correct description. Hence, we replace equation (\ref{internal_gas_energy}) with equation (\ref{total_gas_energy}) in our principal set of governing equations. However, due to the numerical issues described below in connection with equation (\ref{internalfromtotal}), we adopt a dual energy formalism (\cite{BNSO1995}) and evolve equation (\ref{internal_gas_energy}) independently.

\subsection{Momentum and Energy Conservation}

It is useful to examine equation (\ref{momenta}) to understand how each term might contribute to momentum conservation globally. Integrating equation (\ref{momenta}) over all space, and using the divergence theorem, 
\begin{equation}
\int_V \divergence{ \mathbf{v} } \ d \mathrm{V} = \int_S \mathbf{v} \cdot \mathbf{da}
\end{equation}
one can show that, when the density and pressure go to zero at large distances from the coordinate origin, the sum of momentum over all space is not altered by the term $\divergence{\left( \rho \mathbf{u} \mathbf{u} + p \mathbf{I} \right) }$. Using equation (\ref{poisson}) and following \cite{S1992}, we can rewrite the gravitational term as,
\begin{equation}
\label{gravmom}
\rho \gradient{\Phi} \ = \ \frac{\nabla^2 \Phi}{4 \pi G}  \gradient{\Phi} \ = \ \frac{1}{4 \pi G} \divergence{ \left( \mathbf{g} \mathbf{g}  - \frac{1}{2}\left| \mathbf{g} \right|^2 \mathbf{I} \right) },
\end{equation} 
where  the gravitational acceleration is $\mathbf{g} := -\gradient{\Phi}$.
Since equation (\ref{gravmom}) is zero when $\rho = 0$, momentum will be conserved within any volume containing all the mass of an isolated system. Using
equation (\ref{fld}), we rewrite the radiative force term as
\begin{equation}
\frac{\chi}{c} \mathbf{F} \ = \ - \Lambda_E \gradient{E_R} \ = \ - \divergence{ \Lambda_E E_R \mathbf{I} } + E_R \gradient{ \Lambda_E }.
\end{equation}
In the diffusion limit,  $\Theta \rightarrow 0$, $\Lambda_E \rightarrow \frac{1}{3}$, and $\gradient{ \Lambda_E } \rightarrow \mathbf{0}$, therefore the radiation term conserves momentum within any volume in the diffusion limit. Outside of this limit the gas exchanges net momentum with the radiation field. Because we do not evolve the radiative flux $\mathbf{F}$ separately, it is not possible for our method to account for this exchange in a manner which generally conserves momentum. 
Finally, defining the symmetric stress-energy tensor,
\begin{equation}
\mathbf{T} := \rho \mathbf{u} \mathbf{u} + \mathbf{g} \mathbf{g} + \left( p + \Lambda_E E_R - \frac{1}{2} g^2 \right) \mathbf{I},
\end{equation}
we can rewrite equation (\ref{momenta}) as,
\begin{equation}
\label{momenta2}
\ddt{ \rho \mathbf{u} } + \divergence{ \mathbf{T} }  \ = \ 
E_R \gradient{\Lambda_E} - \rho \Omega \times \left( 2 \mathbf{u} + \left(\Omega \times \mathbf{r}\right) \right).
\end{equation} 
Written in this form, all the ``source" terms, that is, all the terms that have been grouped together on the right hand side (RHS), will contribute to net changes in the sum of momentum over all space. Terms on the left hand side (LHS) are in conservative form.

\cite{MTF2002} have argued that a more accurate dynamical treatment will result from the adoption of an entropy tracer.
By defining the entropy tracer, 
\begin{equation}
\tau := e^\frac{1}{\gamma},
\end{equation}
in place of the internal energy density, the $p \divergence{\mathbf{u}}$ term no longer appears as a source. In terms of $\tau$ equation (\ref{internal_gas_energy}) becomes,
\begin{equation}
\label{tracer}
\ddt{ \tau }             + \divergence{ \tau \mathbf{u} } \ = \
\frac{1}{\gamma \tau^{\gamma - 1}} \left(- 4 \pi \kappa_p B_p + c \kappa_E E_R \right).
\end{equation}
As in equation (\ref{momenta2}), we have written equation (\ref{tracer}) in a form that places the  non-conservative source terms on the RHS. The term that appears in this case accounts for the exchange of entropy with the radiation field. As discussed above in the context of equation (\ref{total_gas_energy}), however, in the vicinity of shocks entropy is produced and equation (\ref{tracer}) does not hold. Hence, through a dual energy formalism (see the discussion associated with equation (\ref{internalfromtotal})), we will rely on equation (\ref{total_gas_energy}) instead of equation (\ref{tracer}) in the presence of shocks.

It is worthwhile to ask what expression for the energy density will serve better than $E_G$ to describe total energy conservation when integrated over the volume of our simulated system. Using equation (\ref{continuity}), the gravitational term on the RHS of equation (\ref{total_gas_energy}) can be written as
\begin{equation}
\rho \mathbf{u} \cdot \gradient{\Phi} \ = \
\divergence{ \rho \Phi \mathbf{u} } - \Phi \divergence{ \rho \mathbf{u} } \ = \
\divergence{ \rho \Phi \mathbf{u} } + \Phi \ddt{ \rho } \ = \ 
\ddt{\frac{1}{2} \rho \Phi} + \divergence{ \rho \Phi \mathbf{u} } + \frac{1}{2} \Phi \ddt{ \rho } - \frac{1}{2}  \rho \ddt{ \Phi }.
\end{equation}
Also, using equation (\ref{continuity}) and the definition,
\begin{equation}
	\Phi_{\mathrm{rot}} := - \frac{1}{2} {\left| \Omega \times \mathbf{r} \right|}^2,
\end{equation}
we can rewrite the centrifugal term as,
\begin{equation}
\rho \mathbf{u} \cdot \Omega \times \left( \Omega \times \mathbf{r} \right) \ = \ 
-\rho \mathbf{u} \cdot \gradient{ \frac{1}{2} {\left| \Omega \times \mathbf{r} \right|}^2 } \ = \
\divergence{ \rho \Phi_{\mathrm{rot}} \mathbf{u} } - 
 \Phi_{\mathrm{rot}} \divergence{ \rho \mathbf{u} }  \ = \ 
\ddt{  \rho \Phi_{\mathrm{rot}} }  +
\divergence{ \Phi_{\mathrm{rot}} \rho  \mathbf{u} }.
\end{equation}
Defining,
\begin{equation}
\mathcal{E}_{\mathrm{con}} := E_G + \frac{1}{2} \rho \Phi + \rho \Phi_{\mathrm{rot}},
\end{equation}
equation (\ref{total_gas_energy}) can be written as
\begin{equation}
\label{econ}
\ddt{ \mathcal{E}_{\mathrm{con}} }             + \divergence{ \left(\mathcal{E}_{\mathrm{con}} + p + \frac{1}{2} \rho \Phi \right) \mathbf{u} } + \frac{1}{2} \Phi \ddt{ \rho } - \frac{1}{2}  \rho \ddt{ \Phi }  \ = \ \mathbf{u} \cdot \frac{\chi}{c} \mathbf{F} - 4 \pi \kappa_p B_p + c \kappa_E E_R.
\end{equation}
Note that $\mathcal{E}_{\mathrm{con}}$ represents a sum of kinetic, internal, and potential gas energies. Using equation (\ref{poisson}) and Green's theorem, we see that the integral over all space of the last two terms on the LHS of equation (\ref{econ}) gives,
\begin{multline}
 \int_{V} \left( \frac{1}{2} \Phi \ddt{ \rho } - \frac{1}{2}  \rho \ddt{ \Phi } \right) \ d^3 \mathbf{r} \ = \ 
  \frac{1}{8 \pi G}  \int_{V} \left(\Phi \nabla^2 \ddt{ \Phi } -  \left(\ddt{ \Phi } \right)\nabla^2 \Phi \right) \ d^3 \mathbf{r} \ = \ \\
  \frac{1}{8 \pi G}  \int_{S} \left(\Phi \nabla \ddt{ \Phi } -  \left(\ddt{ \Phi }\right) \nabla \Phi \right) \ \cdot d\mathbf{a}.
\end{multline} 
This quantity will go to zero at large distances from the origin for a finite mass distribution. As in equations (\ref{momenta2}) and (\ref{tracer}), we have written equation (\ref{econ}) with conservative terms on the LHS and non-conservative source terms on the RHS. It should therefore be clear that, within the radiation diffusion limit, $\mathcal{E}_{\mathrm{con}}$ is a conserved quantity. Note that the contribution to $\mathcal{E}_{\mathrm{con}}$ from the gravitational potential is $\frac{1}{2} \rho \Phi$ (instead of  $\rho \Phi$) due to the self interactive nature of the gravitational field (see equation (2-19) in \cite{BT1987}). 

Equation (\ref{total_gas_energy}) can also be written as
\begin{equation}
\label{eloc}
\ddt{ \mathcal{E}_{\mathrm{loc}} }             + \divergence{ \left(\mathcal{E}_{\mathrm{loc}} + p \right) \mathbf{u} } \ =  \  \rho \ddt{ \Phi } + \mathbf{u} \cdot \frac{\chi}{c} \mathbf{F} - 4 \pi \kappa_p B_p + c \kappa_E E_R,
\end{equation}
where we have defined 
\begin{equation}
\label{elocdef}
\mathcal{E}_{\mathrm{loc}} := E_G + \rho \Phi + \rho \Phi_{\mathrm{rot}}.
\end{equation}
For a non-self gravitating fluid, with $\Phi$ fixed in time, and absent the radiation terms, $\mathcal{E}_{\mathrm{loc}}$ will be a globally conserved quantity. We may consider equation (\ref{eloc}) to consist of three parts: (1) the LHS, describing the hydrodynamic flow of a ``locally conserved" energy, $\mathcal{E}_{\mathrm{loc}}$; (2) a contribution to this energy from the first term on the RHS,  $\rho \ddt{ \Phi }$, which is due to the \textit{global} effect of a time varying gravitational potential; and (3) a non-conservative contribution from the remaining terms on the RHS, due to the interaction with the radiation field. We describe $\mathcal{E}_{\mathrm{loc}}$ as ``locally conserved" because it includes the kinetic, internal, and potential energy that is physically carried by the local flow of the fluid. The difference between $\mathcal{E}_{\mathrm{con}}$ and $\mathcal{E}_{\mathrm{loc}}$, $-\frac{1}{2} \rho \Phi$, is carried by the global flow of energy  between non-adjacent fluid elements due to Newtonian gravity.

Defining the total energy density as 
\begin{equation}
\mathcal{E}_{\mathrm{tot}} := \mathcal{E_\mathrm{con}} + E_R,
\end{equation}
we can write the sum of equations (\ref{total_gas_energy}) and (\ref{radiation_energy}) as
\begin{equation}
\ddt{ \mathcal{E}_{\mathrm{tot}} } 
+ \divergence{ \left[ \left(\mathcal{E}_{\mathrm{tot}} + p + \frac{1}{2} \rho \Phi \right) \mathbf{u} + \mathbf{P} \cdot \mathbf{u} \right] } 
+ \frac{1}{2} \Phi \ddt{ \rho } - \frac{1}{2}  \rho \ddt{ \Phi }  \ = \  
\mathbf{u} \cdot \left( \divergence{ \mathbf{P} } - \Lambda_E \gradient{E_R} \right).
\end{equation}
Again, we have placed non-conservative source terms on the RHS and conservative terms on the LHS. Physically, we should expect the quantity $\mathcal{E}_{\mathrm{tot}}$ to be globally conserved, as it is the volume integral  over all space  of all energy densities: kinetic energy; internal heat energy; gravitational potential energy; rotational potential energy; and radiation energy densities. In the diffusion limit the terms on the RHS will cancel one another, resulting in conservation of $\mathcal{E}_{\mathrm{tot}}$. Outside of the diffusion limit the same is not generally true. This is due to the fact that equation (\ref{radiation_energy}) is a zeroth order approximation to the relativistic radiative transport equation. However, since the overwhelming majority of the energy contained in our models will be in the diffusion limit, we do not expect this will have a significant effect.

In a cylindrical coordinate system rotating about the $z$ axis with constant frequency $\Omega$ the components of equation (\ref{momenta2}) are:
\begin{eqnarray}
\label{rad}
\ddt{ s_R } + \divergence{\mathbf{T_R} } & = &  
E_R \frac{\partial}{\partial R} \Lambda_E + \frac{T_{\phi \phi}}{R} + 2 \rho \Omega u_\phi + \rho R \Omega^2; \\
\label{u2}
\ddt{ s_\phi } + \divergence{\mathbf{T_\phi} } & = &  
E_R \frac{1}{R} \frac{\partial}{\partial \phi} \Lambda_E - \frac{T_{R \phi}}{R}  - 2 \rho \Omega u_R; \\
\label{lz}
\ddt{ s_z } + \divergence{\mathbf{T_z} } & = &  
E_R \frac{\partial}{\partial z} \Lambda_E;
\end{eqnarray}
where $s_R := \rho u_R$, $s_\phi := \rho u_\phi$, and $s_z := \rho u_z$. The vertical angular momentum density, $s_z$, is conserved in the radiation diffusion limit. The second terms on the RHS of equations (\ref{rad}) and (\ref{u2}) are coordinate curvature terms that result from applying the divergence operator to $\mathbf{T}$. By using the inertial frame $z$-angular momentum density, 
\begin{equation}
\label{inertial}
l_z = R \rho u_\phi + \rho R^2 \Omega,
\end{equation}
both the coordinate curvature and Coriolis terms in equation (\ref{u2}) can be eliminated. The new equation is
\begin{equation}
\label{lz2}
\ddt{ l_z } + \divergence{ R \mathbf{T_\phi} } \ = \
E_R \frac{\partial}{\partial \phi} \Lambda_E.
\end{equation}
Therefore, in the diffusion limit, $l_z$ is also a conserved quantity. A similar transformation cannot be performed on equation (\ref{rad}). This is because radial momentum is \textit{not} physically conserved. By choosing a curvilinear coordinate system, we are limited to choosing, at most, two conserved generalized momentum components. 

\subsection{Reformulated Governing Equations}
The binary systems we wish to study will begin their evolution in a state of near equilibrium and, for a significant part of their evolution, we expect them to remain in a state of near equilibrium. With the exception of a few computational zones near their surfaces, each star will begin evolution in the radiation diffusion limit. To accurately evolve such a system requires that quantities which are conserved analytically are also conserved numerically. In particular, we require the conservation of $\rho$ and, in the diffusion limit, the conservation of $l_z$, $s_z$, and $\mathcal{E}_\mathrm{tot}$. For adiabatic flow in which the radiation and gas temperatures are the same, we also require local conservation of $\tau$. Above we have manipulated equations (\ref{momenta}), (\ref{total_gas_energy}), (\ref{radiation_energy}), and (\ref{internal_gas_energy}) into a form which highlights the conserved nature of these quantities. Now we develop the same equations, as well as equation (\ref{continuity}), in a form suitable for adaptation to the numerical method described in \S \ref{method_section}. 

Applying the cylindrical divergence operator in equation (\ref{rad}), the radial momentum equation is 
\begin{multline}
\label{rad2}
\ddt{s_R} + \seperatedivergence{ \left( s_R u_R + p \right)}{ p }{ s_R u_\phi}{ s_R u_z } + \Lambda_E  \frac{\partial}{\partial R} E_R \ = \ \\
-\rho \frac{\partial}{\partial R} \Phi  + \rho R \Omega^2 + 2 \rho u_\phi \Omega + \frac{\rho u_\phi^2}{R} + \frac{p}{R}.
\end{multline}
The second, third, and fourth terms on the RHS come from the last three terms on the RHS of equation (\ref{rad}). The pressure term on the LHS can be written $\frac{1}{R}\frac{\partial}{\partial R} R p = \frac{\partial}{\partial R} p + \frac{p}{R}$, resulting in a $\frac{p}{R}$ on both sides of the equation. Though these terms will analytically cancel one another, as noted by \cite{CTL2010}, there is no guarantee they will numerically cancel. Because the LHS of our equations will be handled by an explicit advection scheme, and the RHS will be treated by other methods, we remove the $\frac{p}{R}$ term on both sides of the equation. Using equation (\ref{inertial}), we can write the centrifugal, Coriolis, and remaining coordinate curvature terms as a single term, $\frac{l_z^2}{\rho R^3}$. These changes to equation (\ref{rad2}) are reflected in equation (\ref{n2}).

For our total gas energy equation we will follow equation (\ref{eloc}) and apply the advection scheme to the quantity $\mathcal{E}_\mathrm{loc}$. This is the quantity that is physically transported by advection. As we will show below, treating the gas energy equation in this manner results in numerical conservation of $\mathcal{E}_\mathrm{tot}$ in the diffusion limit. 

The full set of equations, in a form suitable for adaptation to our numerical method, is:
\begin{equation}
\label{n1}
 \ddt{\rho} + \cylindricaldivergence{\rho u_R}{\rho u_\phi}{\rho u_z} \ = \ 0 
\end{equation}
\begin{equation}
\label{n2}
\ddt{s_R} + \cylindricaldivergence{ s_R u_R}{ s_R u_\phi}{ s_R u_z } + \frac{\partial}{\partial R} p + \Lambda_E  \frac{\partial}{\partial R} E_R \ = \
 -\rho \frac{\partial}{\partial R} \Phi  + \frac{l_z^2}{\rho R^3};
\end{equation}
\begin{equation}
\label{n3}
\ddt{l_z} + \cylindricaldivergence{ l_z u_R}{ \left( l_z u_\phi + R p \right) }{ l_z u_z  }  + \Lambda_E \frac{\partial}{\partial \phi } E_R \ = \ 
  -\rho \frac{\partial}{\partial \phi} \Phi;
\end{equation}
\begin{equation}
\label{n4}
\ddt{s_z} + \cylindricaldivergence{ s_z u_R}{ s_z u_\phi}{ \left( s_z u_z + p \right) } + \Lambda_E \frac{\partial}{\partial z} E_R \ = \
-\rho \frac{\partial}{\partial z} \Phi;
\end{equation}
\begin{multline}
\label{n5}
\ddt{ \mathcal{E}_\mathrm{loc} } - \rho \ddt{ \Phi } + 
	\frac{1}{R} \frac{\partial}{\partial R} R { \left( \mathcal{E}_\mathrm{loc} + p \right) u_R} + \frac{1}{R} \frac{\partial}{\partial \phi} { \left( \mathcal{E}_\mathrm{loc} + p \right) u_\phi} +  \frac{\partial}{\partial z} { \left( \mathcal{E}_\mathrm{loc} + p \right) u_z} + \Lambda_E \left( \mathbf{u} \cdot  \gradient{}\right) E_R \ = \\ 
- 4 \pi \kappa_p B_p + c \kappa_E E_R; 
\end{multline}
\begin{equation}
\label{n6}
\ddt{ E_R } + \cylindricaldivergence{ E_R u_R}{ E_R u_\phi}{ E_R u_z }  + \gradient{\mathbf{u}} : \mathbf{P} \ = \ 4 \pi \kappa_p B_p - c \kappa_E E_R -\divergence{ \mathbf{F} }; 
\end{equation}
\begin{equation}
\label{n7}
\ddt{ \tau } + \cylindricaldivergence{ \tau u_R }{ \tau u_\phi }{ \tau u_z } \ = \
- \frac{4 \pi \kappa_p B_p}{\gamma \tau^{\gamma - 1}} + \frac{c \kappa_E E_R}{\gamma \tau^{\gamma - 1}} .
\end{equation}
On the LHS we have placed terms which are handled by the explicit advection scheme. The gravity terms on the RHS of equations (\ref{n2}), (\ref{n3}), and (\ref{n4}) are computed with a first-order explicit differencing scheme, as is the last term of equation (\ref{n2}).  The remaining, radiation related terms on the RHS of equations (\ref{n5}), (\ref{n6}), and (\ref{n7}) are evolved in a separate implicit step.

\section{Numerical Method}
\label{method_section}
Our method is designed to evolve six independent variables in time on a cylindrical mesh rotating with constant and uniform angular frequency $\Omega$ about the coordinate axis: the mass density, $\rho$, the \textit{inertial} frame z-angular momentum density, $l_z$, the radial momentum density, $s_R$, the vertical momentum density, $s_z$, the gas energy density, $E_G$, and the radiation energy density, $E_R$.  The Newtonian gravitational potential, $\Phi$, is solved at each time (sub) step. We evolve a single auxiliary variable, the entropy tracer, $\tau$.

\subsection{Explicit Advection Scheme}
\label{explicitadvect}
We begin our discussion of the explicit advection scheme by applying it to equations (\ref{n1}) through (\ref{n7}) in the limit that $G=0$, $\Omega=0$, $\kappa_E=0$, $\kappa_p=0$, and $\Lambda_E=0$. This has the effect of removing gravitational, rotational, and all radiation terms except the advection of $E_R$. We will  denote the time derivatives of the evolution variables in this limit by prefacing them with $\displaystyle \lim_{\mathrm{fluid}}$.
\begin{eqnarray}
\label{m1}
\displaystyle \lim_{\mathrm{fluid}} \ddt{\rho} + \cylindricaldivergence{\rho u_R}{\rho u_\phi}{\rho u_z} & = & 0;\\
\label{m2}
\displaystyle \lim_{\mathrm{fluid}} \ddt{s_R} + \seperatedivergence{ s_R u_R}{ p }{ s_R u_\phi}{ s_R u_z } & = &  \frac{l_z^2}{\rho R^3};\\
\label{m3}
\displaystyle \lim_{\mathrm{fluid}} \ddt{l_z} + \cylindricaldivergence{ l_z u_R}{ \left( l_z u_\phi + R p \right) }{ l_z u_z  } & = & 0;\\ 
\label{m4}
\displaystyle \lim_{\mathrm{fluid}} \ddt{s_z} + \cylindricaldivergence{ s_z u_R}{ s_z u_\phi}{ \left( s_z u_z + p \right) } & = & 0;\\
\label{m5}
\displaystyle \lim_{\mathrm{fluid}} \ddt{ E_G } + \cylindricaldivergence{ \left( E_G + p \right) u_R}{ \left(E_G + p \right) u_\phi}{ \left( E_G + p \right) u_z} R & = & 0; \\
\label{m6}
\displaystyle \lim_{\mathrm{fluid}} \ddt{ E_R } + \cylindricaldivergence{ E_R u_R}{ E_R u_\phi}{ E_R u_z } & = & 0;\\
\label{m7}
\displaystyle \lim_{\mathrm{fluid}} \ddt{ \tau } + \cylindricaldivergence{ \tau u_R }{ \tau u_\phi }{ \tau u_z } & = & 0.
\end{eqnarray}
Note that $\displaystyle \lim_{\mathrm{fluid}} \mathcal{E}_\mathrm{loc} = \displaystyle \lim_{\mathrm{fluid}}  E_G$. 

The Kurganov-Tadmor (K-T) method (\cite{KT2000}), is a high resolution Godunov type central advection scheme that can be used to solve three-dimensional hyperbolic sets of first-order differential equations of the form
\begin{equation}
\label{form}
\frac{\partial}{\partial t} V + \sum_{i=1}^3 { \frac{\partial}{\partial x_i} \func{H}{V} } \ = \ 0,
\end{equation}
where $V = \left\{ V_0 ... V_N \right\}$ is a set of $N$ conserved quantities and  $\func{H}{V} = \left\{ \func{H_0}{V} ... \func{H_N}{V} \right\}$ a set of $N$ fluxes dependent only on $V$. For the solution to be stable, the matrix ${\partial H}/{\partial V}$ must be hyperbolic. Like previous advection schemes such as the Lax-Friedrichs (\cite{L1954}, \cite{F1954}) and the Nessyahu-Tadmor schemes (\cite{N1990}), the K-T method does not require the use of (approximate) Riemann solvers. It is thus computationally more efficient than Riemann solver based methods. Unlike previous central schemes, the K-T method does not suffer from excessive spatial averaging of the solution, or ``smearing". The K-T method can also be stated in a semi-discrete form, with discretized space and continuous time, allowing it to be coupled to a number of suitable time integration schemes.

Due to the use of cylindrical coordinates, equations (\ref{m1}) through (\ref{m7}) do not quite follow the form of equation (\ref{form}). Instead these equations are of the general form 
\begin{equation}
\label{form2}
\displaystyle \lim_{\mathrm{fluid}} \frac{\partial}{\partial t} V +
\frac{1}{R}\frac{\partial}{\partial R} R \func{H^R}{V} +
\frac{\partial}{\partial R} \func{G}{V} +
\frac{1}{R}\frac{\partial}{\partial \phi} \func{H^\phi}{V} + 
\frac{\partial}{\partial z} \func{H^z}{V} 
\ = \ \func{S}{R,V},
\end{equation}
where $\func{S}{R,V}$ refers to coordinate curvature terms that result from the application of the cylindrical divergence operator. For our particular set of equations,
\begin{multline}
V = \left[\begin{array}{c}
\rho \\
s_R \\
l_z \\
s_z \\
E_G \\
E_R \\
\tau
\end{array}
\right] \ , \
G = \left[\begin{array}{c}
0 \\
p \\
0 \\
0 \\
0 \\
0 \\
0
\end{array}
\right] \ , \ 
S = \left[\begin{array}{c}
0 \\
\frac{l_z^2}{\rho R^3} \\
0 \\
0 \\
0 \\
0 \\ 
0
\end{array}
\right] \ , \
\\ 
H^R = \left[\begin{array}{c}
\rho u_R \\
s_R u_R \\
l_z u_R \\
s_z u_R \\
\left(E_G + p\right) u_R \\
E_R u_R \\
\tau u_R 
\end{array}
\right] \ , \ 
H^\phi = \left[\begin{array}{c}
\rho u_\phi \\
s_R u_\phi \\
l_z u_\phi + R p \\
s_z u_\phi \\
\left(E_G + p\right) u_\phi \\
E_R u_\phi \\
\tau u_\phi 
\end{array}
\right] \ , \ \mathrm{and} \ 
H^z = \left[\begin{array}{c}
\rho u_z \\
s_R u_z \\
l_z u_z \\
s_z u_z + p \\
\left(E_G + p\right) u_z \\
E_R u_z \\
\tau u_z
\end{array}
\right] \ .
\end{multline}

Application of the central-upwind method of \cite{KP2001} to a two-dimensional curvilinear coordinate system is discussed in
\cite{ID2009}. The K-T method differs from that of \cite{KP2001} in that the latter is genuinely multi-dimensional: it requires reconstruction of the evolved variables at cell vertices (edges) as well at cell edges (faces). The multidimensional K-T method is simply the sum of the one-dimensional K-T method applied to each dimension. We choose the simpler method because it requires reconstruction only at six cell faces per cell, as opposed to six faces and and twelve edges, and is thus computationally simpler and more efficient. The key disadvantage is that the multi-dimensional K-T method at most delivers second-order spatial accuracy, regardless of the order of the one-dimensional reconstruction. 

In order to express the K-T method in a more compact form, we define the discrete divergence operator
\begin{multline}
\mathcal{D}\left\{ \func{H}{V} \right\}_{j k l } := \\
  \frac{1}{2 R_j \Delta} \left( 
                      R_{j+\frac{1}{2}}\left(\func{H^R}{V^+_{j+\frac{1}{2} k l}}+\func{H^R}{V^-_{j+\frac{1}{2} k l}}\right) -
                      R_{j+\frac{1}{2}}\left(\func{H^R}{V^+_{j-\frac{1}{2} k l}}+\func{H^R}{V^-_{j-\frac{1}{2} k l}}\right)
\right) \\
+  \frac{1}{2 R_j \Delta} \left(  
                      \func{H^\phi}{V^+_{j k+\frac{1}{2} l}}+\func{H^\phi}{V^-_{j k+\frac{1}{2} l}} -
                      \func{H^\phi}{V^+_{j k-\frac{1}{2} l}}-\func{H^\phi}{V^-_{j k-\frac{1}{2} l}}
\right)  \\
+  \frac{1}{2 \Delta} \left(  
                      \func{H^z}{V^+_{j k l+\frac{1}{2}}}+\func{H^z}{V^-_{j k l+\frac{1}{2}}} -
                      \func{H^z}{V^+_{j k l-\frac{1}{2}}}-\func{H^z}{V^-_{j k l-\frac{1}{2}}}
\right)
\end{multline}
and the discrete viscosity operator
\begin{multline}
\mathrm{\mathcal{V}}\left\{ V \right\}_{j k l } := \\
\frac{1} {2 R_j \Delta } \left\{
	R_{j+\frac{1}{2}} a_{j+\frac{1}{2} k l} \left( V^+_{j+\frac{1}{2} k l} - V^-_{j+\frac{1}{2} k l} \right) -
	R_{j-\frac{1}{2}} a_{j-\frac{1}{2} k l} \left( V^+_{j-\frac{1}{2} k l} - V^-_{j-\frac{1}{2} k l} \right)
\right\}  \\
+  \frac{1} {2 R_j \Delta } \left\{
	a_{j k+\frac{1}{2} l} \left( V^+_{j k+\frac{1}{2} l} - V^-_{j k+\frac{1}{2} l} \right) -
	a_{j k-\frac{1}{2} l} \left( V^+_{j k-\frac{1}{2} l} - V^-_{j k-\frac{1}{2} l} \right)
\right\}  \\
+  \frac{1} {2\Delta } \left\{
	a_{j k l+\frac{1}{2}} \left( V^+_{j k l+\frac{1}{2}} - V^-_{j k l+\frac{1}{2}} \right) -
	a_{j k l-\frac{1}{2}} \left( V^+_{j k l-\frac{1}{2}} - V^-_{j k l-\frac{1}{2}} \right)
\right\}.
\end{multline}
The quantity $\Delta$ is the uniform spacing between grid zones. For our particular implementation, this spacing is the same for each dimension. It is trivial to modify the above expressions for a grid where the spacing is different for each dimension.
The quantities  $V^{\pm}_{j\pm\frac{1}{2} k l}$, $V^{\pm}_{j k\pm\frac{1}{2} l}$, and $V^{\pm}_{j k l\pm\frac{1}{2}}$ are the reconstructed values of $V$ at the faces located at $j\pm\frac{1}{2} k l$, $ j k\pm\frac{1}{2} l$, and $j k l\pm\frac{1}{2}$, respectively. The superscript indicates whether the value is on the left (-) or right (+) side of the face. A value for $V$ at each side of the cell face is required to account for discontinuities in the solution. Both of the above operators reduce to surface integrals when they are summed over a grid volume. Therefore when there is no net flow through such a surface, $V$ is numerically conserved. The signal speeds are defined as
\begin{equation}
a_{j\pm\frac{1}{2} k l} := 
\max \left\{ 
	\lambda_{\mathrm{max}}\left\{ \frac{\partial H^R}{\partial V}\Big{|}_{V = V^+_{j\pm\frac{1}{2} k l}}\right\},
	\lambda_{\mathrm{max}}\left\{ \frac{\partial H^R}{\partial V}\Big{|}_{V = V^+_{j\pm\frac{1}{2} k l}}\right\}
\right\},
\end{equation}
\begin{equation}
a_{j k\pm\frac{1}{2} k} := 
\max \left\{ 
	\lambda_{\mathrm{max}}\left\{ \frac{\partial H^\phi}{\partial V}\Big{|}_{V = V^+_{j k\pm\frac{1}{2} k}}\right\},
	\lambda_{\mathrm{max}}\left\{ \frac{\partial H^\phi}{\partial V}\Big{|}_{V = V^+_{j k\pm\frac{1}{2} k}}\right\}
\right\}
\end{equation}
and
\begin{equation}
a_{j k l\pm\frac{1}{2}} := 
\max \left\{ 
	\lambda_{\mathrm{max}}\left\{ \frac{\partial H^z}{\partial V}\Big{|}_{V = V^+_{j k l\pm\frac{1}{2}}}\right\},
	\lambda_{\mathrm{max}}\left\{ \frac{\partial H^z}{\partial V}\Big{|}_{V = V^+_{j k l\pm\frac{1}{2}}}\right\}
\right\}, 
\end{equation}
where $\lambda_{\mathrm{max}}\left\{ A \right\}$ is the spectral radius operator. For equations (\ref{m1}) through (\ref{m7}),
\begin{equation}
\lambda_{\mathrm{max}}\left\{ \frac{\partial G^n}{\partial V}\right\} = \left| u_n \right| + \sqrt{\frac{\gamma p}{\rho}},
\end{equation}
where $u_n$ refers to one of the vector components of $\mathbf{u}$. Note that for brevity we omit the  dependence on signal speeds in writing $\mathrm{\mathcal{V}}\left\{ V \right\}_{j k l }$. To ensure numerical stability when using an explicit time integrator, the Courant-Friedrichs-Lewy condition (CFL) (\cite{CFL1967}) must be satisfied for the chosen time step. For the K-T method in cylindrical coordinates, this condition is
\begin{equation}
	\Delta t \ \le \ \frac{1}{2} \min_{\mathrm{all} \ j k l} \left\{ \frac{\Delta }{a_{j\pm\frac{1}{2} k l}}, \frac{R \Delta }{a_{j k\pm\frac{1}{2} l}}, \frac{\Delta }{a_{j k l\pm\frac{1}{2}}} \right\}
\end{equation} We also define the non conservative radial component of the discrete gradient,
\begin{equation}
\mathcal{G_R}\left\{ \func{G}{V} \right\}_{j k l } := 
  \frac{1}{2 \Delta} \left(  
                      \func{G}{V^+_{j+\frac{1}{2} k l}}+\func{G}{V^-_{j+\frac{1}{2} k l}} -
                      \func{G}{V^+_{j-\frac{1}{2} k l}}-\func{G}{V^-_{j-\frac{1}{2} k l}}
\right).
\end{equation}

The semi-discrete form of the K-T method in three-dimensional cylindrical coordinates can now be written as
\begin{equation}
\label{method}
\displaystyle \lim_{\mathrm{fluid}} \frac{d}{d t} V_{j k l} + \mathrm{D}\left\{ \func{H}{V} \right\}_{j k l }  + \mathrm{G_R}\left\{ \func{G}{V} \right\}_{j k l } - S_{j k l} \ = \ \mathrm{\mathcal{V}}\left\{ V \right\}_{j k l }.
\end{equation}
Note that here,  and for the remainder of this paper, $\frac{d}{dt}$ does {\it not} refer to the Lagrangian time derivative, $\frac{\partial}{\partial t} + \mathbf{u} \cdot \nabla$, but instead refers to the total time rate of change of the value of a quantity within a grid cell.  The  LHS of equation (\ref{method}) contains the numerical representation of the physical flux components and coordinate curvature terms. The terms on the RHS are, in effect,  artificial viscosity terms. These unphysical terms are required for numerical stability. The magnitude of the viscosity grows larger with increasing maximum signal speed as well as with increasing difference between the values of $V$ at left and right sides of cell faces. When the values of $V$ are identical on both sides of a cell face, the viscosity term at that face becomes zero. 

\begin{figure}
\begin{center}
\includegraphics[angle=-90,scale=0.5]{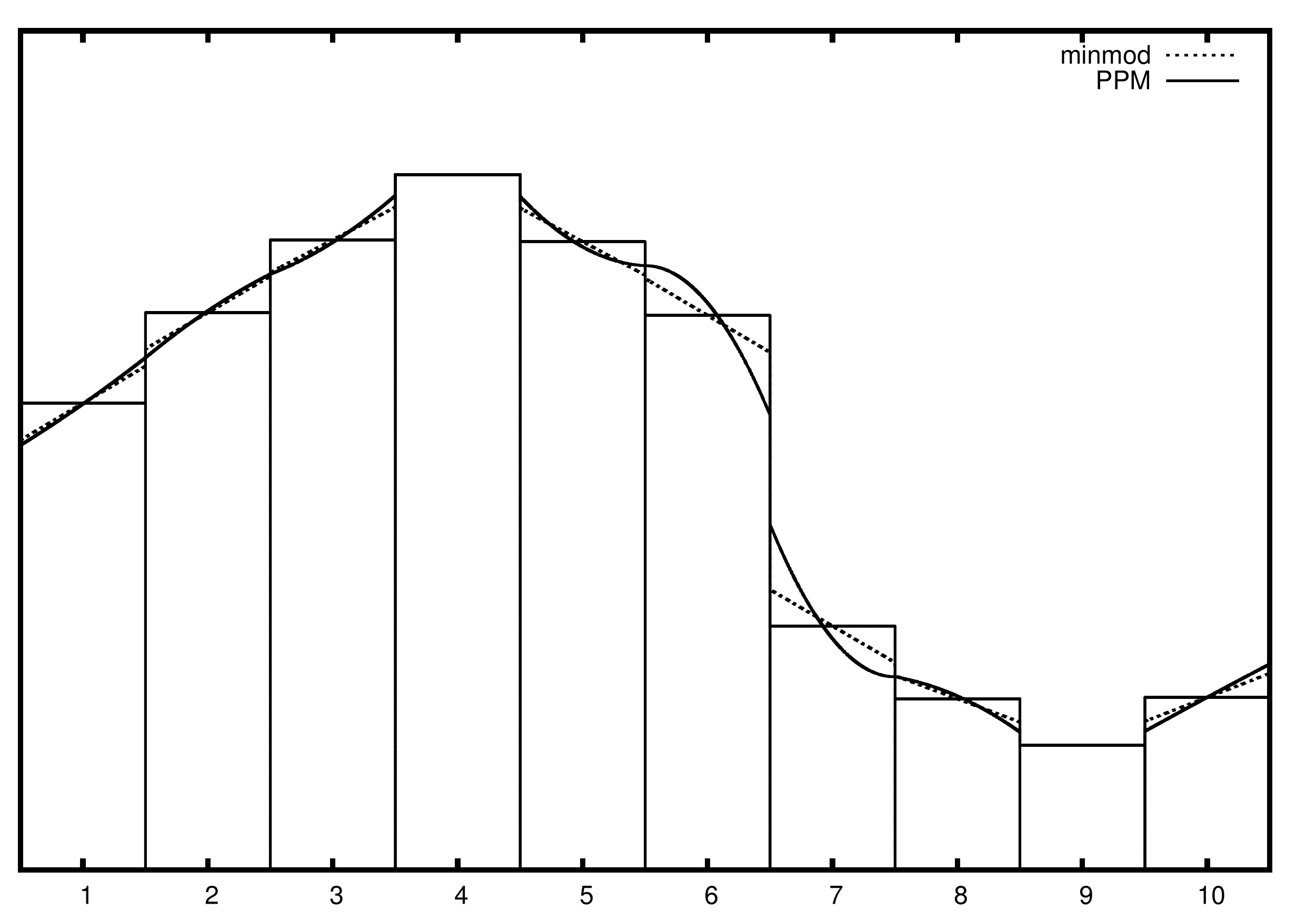}
\caption[Comparison of minmod to PPM reconstruction schemes]{In this comparison of minmod ($\theta = 1$) and PPM reconstruction schemes, the PPM reconstruction  has unequal face values (resulting in artificial viscosity) only at the faces of the  two extrema (the fourth and ninth cells from the left) and at the face of the discontinuity (between the sixth and seventh cells from the left). The minmod reconstruction  has unequal face values at numerous other locations and the difference in face values is larger than PPM for all faces.}
\label{recon1}
\end{center}
\end{figure}

We use the one-dimensional piecewise parabolic (PPM) reconstruction of \cite{CW1984} to compute the cell face values $V^{\pm}_{j\pm\frac{1}{2} k l}$, $V^{\pm}_{j k\pm\frac{1}{2} l}$, and $V^{\pm}_{j k l\pm\frac{1}{2}}$. Although the reconstruction is third order in one dimension, because we do not use a genuinely multidimensional method, the reconstruction reduces to second-order accuracy. However, we still retain another advantage of using a high order reconstruction: in sufficiently smooth regions, left and right face values will be exactly equal to one another, completely eliminating the artificial viscosity. In Figure \ref{recon1} 
we compare the PPM reconstruction to the ubiquitous minmod linear reconstruction. For PPM, the left and right face values differ only at the two extrema and at the discontinuity, whereas the minmod reconstruction has unequal face values at numerous locations. Rather than applying the reconstruction to the conserved variables, as in \cite{KT2000}, we apply the reconstruction to the variables $\rho$, $\frac{s_R}{\rho}$, $\frac{s_z}{\rho}$, $\frac{l_z}{\rho}$, $\frac{\tau}{\rho}$, $\frac{E_G}{\rho}$, and $\frac{E_R}{\rho}$, and then transform these quantities back to the conserved variables. Reconstructing face values in this manner has two advantages: (1) the velocity values obtained at cell faces transform correctly under a Galilean transformation and (2) the magnitude of the sound speed and velocity at cell faces will not exceed their respective values at cell centers. 
	
As mentioned above, we evolve the entropy tracer, $\tau$, independently of $E_G$. When the internal energy is a small fraction of  $E_G$, the expression
\begin{equation}
\label{internalfromtotal}
e = \left(E_G-\frac{1}{2}\rho u^2\right)
\end{equation}
can suffer from numerical difficulties. If the minuend and subtrahend of a difference are nearly equal, the result can lose significant numerical precision when determined by a computer. To account for this, we use the dual energy formalism of \cite{BNSO1995}. The pressure is computed according to
\begin{equation}
p = \left\{\begin{array}{cc}
\left( \gamma - 1 \right) \left(E_G - \frac{1}{2}\rho u^2\right) & \mathrm{if} \left(E_G - \frac{1}{2}\rho u^2\right) > \epsilon_1  E_G \\
\left( \gamma - 1 \right) \tau^\gamma & \mathrm{else} \\ 
\end{array}\right.,
\end{equation}
where $0 < \epsilon_1 \ll 1$. Additionally, at the end of each computational time step, the entropy tracer is updated according to
\begin{equation}
\tau \rightarrow \left\{\begin{array}{cc}
 \left(E_G - \frac{1}{2}\rho u^2\right)^\frac{1}{\gamma} & \mathrm{if} \left(E_G - \frac{1}{2}\rho u^2\right) > \epsilon_2 E_G \\
\tau & \mathrm{else} \\ 
\end{array}\right.,
\end{equation}
where $0 < \epsilon_1 < \epsilon_2 \ll 1$. For the simulations discussed in this paper we use $\epsilon_1 = 0.001$ and $\epsilon_2 = 0.1$.

\subsection{Extension to Gravity}
\label{gravitysection}

We now extend the K-T method in cylindrical coordinates to include a potential formed by Newtonian gravitation and/or rotation. We take equations (\ref{n1}) through (\ref{n7}) in the limit that $\kappa_E=0$, $\kappa_p=0$, and $\Lambda_E=0$. We refer to this limit  by prefacing time derivatives with $\displaystyle \lim_{\mathrm{grav}}$.
\begin{equation}
\label{o1}
\displaystyle \lim_{\mathrm{grav}} \ddt{\rho} + \cylindricaldivergence{\rho u_R}{\rho u_\phi}{\rho u_z} \ = \ 0 ; 
\end{equation}
\begin{equation}
\label{o2} 
\displaystyle \lim_{\mathrm{grav}} \ddt{s_R} + \seperatedivergence{ s_R u_R}{ p }{ s_R u_\phi}{ s_R u_z } \ = \ 
-\rho \frac{\partial}{\partial R} \Phi  + \frac{l_z^2}{\rho R^3}; 
\end{equation}
\begin{equation}
\label{o3}
\displaystyle \lim_{\mathrm{grav}} \ddt{l_z} + \cylindricaldivergence{ l_z u_R}{ \left( l_z u_\phi + R p \right) }{ l_z u_z  }  \ = \ 
-\rho \frac{\partial}{\partial \phi} \Phi; 
\end{equation}
\begin{equation}
\label{o4}
\displaystyle \lim_{\mathrm{grav}} \ddt{s_z} + \cylindricaldivergence{ s_z u_R}{ s_z u_\phi}{ \left( s_z u_z + p \right) }\ = \ 
-\rho \frac{\partial}{\partial z} \Phi; 
\end{equation}
\begin{equation}
\label{o5}
\displaystyle \lim_{\mathrm{grav}} \ddt{ \mathcal{E}_\mathrm{loc} } - \rho \ddt{ \Phi  } + \cylindricaldivergence{ \left( \mathcal{E}_\mathrm{loc} + p \right) u_R}{ \left( \mathcal{E}_\mathrm{loc} + p \right) u_\phi}{ \left( \mathcal{E}_\mathrm{loc} + p \right) u_z}  \ = \  0; 
\end{equation}
\begin{equation}
\label{o6}
\displaystyle \lim_{\mathrm{grav}} \ddt{ E_R } + \cylindricaldivergence{ E_R u_R}{ E_R u_\phi}{ E_R u_z } \ = \ 0; 
\end{equation}
\begin{equation}
\label{o7}
\displaystyle \lim_{\mathrm{grav}} \ddt{ \tau } + \cylindricaldivergence{ \tau u_R }{ \tau u_\phi }{ \tau u_z } \ = \ 0. 
\end{equation}

To solve equation (\ref{poisson}) for the gravitational potential, $\Phi$, we solve the discrete equation
\begin{multline}
\label{poisson_discrete}
R_{j+\frac{1}{2}} \Phi_{j+1 k l} + R_{j-\frac{1}{2}} \Phi_{j-1 k l} + 
 \Phi_{j k+1 l} +  \Phi_{j k-1 l} + \\
R_j \Phi_{j k l+1} + R_j \Phi_{j k l-1} - \left(4 R_j + 2\right) \Phi_{j k l} \ = \  4 \pi G R_j \Delta^2 \rho,
\end{multline}
using a conjugate gradient solver for the interior (\cite{H1952}). The boundary cell values for $\Phi$ are computed using the solver of \cite{CT1999}.

Equation (\ref{o5}) differs from equation (\ref{m5}) only in the addition of an extra time derivative term, $\rho \ddt{\Phi}$, and the use of $\mathcal{E}_\mathrm{loc}$ in place of $E_G$. The semi-discrete form of this equation is
\begin{equation}
\label{l001}
\displaystyle \lim_{\mathrm{grav}} \frac{d}{d t}{\mathcal{E}_{\mathrm{loc}, j k l}} - \rho_{j k l} \frac{d}{d t}{\Phi_{j k l}} + 
\mathcal{D}\left\{ \left( \mathcal{E}_\mathrm{loc} + p \right) \mathbf{u} \right\}_{j k l} \ = \
\mathcal{V} \left\{ \mathcal{E}_\mathrm{loc} \right\}_{j k l}.
\end{equation}
Because time is continuous in the semi-discrete form, we may use equation  (\ref{elocdef}) to rewrite  equation (\ref{l001}) as
\begin{equation}
\label{estar_}
\displaystyle \lim_{\mathrm{grav}} \frac{d}{d t}{E_{G,j k l}}
+ \mathcal{D}\left\{ \left( \mathcal{E}_\mathrm{loc} + p \right) \mathbf{u} \right\}_{j k l}  - \left[\Phi_{\mathrm{eff}}\right]_{j k l} \frac{d}{d t}{\rho_{j k l}}   \ = \
\mathcal{V} \left\{ \mathcal{E}_\mathrm{loc} \right\}_{j k l}.
\end{equation}
where $\Phi_\mathrm{eff} := \Phi + \Phi_\mathrm{rot}$ is the ``effective potential". Note that the term $\frac{d}{d t}{\rho_{j k l}}$ is obtained by applying equation (\ref{method}) to equation (\ref{o1}). The quantity $\mathcal{E}_\mathrm{con}$ will be nearly globally conserved under application of equation (\ref{estar_}). To prove this, we rewrite equation (\ref{l001}) as
\begin{equation}
\label{l002}
\displaystyle \lim_{\mathrm{grav}} \frac{d}{d t}{\mathcal{E}_{\mathrm{con}, j k l}} + \frac{1}{2} \left( \rho_{j k l} \frac{d}{d t}{\Phi_{j k l}} - \Phi_{j k l} \frac{d}{d t}{\rho_{j k l}} \right) + 
\mathcal{D}\left\{ \left( \mathcal{E}_\mathrm{loc} + p \right) \mathbf{u} \right\}_{j k l} \ = \
\mathcal{V} \left\{ \mathcal{E}_\mathrm{loc} \right\}_{j k l}.
\end{equation}
The discrete divergence and viscosity operators are conservative. Using equation (\ref{poisson_discrete}), we can rewrite the sum over volume of the middle term on the LHS,
\begin{multline}
\label{gsum}
\sum_{j k l } \frac{1}{2} \left( \rho_{j k l} \frac{d}{d t}{\Phi_{j k l}} - \Phi_{j k l} \frac{d}{d t}{\rho_{j k l}} \right) R_j \Delta^3 \ = \
\frac{ \Delta}{8 \pi G} 
\sum_{j k l } \left\{ \left(R_{j+\frac{1}{2}} \dot{\Phi}_{j+1 k l} + R_{j-\frac{1}{2}} \dot{\Phi}_{j-1 k l} + \right. \right. \\  \left. \left. \dot{\Phi}_{j k+1 l} +  \dot{\Phi}_{j k-1 l} - 
          R_j \left( \dot{\Phi}_{j k l+1} + \dot{\Phi}_{j k l-1} \right) - \left(4 R_j + 2\right) \dot{\Phi}_{j k l}\right) \Phi_{j k l}  \ - \ \right. \\
\left.
\left[R_{j-\frac{1}{2}} {\Phi}_{j-1 k l} + R_{j+\frac{1}{2}} {\Phi}_{j+1 k l} +  {\Phi}_{j k-1 l}  + {\Phi}_{j k+1 l} \right.\right.
         + \\
          \left.\left. R_j \left( {\Phi}_{j k l-1} +  {\Phi}_{j k l+1} \right) - \left(4 R_j + 2\right) {\Phi}_{j k l}\right] \dot{\Phi}_{j k l} \right\},
\end{multline}
where $\dot{\Phi}_{j k l} := \frac{d}{d t}{\Phi_{j k l}}$. 
Expression (\ref{gsum})  sums to zero for interior grid points, depending only on the values of $\Phi_{j k l}$ and $\dot{\Phi}_{j k l}$ along a two-cell-wide boundary at the surface of the computational grid. To be physically correct, these terms must also sum to zero (so long as there is no mass leaving the grid), however, this is not numerically guaranteed. The extent to which equation (\ref{poisson_discrete}) is numerically satisfied will also affect conservation. As shown below, these non-conservative effects are minimized when the center of mass of the system is coincident with the center of the coordinate system. 

\begin{figure}
\begin{center}
\includegraphics[angle=-90,scale=0.5]{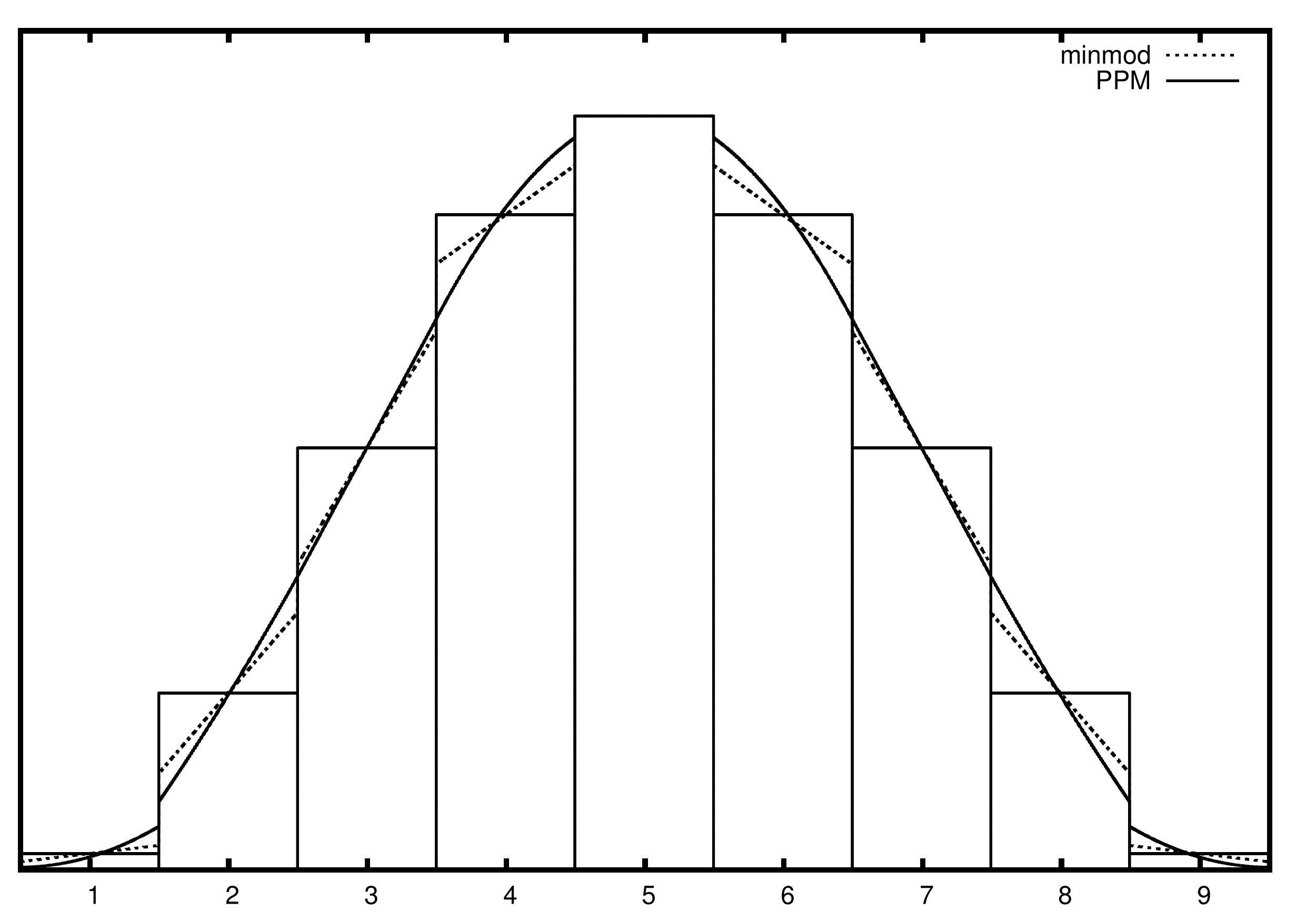}
\caption[PPM reconstruction of a polytrope]{ In this one-dimensional PPM reconstruction of the mass density of an $n = \frac{3}{2}$ polytrope, there are discontinuities in the reconstruction at the center and near the surface. Artificial viscosity will be applied to the cells next to these discontinuities. As a result, mass will move from the center cell to the two cells next to it, and from the second (eighth) cell to the first (ninth) cell.}
\label{recon2}
\end{center}
\end{figure}

The application of the viscosity operator, $\mathrm{\mathcal{V}}\left\{ V \right\}_{j k l }$, to the mass density, $\rho$, for any cell in which left and right face values are unequal (non-smooth regions), results in the flow of mass from cells of higher density to cells of lower density. In the absence of a potential, this will not alter energy conservation. When a potential force is applied, however, this non-physical movement of mass will violate energy conservation unless it is properly accounted for. To illustrate this effect, consider the one-dimensional PPM reconstruction of an equilibrium  $n=\frac{3}{2}$ polytrope in Figure (\ref{recon2}). 
Even with the PPM reconstruction, there are discontinuities in the reconstruction of $\rho$ at the faces of the center cell and at the outer cells. Application of the the K-T method will therefore cause mass to move up the gravitational potential, from the center cell to the surrounding cells and from the cells just below the surface cells of the star to the surface cells. If not properly accounted for, this added potential energy comes at no cost to either the kinetic energy or the internal gas energy of the fluid and $\mathcal{E}_\mathrm{con}$ will not be conserved. For models of gravitationally bound objects in near equilibrium, this effect will accumulate over time, and the object can become  gravitationally unbound and dissipate.  Equation (\ref{estar_}) accounts for this spontaneous potential energy generation by removing it from the total gas energy. Because we do not alter how the kinetic energy is calculated, the difference is effectively removed from the internal energy. Matter which moves up (down) a potential by  application of the viscosity operator will lose (gain) internal energy. This presents a problem for  a zero temperature fluid. With no internal energy to lose, application of equation (\ref{estar_}) will result in values for $E_G$ which yield negative internal energies under application of equation (\ref{internalfromtotal}). Regions with non-positive pressure have nothing to prevent them from collapsing due to their own gravity, leading to numerically unstable conditions. We avoid this issue by using the dual energy formalism, which guarantees a positive pressure so long as $\tau$ is positive.

	There are two applications of the viscosity operator  resulting from the potential energy in equation (\ref{estar_}). The $\frac{d}{d t}{\rho_{j k l}}$ term contains $\mathrm{\mathcal{V}}\left\{ \rho \right\}_{j k l }$. It is this term that cancels the spontaneous gains or losses in potential energy. The second source is the contribution from the potential energy to $\mathcal{V} \left\{ \mathcal{E}_\mathrm{loc} \right\}_{j k l}$. This term causes the correction in energy due to $\frac{d}{d t}{\rho_{j k l}}$  to flow with the fluid. Without it, this correction would be applied to the cell the fluid is flowing out of instead of to the cell into which it is flowing. We refer to these extra viscosity terms in the  energy equation as the ``E*" correction. Appendix \ref{energy_schemes} outlines two additional methods of treating the energy equation. These methods are compared to the ``E*" correction in some of the verification problems presented below.

Equations (\ref{o2}), (\ref{o3}), and (\ref{o4}) differ from equations (\ref{m2}), (\ref{m3}), and (\ref{m4}) in that they each contain a gravitational term on the RHS. We require that $l_z$ and $s_z$ be nearly conserved, therefore the numerical form of these terms for equations (\ref{o3}) and (\ref{o4}) must be numerically conservative. This can be accomplished using second-order differencing for $\gradient{\Phi}$ and the cell-centered values of $\rho$. The resulting equations for $l_z$ and $s_z$ are, respectively,
\begin{equation}
\label{m0000}
\displaystyle \lim_{\mathrm{grav}} \frac{d}{d t}{l_z}
+ \mathcal{D}\left\{ l_z + \mathbf{u} + p \mathbf{\hat{\phi}} \right\}_{j k l}  + \rho_{j k l}\frac{1}{2 \Delta}\left(\Phi_{j k+1 l}-\Phi_{j k-1 l}\right) \ = \
\mathcal{V} \left\{l_z \right\}_{j k l}
\end{equation}
and 
\begin{equation}
\label{m0001}
\displaystyle \lim_{\mathrm{grav}} \frac{d}{d t}{s_z}
+ \mathcal{D}\left\{ s_z \mathbf{u}  + p  \mathbf{\hat{z}} \right\}_{j k l}  + \rho_{j k l}\frac{1}{2 \Delta}\left(\Phi_{j k l-1}-\Phi_{j k l-1}\right) \ = \
\mathcal{V} \left\{s_z \right\}_{j k l}.
\end{equation}
Just as with the total energy equation, by using equation (\ref{poisson_discrete}) to remove $\rho_{j k l}$ from the gravitational terms, it is possible to show that equations (\ref{m0000}) and (\ref{m0001}) nearly conserve angular and vertical momentum over the interior of the computational grid. There is a similar non-conservative effect from the limited precision of the Poisson solver and numerical boundary conditions. Using the same second-order differencing, the radial momentum equation is 
\begin{equation}
\label{m000}
\displaystyle \lim_{\mathrm{grav}} \frac{d}{d t}{s_R}
+ \mathcal{D}\left\{  s_R \mathbf{u} \right\}_{j k l} + \mathcal{G_R}\left\{ p \right\}_{j k l } 
 + \rho_{j k l}\frac{1}{2 \Delta}\left(\Phi_{j+1 k l}-\Phi_{j-1 k l}\right) - \frac{l_{z,j k l}^2}{\rho_{j k l} R_{j}^3} \ = \
\mathcal{V} \left\{s_R \right\}_{j k l}.
\end{equation}

 Without the addition of gravity, the one-dimensional Kurganov-Tadmor method will satisfy the maximum principle. Positive scalars at $t=0$ remain positive throughout the evolution. With multiple dimensions, it is also  possible to satisfy the maximum principle by using a small enough time-step (see \S 5 of \cite{KT2000}). The addition of gravity complicates matters, however, and we have not found a general method to guarantee that positive scalars remain positive without using overly small time-steps. There are four evolved quantities which are physically expected to be positive: the mass density, $\rho$; the entropy tracer, $\tau$; the radiation energy density, $E_R$; and the total gas energy density, $E_G$. We adopt ``floor" values for the first three of these quantities. At the beginning of each time sub-step and for each grid cell, 
$\rho$, $\tau$, and $E_R$ are all set to the maximum of themselves or a predefined floor value. The total gas energy, $E_G$, is not altered. When it is negative, the dual energy formalism will use $\tau$ to determine the internal energy. The floor values we use are simulation dependent. Gravity also has the potential to complicate the CFL requirement. For our particular purposes, we have found this to only be a problem during the initial stages of the evolution. Our simulations generally begin with zero or near-zero velocities. If the time-step is limited to only the CFL time-step limit, velocities (especially in low density regions) can grow very large within the first time-step, leading to immediate numerical instability. For our evolutions with gravity, we begin the evolution with a time-step size that has empirically been shown to not lead to immediate instability. Then we gradually increase the time-step size over the next several hundred time-steps, until it is equal to the maximum imposed by the CFL condition. 

\subsection{Radiation Transport - Explicit Step} 
\label{rad_explicit_section}
In a manner similar to \cite{KKMB2007}, we take equations (\ref{n1}) through (\ref{n7}) and split them into explicit and implicit parts,
\begin{equation}
\label{form3}
\frac{\partial}{\partial t} V + \mathbf{q}_\mathrm{explicit} = \mathbf{q}_\mathrm{implicit},
\end{equation}
where
\begin{multline}
\mathbf{q}_\mathrm{explicit} = \\ 
\\
\left[\begin{array}{l} \cylindricaldivergence{\rho u_R}{\rho u_\phi}{\rho u_z} \\
\seperatedivergence{ s_R u_R}{ p }{s_R u_\phi}{s_R u_z} + \Lambda_E  \frac{\partial}{\partial R} E_R + \rho \frac{\partial}{\partial R} \Phi  - \frac{l_z^2}{\rho R^3} \\
\cylindricaldivergence{ l_z u_R}{ \left( l_z u_\phi + R p \right)}{l_z u_z} + \Lambda_E \frac{\partial}{\partial \phi } E_R + \rho \frac{\partial}{\partial \phi} \Phi \\
\cylindricaldivergence{ s_z u_R}{ s_z u_\phi}{ \left( s_z u_z + p \right) } + \Lambda_E \frac{\partial}{\partial z} E_R  + \rho \frac{\partial}{\partial z} \Phi \\
\cylindricaldivergence{\left( \mathcal{E}_\mathrm{loc} + p \right) u_R}{ \left( \mathcal{E}_\mathrm{loc} + p \right) u_\phi}{ \left( \mathcal{E}_\mathrm{loc} + p \right) u_z} + \Lambda_E \left( \mathbf{u} \cdot  \gradient{}\right) E_R  - \rho \ddt{ \Phi } \\
 \cylindricaldivergence{ E_R u_R}{ E_R u_\phi}{ E_R u_z }  + \gradient{\mathbf{u}} : \mathbf{P} \\
 \cylindricaldivergence{ \tau u_R }{ \tau u_\phi }{ \tau u_z }
\end{array}
\right],
\end{multline}
and
\begin{equation}
\mathbf{q}_\mathrm{implicit} = \left[\begin{array}{c} 
0 \\
0 \\
0 \\
0 \\
-4 \pi \kappa_p B_p + c \kappa_E E_R \\
-\divergence{ \mathbf{F} }  + 4 \pi \kappa_p B_p - c \kappa_E E_R \\
\frac{1}{\gamma \tau^{\gamma - 1}} \left(- 4 \pi \kappa_p B_p + c \kappa_E E_R \right) \\
\end{array}
\right].
\end{equation}
The expression $\mathbf{q}_\mathrm{explicit}$ contains the terms in the $\displaystyle \lim_{\mathrm{grav}}$ case described in \S \ref{gravitysection}, as well as $\Lambda_E \gradient{E_R}$ terms in the momentum equations,  the $\Lambda_E \mathbf{u} \cdot \gradient{E_R}$ term in the total gas energy equation, and  the $\mathbf{P} : \gradient{\mathbf{u}}$ term in the radiation energy equation. These terms are calculated using cell-centered quantities, $V_{i j k }$, and the first-order differences, 
\begin{equation}
\label{diff0}
\frac{1}{2}\left(\left(V^-_{j+\frac{1}{2} k l } + V^+_{j+\frac{1}{2} j k }\right) -\left(V^-_{j-\frac{1}{2} j k } + V^+_{j-\frac{1}{2} k l }\right)\right),
\end{equation}
for derivatives in the radial direction,
\begin{equation}
\label{diff1}
\frac{1}{2}\left(\left(V^-_{j k+\frac{1}{2} l } + V^+_{j k+\frac{1}{2} l }\right) -\left(V^-_{j k-\frac{1}{2} l } + V^+_{j k-\frac{1}{2} l }\right)\right),
\end{equation} 
for derivatives in the azimuthal direction, and 
\begin{equation}
\label{diff2}
\frac{1}{2}\left(\left(V^-_{j k l+\frac{1}{2} } + V^+_{j k l+\frac{1}{2} }\right) -\left(V^-_{j k l-\frac{1}{2} } + V^+_{j k l-\frac{1}{2} }\right)\right),
\end{equation} for derivatives in the vertical direction. With radiation, the  characteristic  speeds are calculated using
\begin{equation}
\label{rada}
a_i = |u_i| + \sqrt{\frac{\gamma p + \left(f_{i i} + 1\right) \Lambda_E E_{R}}{\rho}}.
\end{equation}
The subscripts for $a$, $u$, and $f$ refer to the $i^{th}$ vector and $i i^{th}$ tensor component of those quantities. This equation is exact in the diffusion limit, where $f_{i i} \rightarrow \frac{1}{3}$ and $\Lambda_E \rightarrow \frac{1}{3}$. In the free-streaming limit, equation (\ref{rada}) is only an approximation. It was chosen so that as $\Lambda_E$ goes to zero, the contribution of radiation to the sound speed also goes to zero. 

To compute the explicit step, we solve equation (\ref{form3})  with $\mathbf{q}_\mathrm{implicit}$  set to zero. The solution is computed this way over all time sub-steps of the integration. The set of semi-discrete equations is:
\begin{equation}
\label{discrete1}
\frac{d}{d t} \rho
+ \mathcal{D}\left\{ \rho \mathbf{u} \right\}_{j k l} \ = \ \mathcal{V} \left\{\rho \right\}_{j k l};
\end{equation}
\begin{multline}
\label{discrete2}
\frac{d}{d t}{s_R}
+ \mathcal{D}\left\{  s_R \mathbf{u} \right\}_{j k l} + \mathcal{G_R}\left\{ p \right\}_{j k l } 
 + \rho_{j k l}\frac{1}{2 \Delta}\left(\Phi_{j+1 k l}-\Phi_{j+1 k l}\right) + \\
 + \Lambda_{j k l}\frac{1}{\Delta}\left(E_{R,j+\frac{1}{2} k l}-E_{R,j-\frac{1}{2} k l}\right) - \frac{l_{z,j k l}^2}{\rho_{j k l} R_{j}^3} \ = \
\mathcal{V} \left\{s_R \right\}_{j k l};
\end{multline}
\begin{multline}
\label{discrete3}
\frac{d}{d t}{l_z}
+ \mathcal{D}\left\{  l_z \mathbf{u} \mathbf{u}  + p \mathbf{\hat{\phi}} \right\}_{j k l}
 + \rho_{j k l}\frac{1}{2 \Delta}\left(\Phi_{j k+1 l}-\Phi_{j k-1 l}\right) + \\
 + \Lambda_{j k l}\frac{1}{ \Delta}\left(E_{R,j k+\frac{1}{2}  l}-E_{R,j k-\frac{1}{2} l}\right) \ = \
\mathcal{V} \left\{l_z \right\}_{j k l};
\end{multline}
\begin{multline}
\label{discrete4}
\frac{d}{d t}{s_z}
+ \mathcal{D}\left\{  s_z \mathbf{u} \mathbf{u} + p \mathbf{\hat{z}} \right\}_{j k l} 
 + \rho_{j k l}\frac{1}{2 \Delta}\left(\Phi_{j k l+1}-\Phi_{j k l-1}\right) + \\
 + \Lambda_{j k l}\frac{1}{ \Delta}\left(E_{R,j k l+\frac{1}{2} }-E_{R,j k l-\frac{1}{2}}\right) \ = \
\mathcal{V} \left\{s_z \right\}_{j k l};
\end{multline}
\begin{multline}
\label{discrete5}
\frac{d}{d t}{E_{G,j k l}}
+ \mathcal{D}\left\{ \left( \mathcal{E}_\mathrm{loc} + p \right) \mathbf{u} \right\}_{j k l}  - \Phi_{\mathrm{eff},j k l} \frac{d}{d t}{\rho_{j k l}}
 + 
         u_{R,j k l}        \Lambda_{j k l}\frac{1}{ \Delta}\left(E_{R,j+\frac{1}{2} k l }-E_{R,j-\frac{1}{2} k l}\right)  + 
\\ \frac{u_{\phi,j k l}}{R} \Lambda_{j k l}\frac{1}{ \Delta}\left(E_{R,j k+\frac{1}{2}  l}-E_{R,j k-\frac{1}{2} l}\right)  + 
         u_{z,j k l}        \Lambda_{j k l}\frac{1}{ \Delta}\left(E_{R,j k l+\frac{1}{2} }-E_{R,j k l-\frac{1}{2} }\right)
 \ = \
\mathcal{V} \left\{ \mathcal{E}_\mathrm{loc} \right\}_{j k l};
\end{multline}
\begin{multline}
\label{discrete6}
\frac{d}{d t}{E_{R,j k l}}
+ \mathcal{D}\left\{ E_R \mathbf{u} \right\}_{j k l}  + \\
   P_{R R, j k l}        \frac{1}{ \Delta}\left(u_{R,   j+\frac{1}{2} k l} -u_{R,   j-\frac{1}{2} k l}\right)  +
   P_{R \phi, j k l}     \frac{1}{ \Delta}\left(u_{\phi,j+\frac{1}{2} k l} -u_{\phi,j-\frac{1}{2} k l}\right)  + \\
   P_{R z, j k l}        \frac{1}{ \Delta}\left(u_{z,   j+\frac{1}{2} k l} -u_{z,   j-\frac{1}{2} k l}\right)  + 
   P_{\phi R, j k l}     \frac{1}{ \Delta}\left(u_{R,   j k+\frac{1}{2} l }-u_{R,   j k-\frac{1}{2} l}\right)  + \\
   P_{\phi \phi, j k l}  \frac{1}{ \Delta}\left(u_{\phi,j k+\frac{1}{2} l }-u_{\phi,j k-\frac{1}{2} l}\right)  + 
   P_{\phi z, j k l}     \frac{1}{ \Delta}\left(u_{z,   j k+\frac{1}{2} l }-u_{z,   j k-\frac{1}{2} l}\right)  + \\
   P_{z R, j k l}        \frac{1}{ \Delta}\left(u_{R,   j k l+\frac{1}{2} }-u_{R   ,j k l-\frac{1}{2}}\right)   + 
   P_{z \phi, j k l}     \frac{1}{ \Delta}\left(u_{\phi,j k l+\frac{1}{2} }-u_{\phi,j k l-\frac{1}{2}}\right)  + \\
   P_{z z, j k l}        \frac{1}{ \Delta}\left(u_{z,   j k l+\frac{1}{2} }-u_{z   ,j k l-\frac{1}{2}}\right)   
 \ = \
\mathcal{V} \left\{ E_R \right\}_{j k l};
\end{multline}
\begin{equation}
\label{discrete7}
\frac{d}{d t} \tau
+ \mathcal{D}\left\{ \tau \mathbf{u} \right\}_{j k l} \ = \ \mathcal{V} \left\{\tau \right\}_{j k l}.
\end{equation}
The average of the left- and right-face quantities  is $V_{j+\frac{1}{2} k l} := \frac{1}{2}\left(V^+_{j+\frac{1}{2} k l} + V^-_{j+\frac{1}{2} k l}\right)$. Equations (\ref{discrete1}) through (\ref{discrete7}) are transformed into fully discrete form by using the third-order Runge Kutta (RK) time integrator of \cite{SO77}. After the RK update is performed, the evolution variables are in a state between the last, $n^\mathrm{th}$, and the next, $n+1^\mathrm{th}$, time-step. We refer to this as the $n+a^\mathrm{th}$ time step.

\subsection {Radiation - Implicit Update}
\label{rad_implicit_section}

We solve for the implicit terms by taking equation (\ref{form3}) with $\mathbf{q}_\mathrm{explicit}$ set to zero. The resulting equation set is
\begin{equation}
\label{implicit_total_energy}
\displaystyle \lim_{\mathrm{imp}} \frac{\partial}{\partial t}  E_G = - 4 \kappa_p B_p + c \kappa_E E_R,
\end{equation}
\begin{equation}
\label{implicit_radiation}
\displaystyle \lim_{\mathrm{imp}} \frac{\partial}{\partial t} E_R + \nabla \cdot \mathbf{F}  = 
4 \kappa_p B_p - c \kappa_E E_R,
\end{equation}
and
\begin{equation}
\label{implicit_tracer}
\displaystyle \lim_{\mathrm{imp}} \frac{\partial}{\partial t} \tau  = \frac{1}{\gamma \tau^{\left(\gamma-1\right)}} \left( -4 \kappa_p B_p + c \kappa_E E_R \right).
\end{equation}
The implicit step is computed as if there are no contributions from explicit terms over an entire time-step. We have applied the prefix $\displaystyle \lim_{\mathrm{imp}}$ to indicate that we are referring only to the time variance of these quantities due to the implicit terms. Since the fluid momentum does not change due to implicit terms, $\displaystyle \lim_{\mathrm{imp}} \frac{\partial}{\partial t}  E_G = \displaystyle \lim_{\mathrm{imp}} \frac{\partial}{\partial t}  e$, and using $e := \tau^{\gamma}$, we can eliminate equation (\ref{implicit_tracer}) and solve only equations (\ref{implicit_total_energy}) and (\ref{implicit_radiation}).  We compute the solution to these  equations using a backward Euler time step and first-order spatial differencing. The fully discrete equations are:
\begin{equation}
\label{root1}
E^{n+1}_{R,j k l} - E^{n+a}_{R,j k l} - \Delta t \left( 4 \kappa^{n+1}_{p,j k l} B^{n+1}_{p,j k l} - c \kappa^{n+1}_{E,j k l} E^{n+1}_{R,j k l} - \left[\nabla \cdot \mathbf{F}\right]^{n+1}_{j k l}\right) = 0
\end{equation}
and
\begin{equation}
\label{root2}
E^{n+1}_{G,j k l} - E^{n+a}_{G,j k l} + \Delta t \left( 4 \kappa^{n+1}_{p,j k l} B^{n+1}_{p,j k l} - c \kappa^{n+1}_{E,j k l} E^{n+1}_{R,j k l}\right) = 0,
\end{equation}
The term $\left[\nabla \cdot \mathbf{F}\right]^{n+1}_{j k l}$ is defined as
\begin{eqnarray}
\notag
& \left[\nabla \cdot \mathbf{F}\right]^{n+1}_{j k l} := \\ 
&-\frac{1}{R_j \Delta^2}\left[ R_{j+\frac{1}{2}} D^{n+a}_{j+\frac{1}{2} k l} \left(E^{n+1}_{R,j+1 k l}-E^{n+1}_{R,j k l}\right) - R_{j-\frac{1}{2}} D^{n+a}_{j-\frac{1}{2} k l} \left(E^{n+1}_{R,j k l}-E^{n+1}_{R,j-1 k l}\right) \right] -  \notag \\
&\frac{1}{R_j^2 \Delta^2}\left[D^{n+a}_{j k+\frac{1}{2} l} \left(E^{n+1}_{R,j k+1 l}-E^{n+1}_{R,j k l}\right) - D^{n+a}_{j k-\frac{1}{2} l} \left(E^{n+1}_{R,j k l}-E^{n+1}_{R,j k-1 l}\right)\right] -  \notag \\
& \frac{1}{\Delta^2}\left[D^{n+a}_{j k l+\frac{1}{2}} \left(E^{n+1}_{R,j k l+1}-E^{n+1}_{R,j k l}\right) - D^{n+a}_{j k l-\frac{1}{2}} \left(E^{n+1}_{R,j k l}-E^{n+1}_{R,j k l-1}\right)\right] ,
\end{eqnarray} 
where
\begin{equation}
D^{n+a}_{j\pm\frac{1}{2} k l } = \frac{c \Lambda^{n+a}_{E, j\pm\frac{1}{2} k l }}{\chi^{n+a}_{j\pm\frac{1}{2} k l }}, 
\end{equation}
\begin{equation}
D^{n+a}_{j k\pm\frac{1}{2} l } = \frac{c \Lambda^{n+a}_{E, j k\pm\frac{1}{2} l }}{\chi^{n+a}_{j k\pm\frac{1}{2} l }},
\end{equation}
and
\begin{equation}
D^{n+a}_{j k l\pm\frac{1}{2} } = \frac{c \Lambda^{n+a}_{E, j k l\pm\frac{1}{2} }}{\chi^{n+a}_{j k l\pm\frac{1}{2} }}. 
\end{equation}
 We compute the $\Lambda_E$'s and $\chi$'s using the cell averaged quantities and the first-order  differences in equations (\ref{diff0}) through (\ref{diff2}). As in \cite{HNF2006}, we obtain the numerical solution to the nonlinear equations (\ref{root1}) and (\ref{root2}) with a linear iterative solver coupled to a Newton-Raphson solver. Unlike \cite{HNF2006}, our method has fewer implicit terms to compute. \cite{KKMB2007} argue it is only necessary to implicitly compute the terms found in equations (\ref{implicit_radiation}) and (\ref{implicit_total_energy}). Explicit gas pressure terms also fit more easily into the framework of the K-T method.

\subsection{Implementation }
Our method is implemented in FORTRAN-90 with the Message Passing Interface (MPI) standard for execution on high performance computing clusters. We have relied heavily on the Hyper Adaptive Mesh Refinement (AMR) Driver (HAD) of \cite{L2002}, which we have modified to suit our particular needs. At the time of this writing, we have not yet implemented the AMR feature in HAD for use in our code. 

\section{Verification Testing }
\label{verification_section}
Here we present the results of tests that have been used to verify the accuracy of our new code. As a test of basic hydrodynamics, we simulate a Sod shock tube. To test the radiation diffusion solver we use the Marshak problem.  To verify that the radiation energy is properly coupled to the fluid energy and momentum, we simulate two cases of a radiating wall shock and compare the results to another radiation hydrodynamics code. Lastly, we investigate the degree to which our E* energy scheme is able to preserve an equilibrium polytrope.

\subsection {Sod Shock Tube}
\begin{figure}
\begin{center}
\begin{tabular}{|c|c|}
\tableline
\includegraphics[angle=-90,scale=0.30]{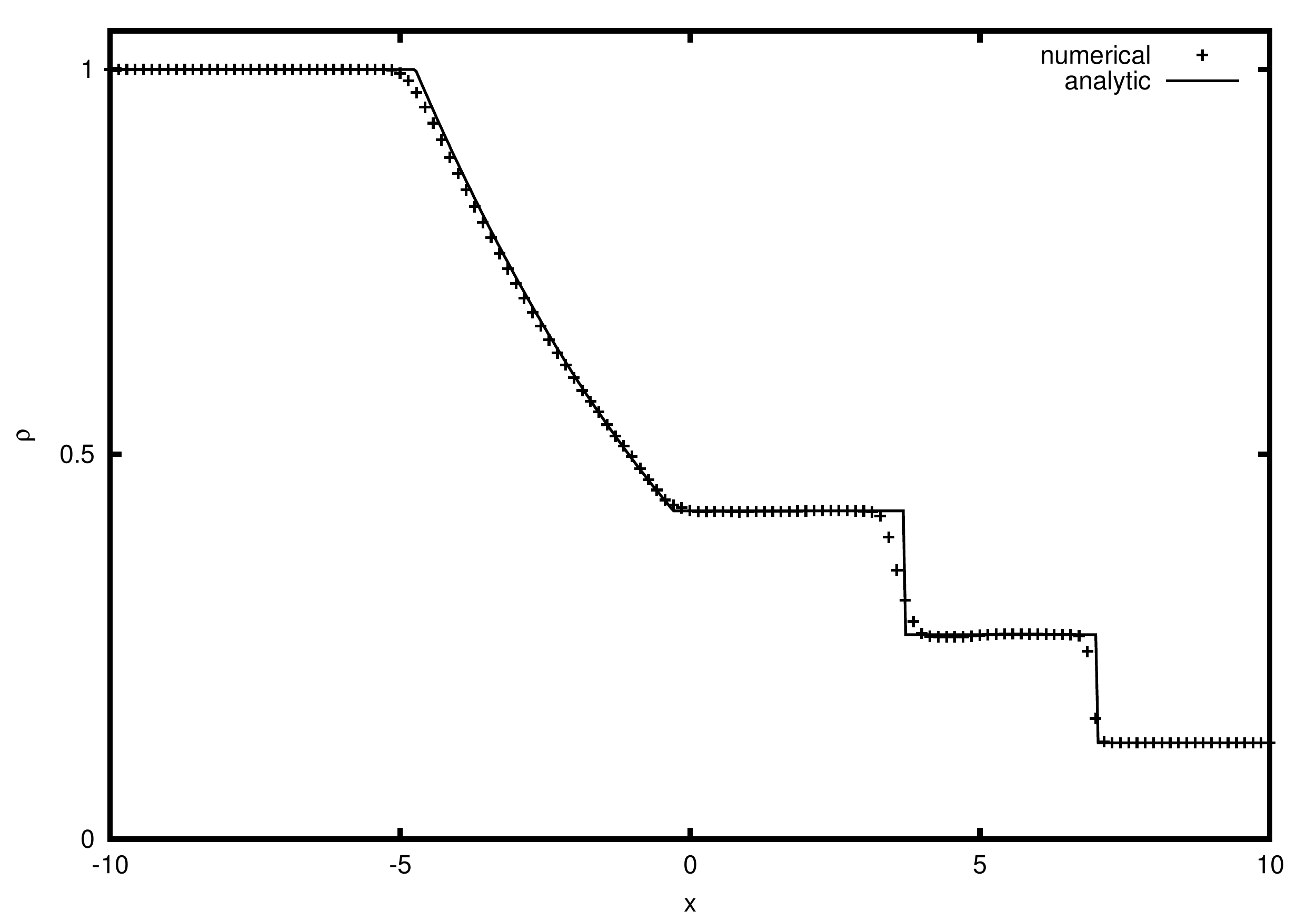} &
\includegraphics[angle=-90,scale=0.30]{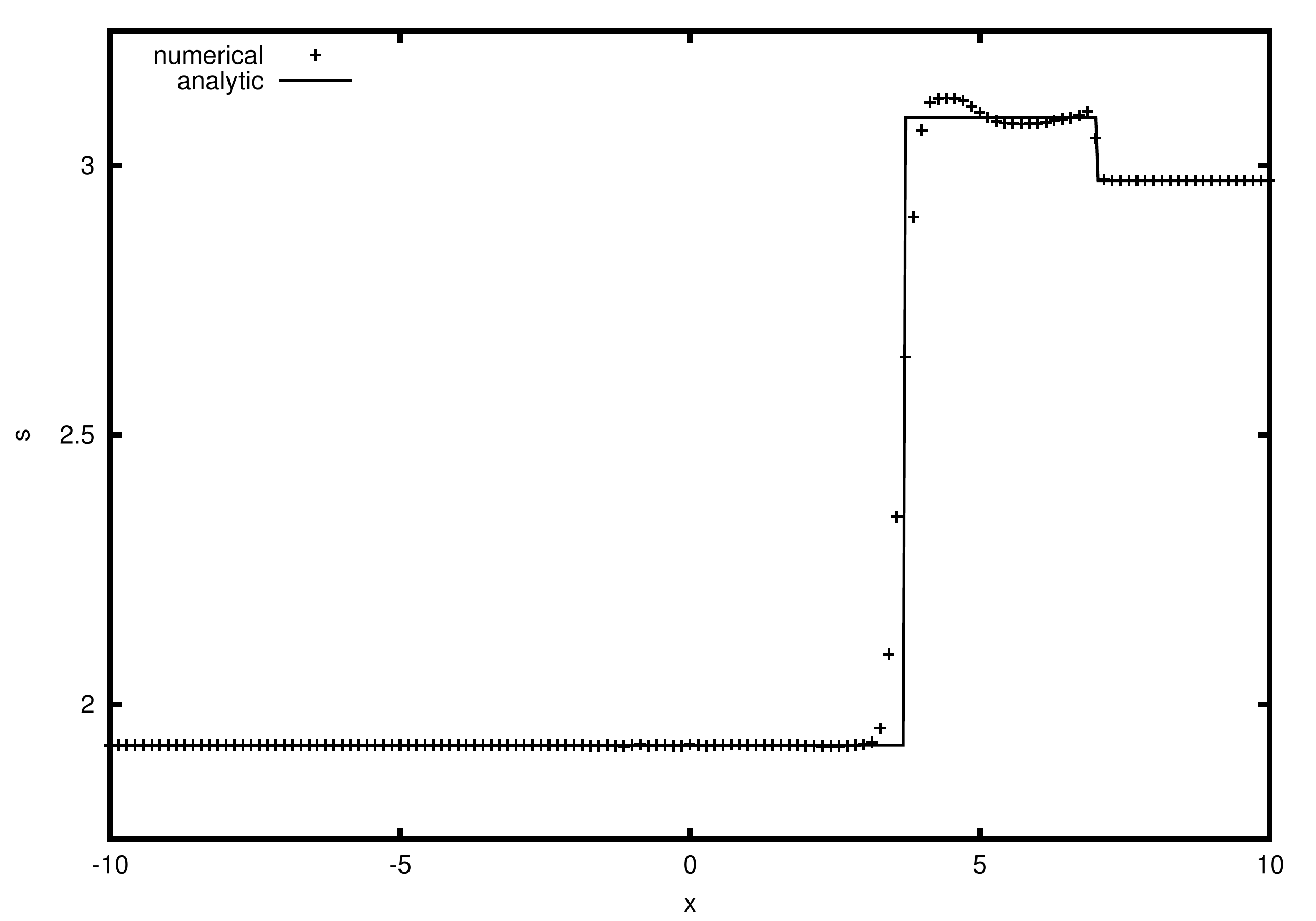} \\
\hline
\includegraphics[angle=-90,scale=0.30]{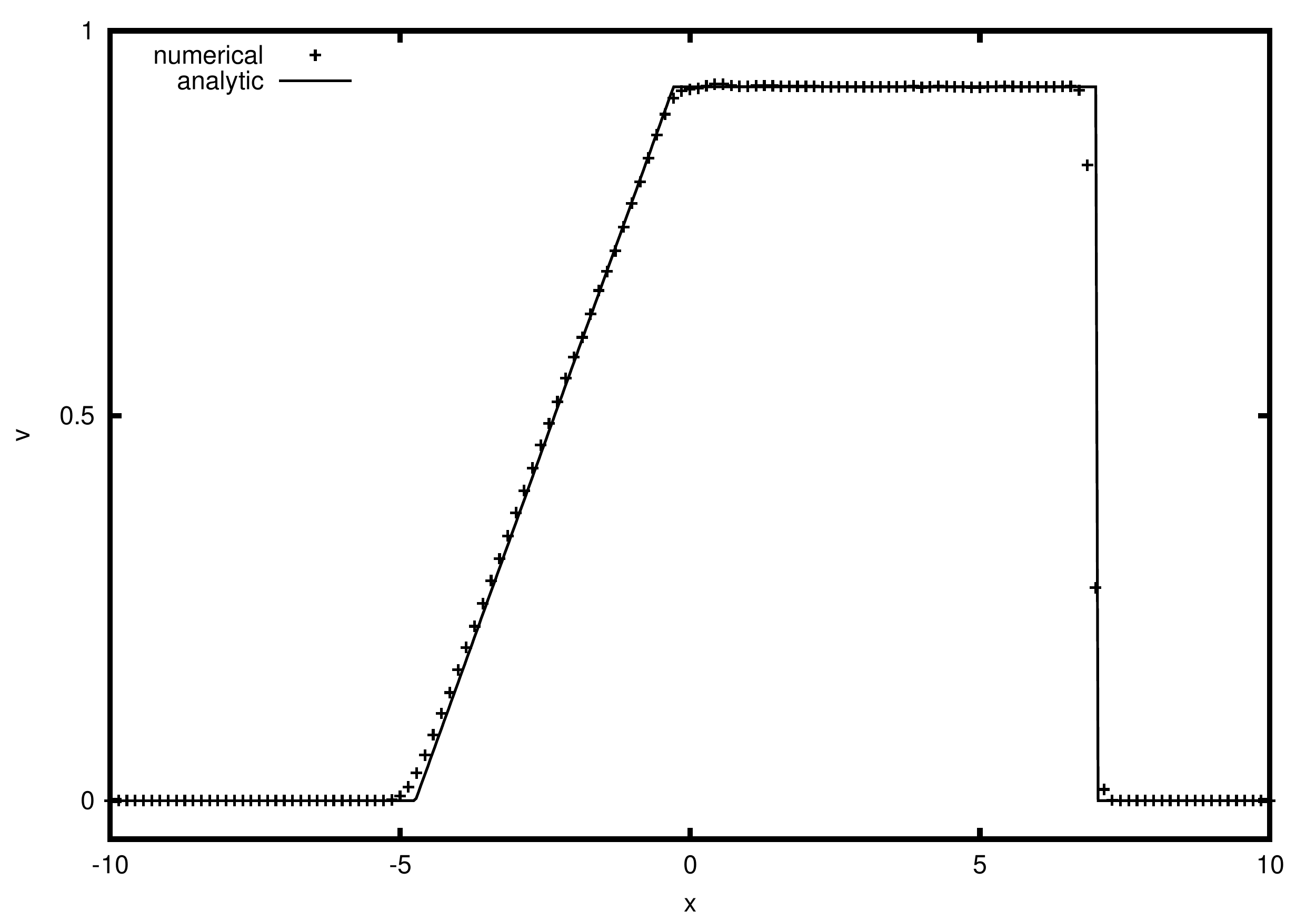} &
\includegraphics[angle=-90,scale=0.30]{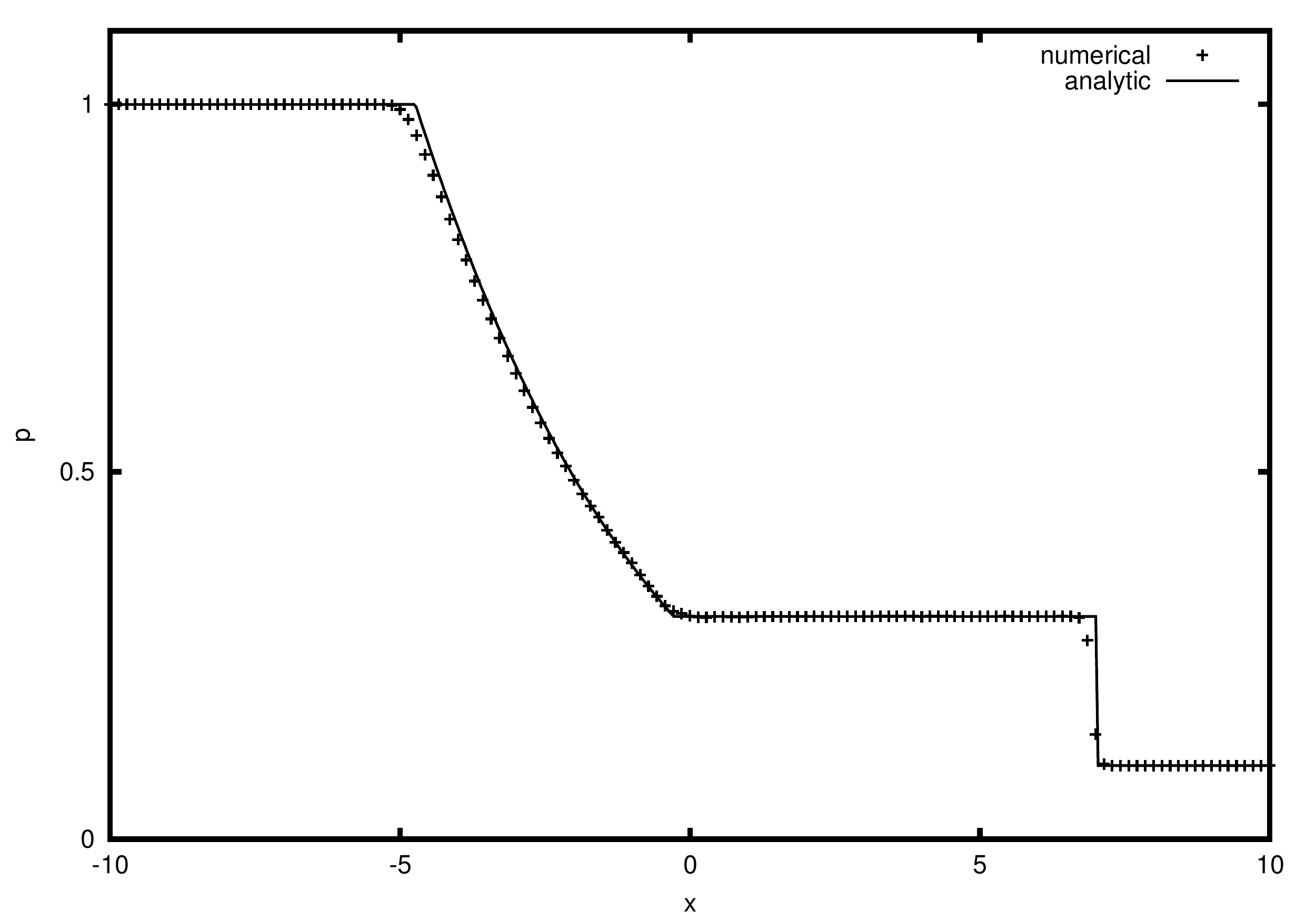} \\
\hline
\end{tabular}
\caption[Sod's shock tube]{ Our code's results for the Sod shock tube are compared to analytic results at $t=4.001$.
{\it Top Left:} Density.
{\it Top Right:} Specific entropy.
{\it Bottom Left:} Velocity.
{\it Bottom Right:} Pressure. }
\label{sod1}
\end{center}
\end{figure}

The Sod shock tube problem is a Riemann problem which includes the three fundamental hydrodynamic waves: shock waves, contact discontinuities, and rarefaction waves (\cite{S1978}). There are known analytic solutions to the problem, making it suitable as a test of basic hydrodynamics. The initial conditions are defined as 
\begin{equation}
\rho = \left\{\begin{array}{cc}
1.0 & z > 0 \\
0.125 & \mathrm{else} \\ 
\end{array}\right., \
\end{equation}
\begin{equation}
E_G = \left\{\begin{array}{cc}
2.5 & z > 0 \\
0.25 & \mathrm{else} \\ 
\end{array}\right., \ 
\end{equation}
and
\begin{equation}
\mathbf{u} = 0.
\end{equation}
We set $\gamma = 1.4$ and turn off reflection along the z plane. The  simulation was run with $34$ radial, $34$ azimuthal, and $144$ vertical interior zones. Figure (\ref{sod1})
 depicts , from left to right and top to bottom, the mass density, specific entropy, velocity, and pressure at time $t=4.001$ for zones along the vertical line occupying the $22^{\mathrm{nd}}$ radial and azimuthal locations. Comparing these results with those of our previous code in Figure 5 of \cite{MTF2002}, we see that the present method represents the shock with greater accuracy. The width of the shock is narrower and there is an entropy jump across the shock. The method of \cite{MTF2002} did not include an equation for 
total gas energy and thus could not properly account for the shock jump conditions. The numerical solutions in both codes disagree slightly at the tail end of the
rarefaction wave. The one way in which the present code is less accurate is at the contact discontinuity.  The results of our code at the shock also compare favorably with those of other codes, including the ZEUS-2D code of \cite{SN1992} and most of the codes mentioned in \cite{TBMMHPBT2008}. Other codes, such as FLASH, however, resolve the contact discontinuity better. 

\subsection {Marshak Wave}
There are few problems involving radiative transport that admit analytic solutions. Certain forms of the Marshak wave are an exception. In this problem, radiation is incident along the boundary of a semi-infinite slab of uniform optical opacity. Initially the slab is at zero temperature. In the original problem, described by \cite{M1958}, the radiation and fluid temperatures were taken equal to one another throughout the evolution. \cite{P1979} extended the problem to allow the radiation and fluid temperatures  to evolve separately and presented a semi-analytic solution for the case where the speed of light is taken to be infinite. \cite{SO1996} developed a semi-analytic solution for the case with a finite speed of light. To make an analytic solution possible, it is necessary to alter the heat capacity such that 
\begin{equation}
e = c_0 T^4,
\end{equation}
where $c_0$ is a constant. This linearizes the RHS of the radiation energy density equation and total gas energy equation. The hydrodynamics part of our code is disabled for this test problem. We use a grid of 20 radial zones, 20 azimuthal zones, and 194 vertical interior zones. The Marshak boundary condition identified by \cite{SO1996},
\begin{equation}
E_R\left(z=z_0,t\right) - \frac{2}{3 \kappa} \frac{\partial}{\partial z} E_R\left(z=z_0,t\right) = \frac{4}{c} F_{\mathrm{in}},
\end{equation}
 is imposed at all zones along the upper vertical boundary. The inflowing radiative flux, $F_{\mathrm{in}}$, is taken as $1$. Outflow conditions are imposed at the lower vertical and outer radial boundaries. 

\begin{figure}
\begin{center}
\begin{tabular}{|c|c|}
\hline
\includegraphics[angle=-90,scale=0.39]{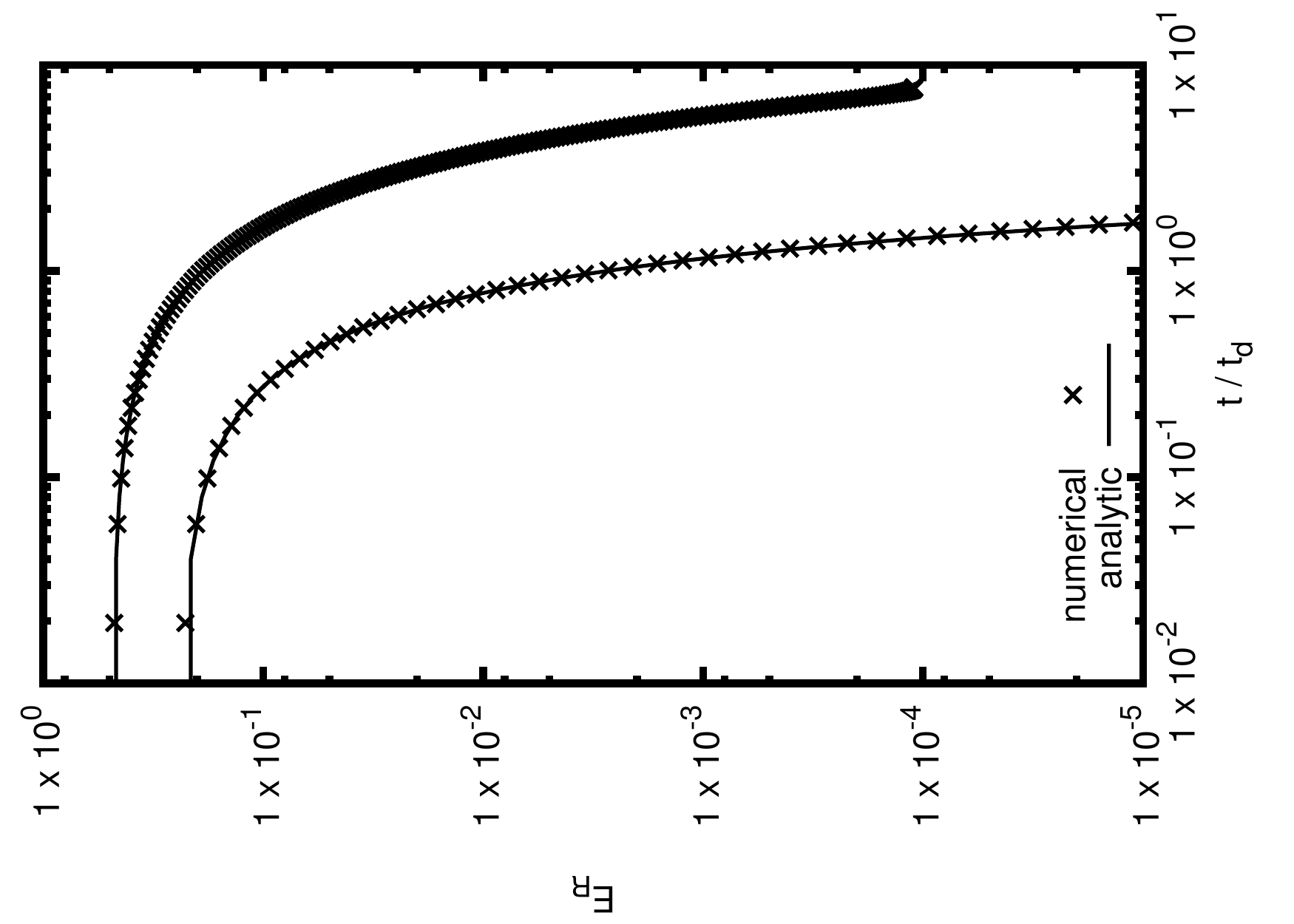} &
\includegraphics[angle=-90,scale=0.39]{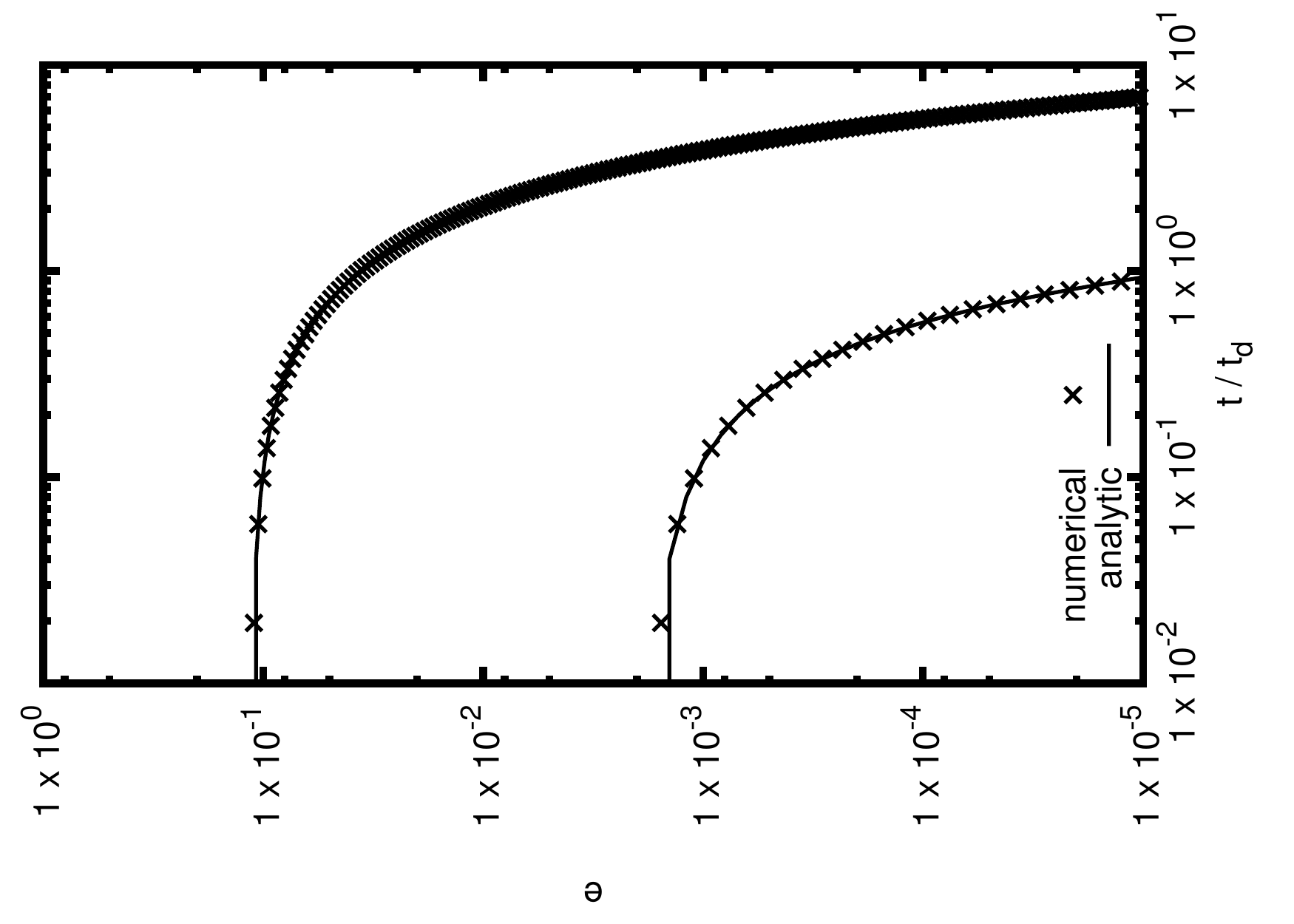} \\
\hline
\end{tabular}
\caption[Marshak wave]{ Simulation for the Marshak wave are compared to analytic results. The top curve in both plots is at $t=0.3$ and the bottom curve is at $t=0.01$.
{\it Left:} The radiation energy density versus vertical coordinate.
{\it Right:} The internal gas energy density  versus vertical coordinate.
}
\label{marshak1}
\end{center}
\end{figure}

In Figure \ref{marshak1}
we compare our results to the  semi-analytic results of \cite{SO1996} in a format similar to their Figure 3 (\cite{HNF2006} also uses a similar format). Depicted are the radiation and material energy densities at two sample times in the evolution. With the exception of a slight disagreement at the inflow boundary point, our results are consistent with the analytic results. Because the analytic solutions apply to a semi-infinite slab geometry, numerical results are not expected to be accurate once the wave hits the grid's outer boundary, and therefore the run is terminated close to that point.

\subsection {Radiating Shock Waves }
Coupled radiation and hydrodynamics does not readily admit test problems with analytic solutions. Nonetheless, the numerical results of different codes can be compared with one another for consistency. A problem suitable for these purposes, described in \cite{E1994}, is the radiating shock wave. The problem in its one-dimensional form consists of a reflecting boundary condition on one end of the computational domain and an inflow boundary condition at the other. As the inflowing gas strikes the reflecting boundary, a shock wave is formed. In the purely hydrodynamical case, this wave forms a step discontinuity in mass density, velocity, and energy density. When radiation transport is present, the heat created at the shock front is radiated into the incoming gas stream, preheating it. When the radiation preheating is intense enough to heat the incoming material to the same temperature as the post-shock material, the shock is said to be super-critical. When the temperature of the incoming stream is below the temperature of the post-shock material, the shock is said to be sub-critical. 

\begin{figure}
\begin{center}
\begin{tabular}{|c|c|}
\hline
\includegraphics[angle=-90,scale=0.30]{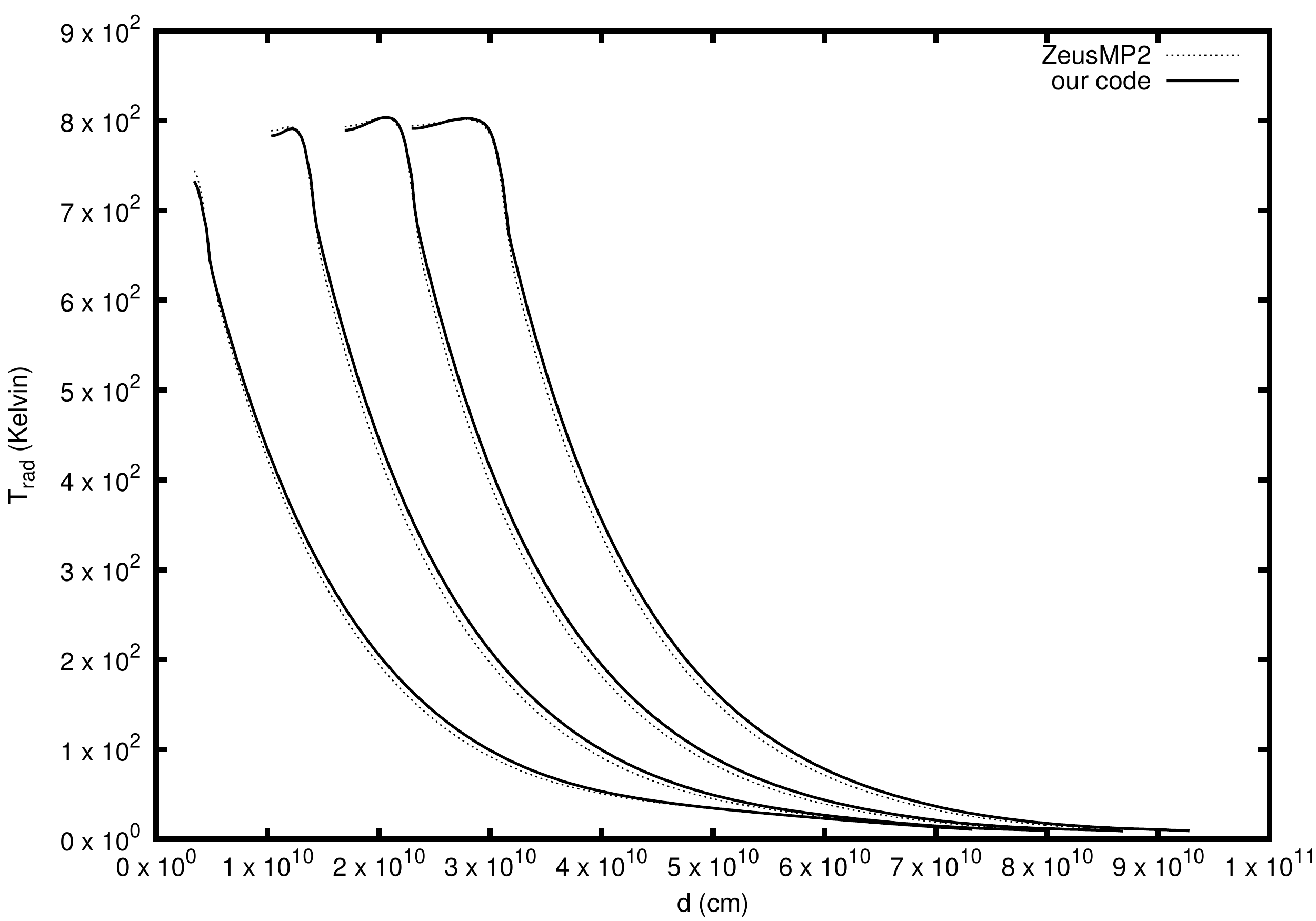} & 
\includegraphics[angle=-90,scale=0.30]{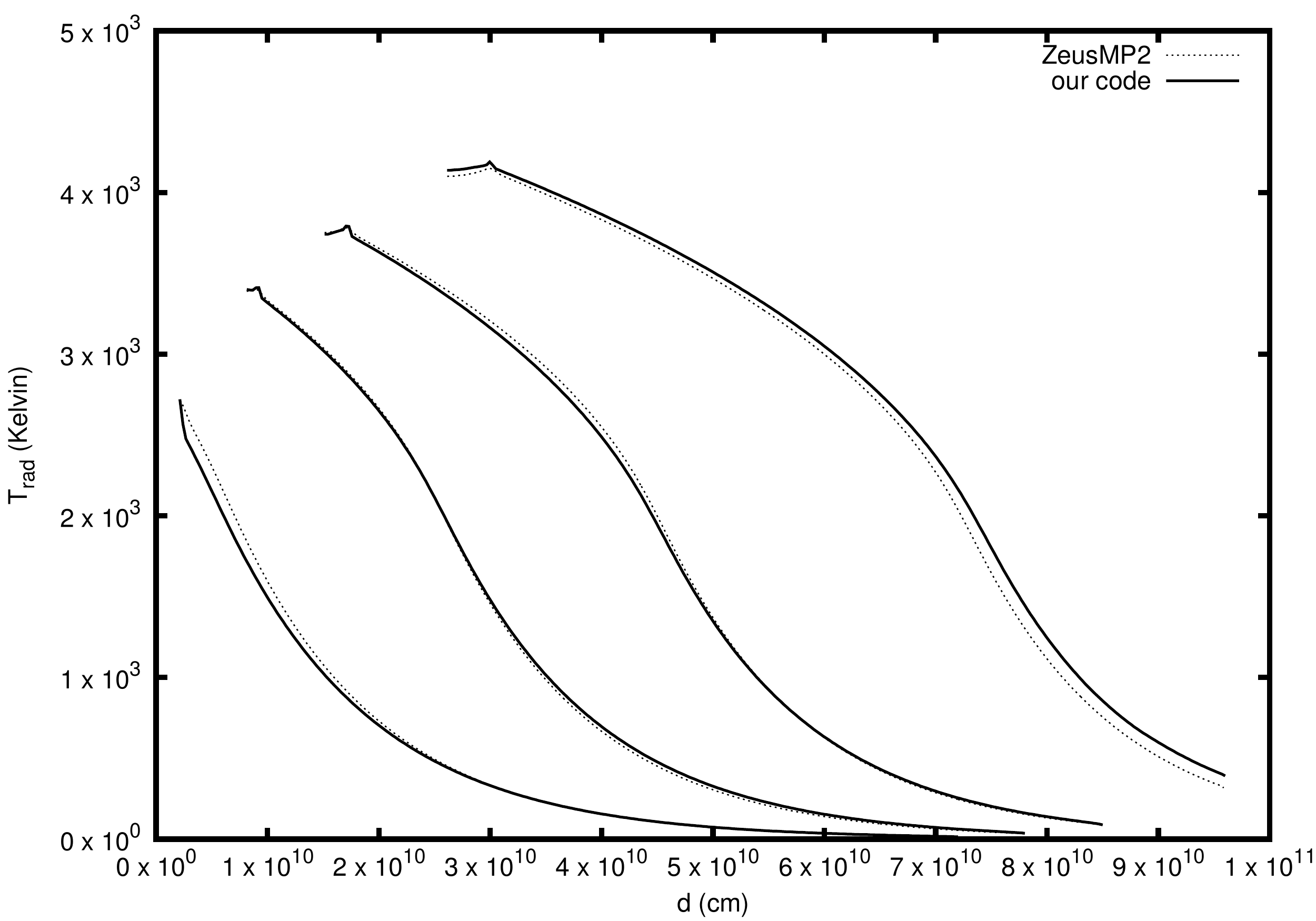} \\
\hline
\includegraphics[angle=-90,scale=0.30]{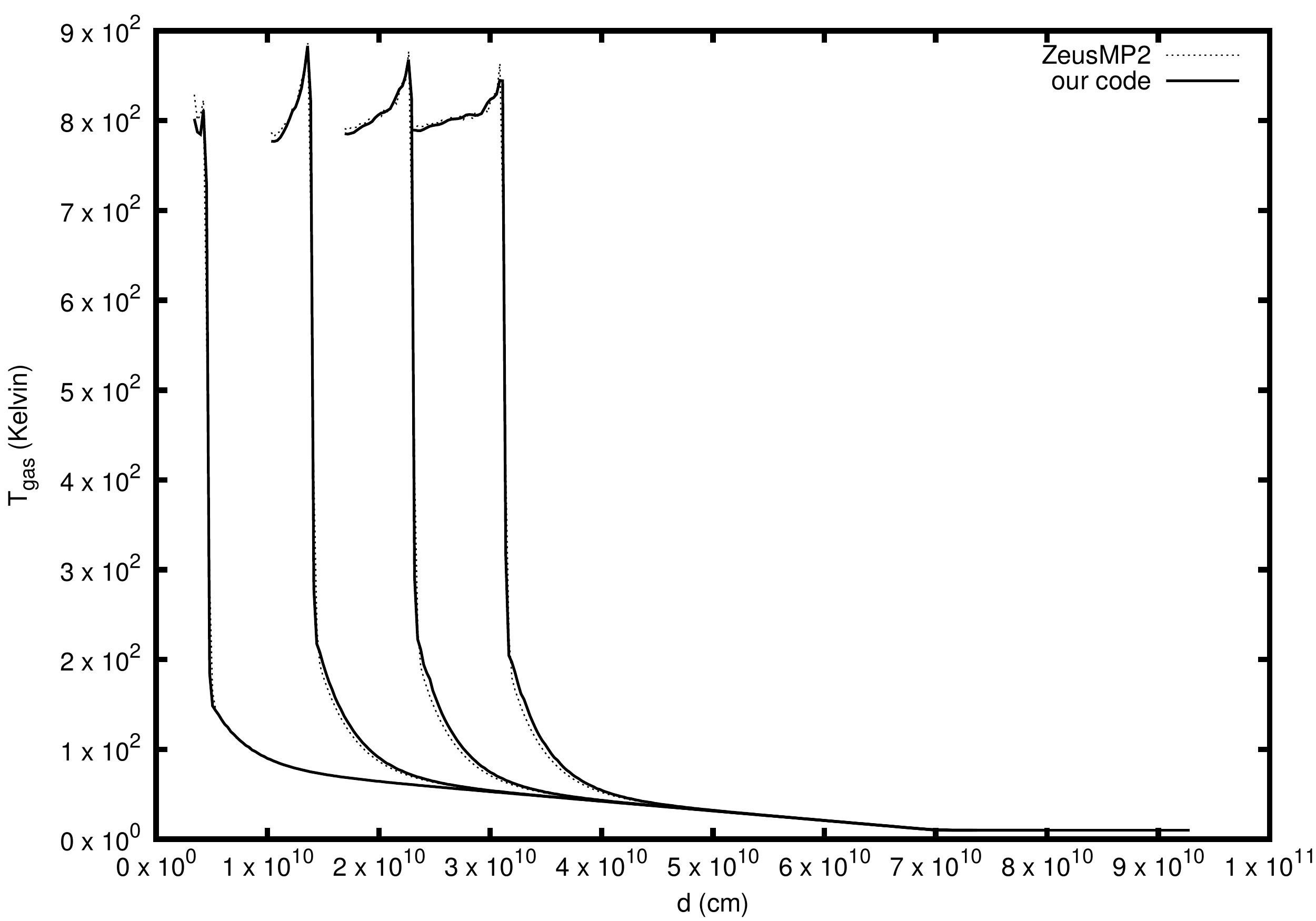} & 
\includegraphics[angle=-90,scale=0.30]{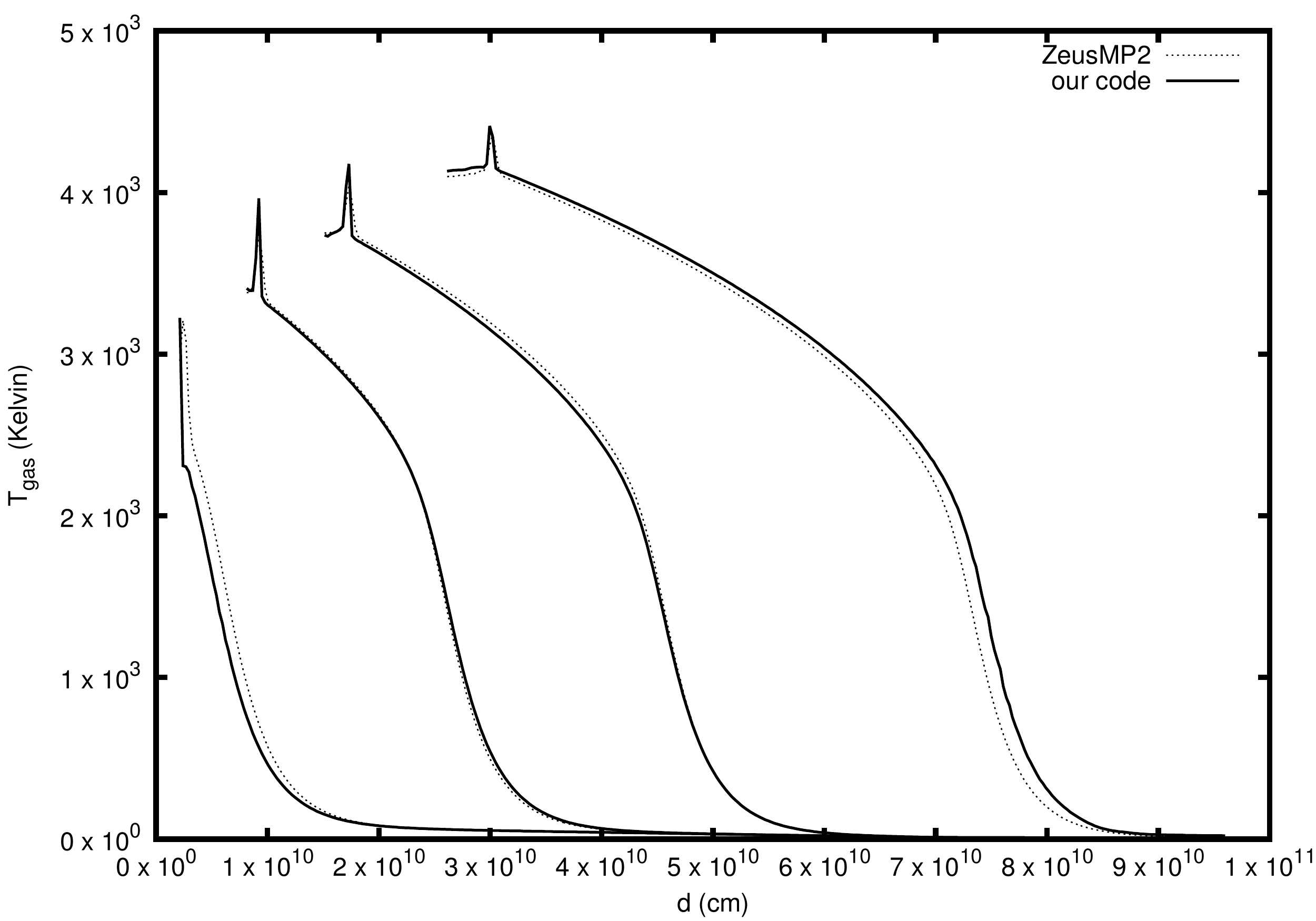} \\
\hline
\end{tabular}
\caption[Radiating shock wave]{ Our code's results for the radiating shock wave are compared to results generated using ZeusMP2. The profiles are shown in the frame which is at rest relative to the inflowing gas. For the sub-critical shock ({\it left}), the times shown are $t=5.5 \times 10^3 \ \mathrm{s}$, $t=1.7 \times 10^4  \ \mathrm{s} $, $t=1.7 \times 10^4 \ \mathrm{s}$, $t=2.8 \times 10^4 \ \mathrm{s}$, and $t=3.8 \times 10^4 \ \mathrm{s}$. For the super-critical shock ({\it right}), the times shown are $t=1.0 \times 10^3 \ \mathrm{s}$, $t=4.0 \times 10^3 \ \mathrm{s}$, $t=7.5 \times 10^3  \ \mathrm{s}$, and $t=1.3 \times 10^4 \ \mathrm{s}$.
{\it Top Left}: Radiation temperature profile for sub-critical shock.
{\it Top Right}: Radiation temperature  profile for super-critical shock.
{\it Bottom Left}: Gas temperature  profile for sub-critical shock.
{\it Bottom Right}: Gas  profile for super-critical shock.} 
\label{subcritical}
\end{center}
\end{figure}

Figure \ref{subcritical} 
 compares the results obtained with our code with results from the ZeusMP2 code of \cite{HNF2006}. Our computations were conducted using $10$ radial, $26$ azimuthal, and $256$ vertical interior zones, with the in-falling gas injected at the upper vertical boundary. The ZeusMP2 runs were performed using a one-dimensional $256$ zone mesh. The displayed profiles are averages of the plotted quantities over all radial and azimuthal zones for a given vertical displacement. In the problem as posed by \cite{E1994}, the lab frame is co-moving with the in-falling gas and the reflecting boundary condition, acting as a piston, moves relative to the gas.  Our Figure \ref{subcritical} is plotted in this frame, with the profile curves moving to the right in time. The left panels of Figure \ref{subcritical} are for the sub-critical case and the right panels for the super-critical case. The radiation temperature is displayed in the upper panels and the gas temperature in the lower panels.  There is generally good agreement with  ZeusMP2. The biggest disagreements are at times $t=10^3  \ \mathrm{s}$ and $t=1.3\times10^4 \ \mathrm{s}$. In the former case, the difference is likely due to the different ways in which the codes handle shocks, and in the latter, likely due to differences in the outer $z$-boundary condition. The two codes also handle the gas pressure terms differently.

\subsection {Single Polytrope }
\label{singlesection}
We have run a series of test simulations involving a single, spherically symmetric, polytropic star with polytropic index $n = \frac{3}{2}$. The initial density is computed by solving the Lane-Emden equation with a fourth-order Runge Kutta solver. The initial internal energy is then determined from  equation (\ref{idealgas}) and the polytropic equation of state, 
\begin{equation}
P_\mathrm{poly} := K \rho^{1 + \frac{1}{n}},
\label{polytropiceos}
\end{equation} where $K$ is the polytropic constant. 
 We have run four simulations, without radiation transport, at two resolutions. Each resolution is run with the E* energy correction and without it. The center of mass of the polytrope is initially coincident with the coordinate origin. The center of mass correction, described in Appendix \ref{com_correct}, was turned on for each run. The high (low) resolution run was computed with 94 (44) radial, 128 (64) azimuthal, and 44 (20) vertical interior zones. The polytrope has a radius of 33 (17) cells at high (low) resolution.

\begin{figure}
\begin{center}
\begin{tabular}{|c|c|}
\hline
\includegraphics[angle=-90,scale=0.39]{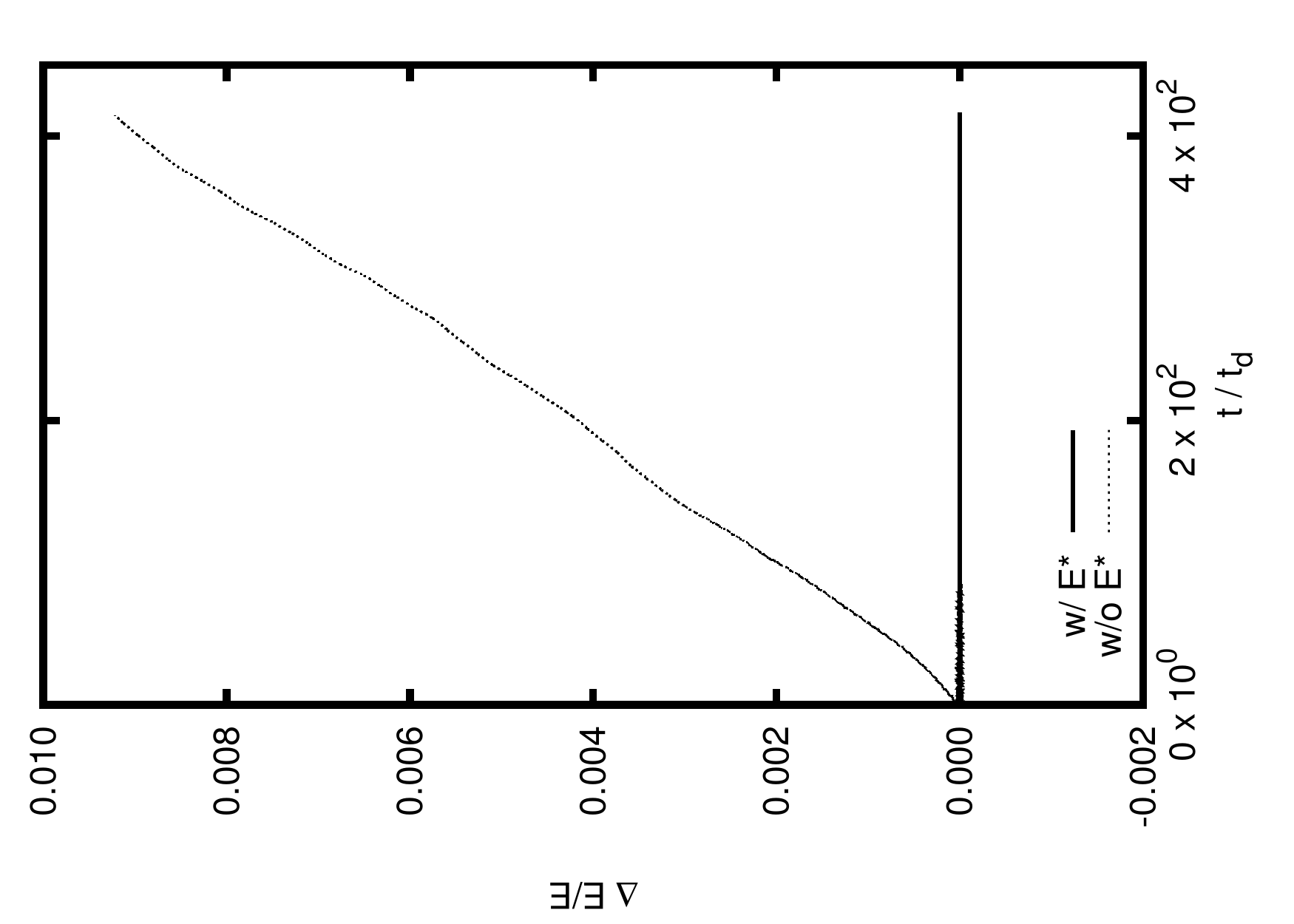} & 
\includegraphics[angle=-90,scale=0.39]{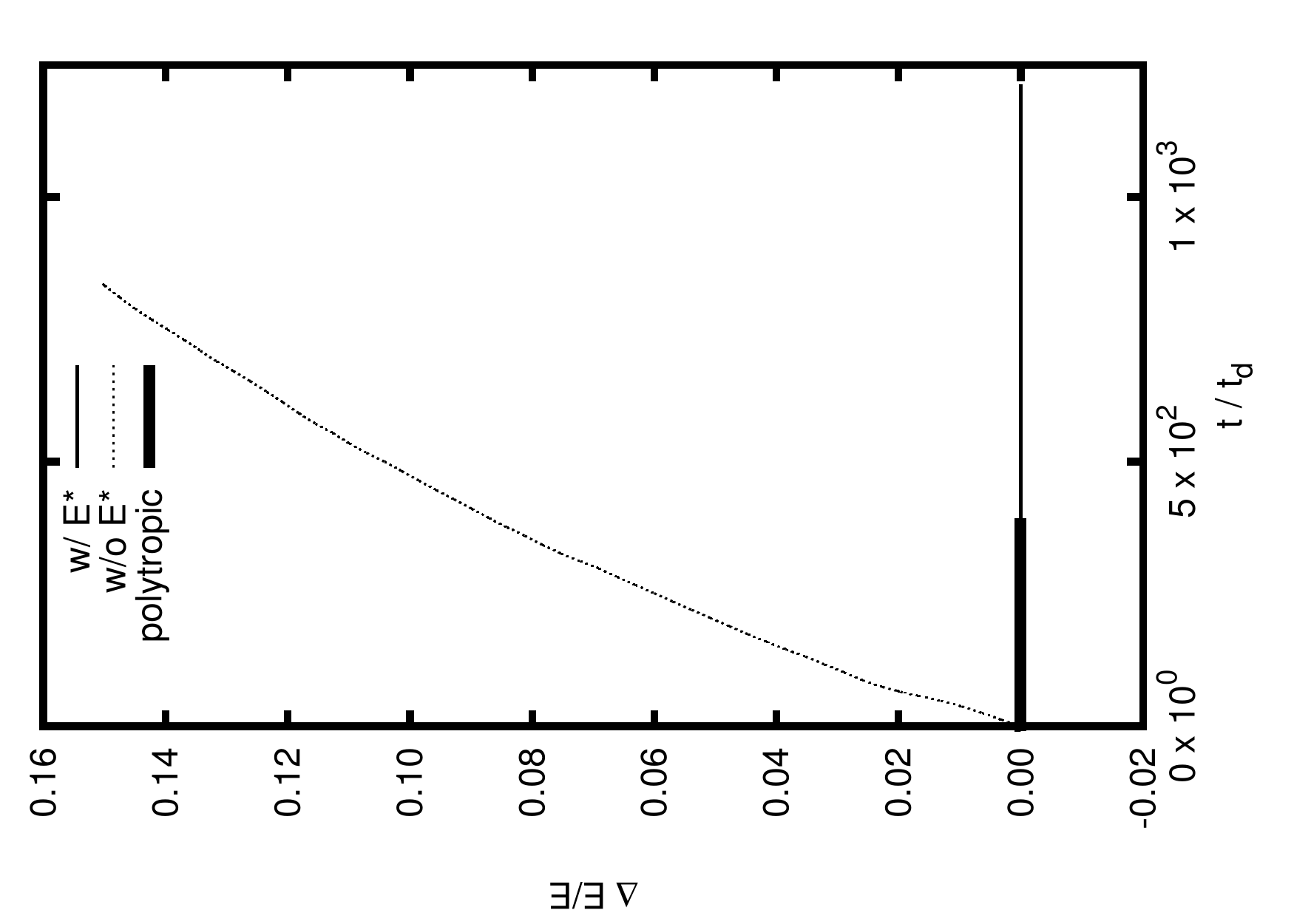} \\ 
\hline
\includegraphics[angle=-90,scale=0.39]{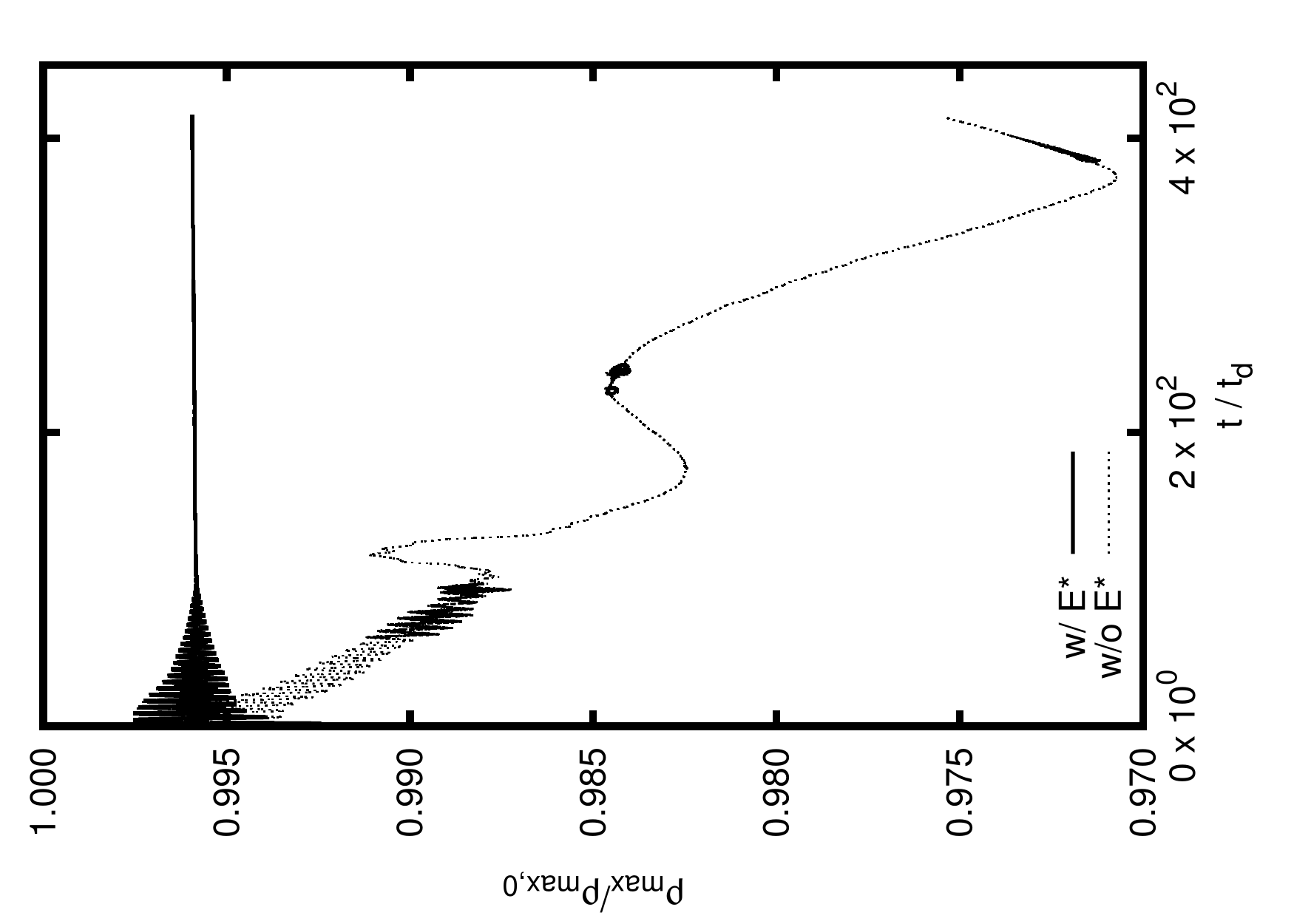}  &
\includegraphics[angle=-90,scale=0.39]{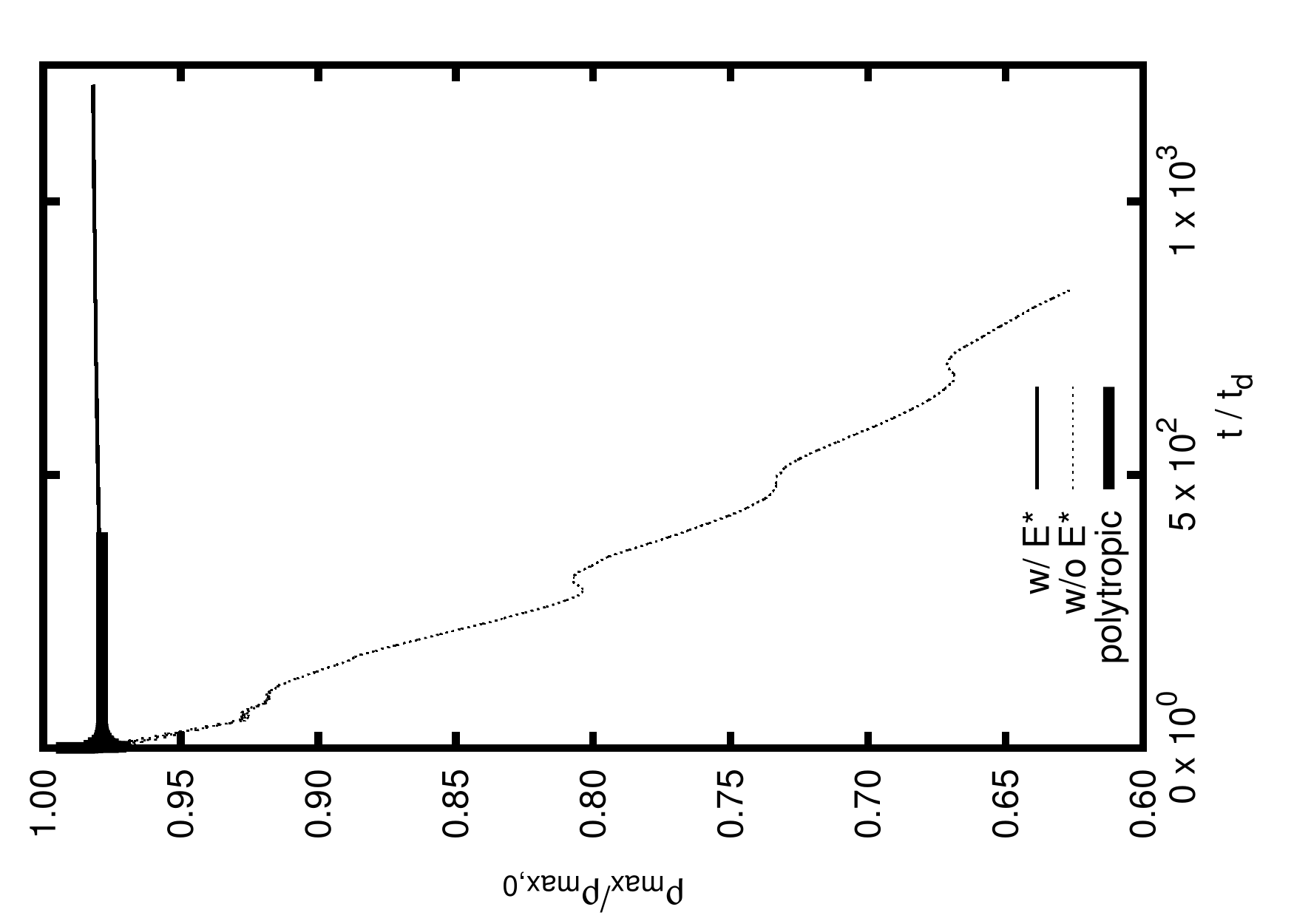} \\
\hline
\end{tabular}
\caption[Singe polytrope]{ Single Polytrope.
{\it Top Left}: Relative change in total energy from initial value for the high resolution run.
{\it Top Right}: Relative change in total energy from initial value for the low resolution run.
{\it Bottom Left}: Maximum density normalized to its initial value for the high resolution run.
{\it Bottom Right}: Maximum density normalized to its initial value for the low resolution run.
 }
\label{poly1}
\end{center}
\end{figure}

This initial configuration is placed in the cylindrical grid of our code and evolved for many dynamical time scales to test how well the code preserves the star's original structure. A dynamical time is given by
\begin{equation}
\label{tdyn2}
t_\mathrm{d} := \sqrt{\frac{R^3}{2 G M}}, 
\end{equation}
where $R$ is the radius and $M$ is the mass of the polytrope. Figure \ref{poly1}
 depicts the sum of $\mathcal{E}_{\mathrm{con}}$ (top panels) over the grid and the maximum density (bottom panels) on the grid for the high (left panels) and low (right panels) resolution runs. The maximum density is equivalent to the central density for these particular simulations. With the E* correction turned on, the sum of $\mathcal{E}_{\mathrm{con}}$ remains nearly constant. Without the E* correction it increases monotonically with time. Similarly, the central density of the polytrope without the E* correction secularly decreases over time, while the central density of the polytrope with the E* correction oscillates for a few dozen dynamical timescales before it settles to a constant near its initial value. 

\begin{figure}
\begin{center}
\includegraphics[angle=-90,scale=0.39]{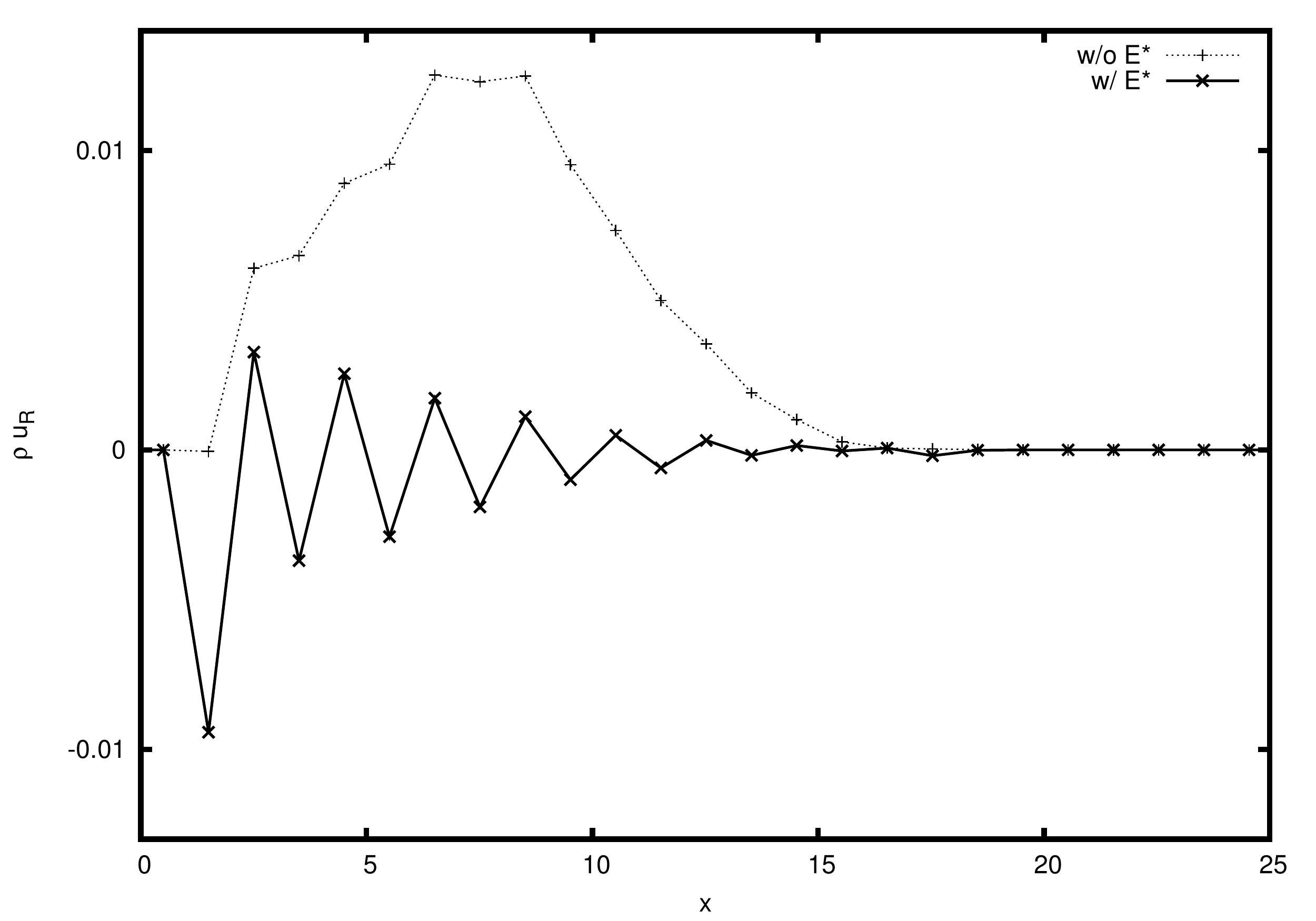}
\caption[Centered polytrope radial momentum profile]{ The radial momentum profiles for the low resolution, centered polytrope with the E* correction (sold curve) and without it (dotted curve) at $t=900$ dynamical timescales. This is several hundred dynamical timescales after the E* corrected polytrope settles into the depicted steady-state configuration. }
\label{poly5}
\end{center}
\end{figure}

The initial equilibrium numerical model with no internal velocity structure, as computed with the Lane-Emden equation, is initially {\it not} in equilibrium when placed in our dynamical solver. The model with the E* correction, when left to itself, is capable of forming a steady-state configuration within a few dozen dynamical timescales. The resulting steady-state configuration has the sawtooth radial momentum profile depicted by the solid curve in Figure \ref{poly5}.
 Despite the non-zero velocities implied by this profile, the structure of the polytropic star is time-invariant. The physical fluxes resulting from the reconstructed evolution variables are canceled by the application of the  viscosity operator in the K-T method. In the case without the E* correction, depicted by the dotted curve, the momentum is directed outward from the center of the polytrope for all but the inner two cells. The result is that the polytrope without the E* correction does not reach equilibrium within the several hundred dynamical timescales over which we have run the simulation. If an equilibrium point is ever reached, the resulting evolution variable profiles will likely look nothing like the initial model. In contrast, the structure of the polytrope with the E* correction only deviates slightly from the initial conditions. Note that while the resulting radial momentum profile is oscillatory for the E* correction, the reconstructed total (physical plus viscous) fluxes at cell faces are not, as they sum to zero.

	Our code must have the ability to evolve stars for many hundreds of dynamical times because the donor of any DWD system will undergo on the order of dozens of 
	dynamical times for each orbital period. \cite{K1959} gives an approximation for the radius of a Roche lobe filling $n=\frac{3}{2}$ polytrope,
	\begin{equation}
		\label{tdyn1}
		\frac{R_2}{a} \approx \frac{2}{3^\frac{3}{4}} \left(\frac{M_\mathrm{acc}}{M_\mathrm{acc}+M_\mathrm{don}}\right)^\frac{1}{3},
		\end{equation}
		where $M_\mathrm{don}$ is the mass of the star which is losing mass (the ``donor"), $M_\mathrm{acc}$ the mass of the star which is gaining mass (the 			``accretor"),
	$R_2$ is the radius of the donor, and $a$ is the orbital separation.
	The orbital period of the binary  is given by
	\begin{equation}
		\label{tdyn3}
		t_p := 2 \pi \sqrt{\frac{a^3}{G M}},
		\end{equation}
	where $M = M_\mathrm{don} + M_\mathrm{acc}$ is the total mass of the system. Using equations (\ref{tdyn2}), (\ref{tdyn1}), and (\ref{tdyn3}) we can derive an 
	expression for the ratio of the orbital period to the dynamical time of a Roche lobe filling star, 
	\begin{equation}
		\frac{t_p}{t_d} \approx 9 \pi \sqrt{ 1 + q^{-1} },
	\end{equation}
	where the $q := \frac{M_\mathrm{don}}{M_\mathrm{acc}}$ is the mass ratio. For the $q=0.7$ simulations described in \S \ref{binary_section}, this ratio is 
	approximately $37$. The simulations are run for about $25$ orbits, therefore the donor undergoes nearly $1000$ dynamical times during the run. Although the grid 
	sizes used in our binary evolutions are larger than that used for the high resolution single polytrope runs, the binary components occupy only a slightly larger 
	number of grid cells. Hence without the E* correction, they will tend to lose energy and dissipate on approximately the same time-scale as the high resolution single 
	polytropes presented here.

\begin{figure}
\begin{center}
\begin{tabular}{|c|c|c|}
\hline
\includegraphics[angle=-90,scale=0.39]{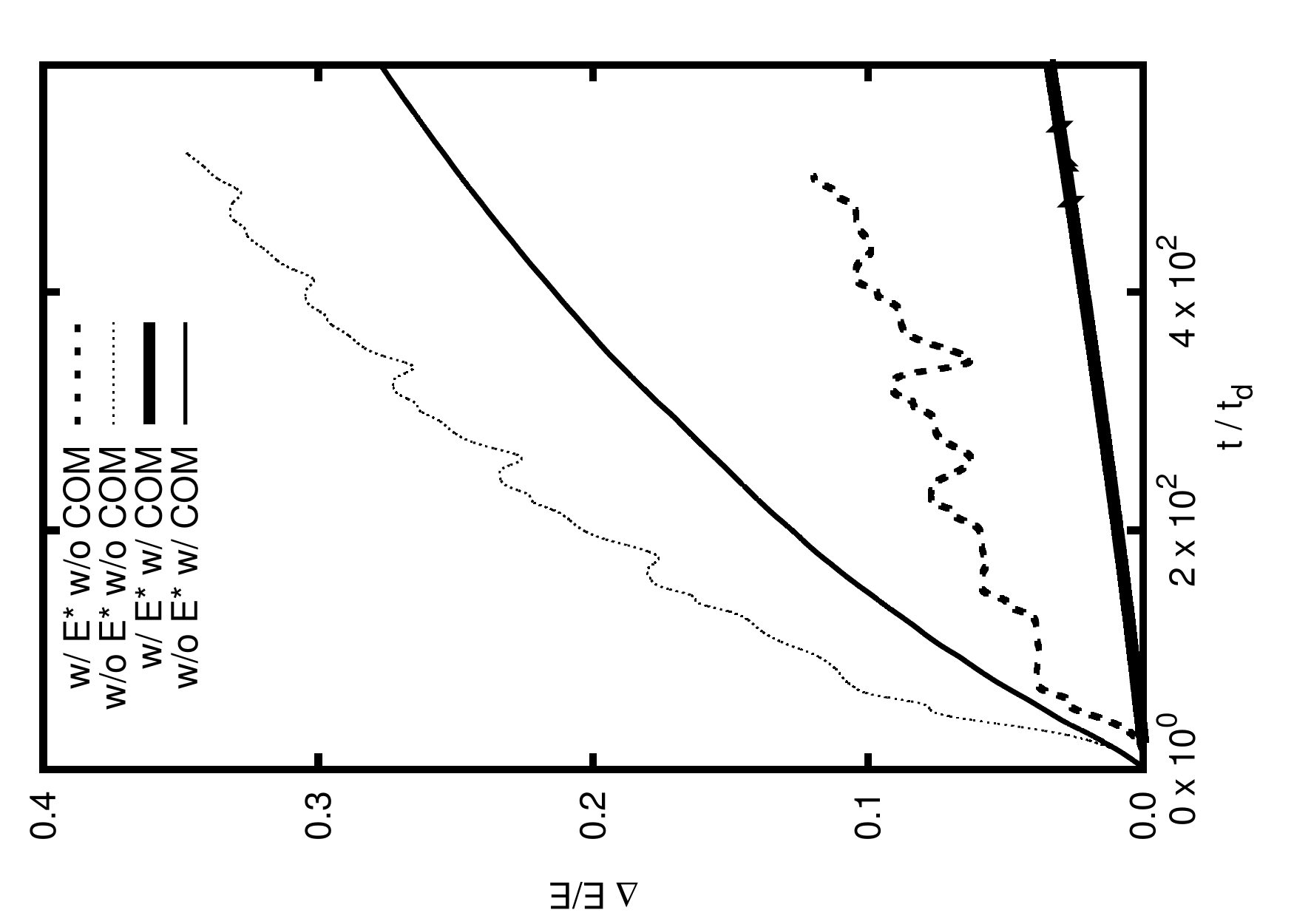} &
\includegraphics[angle=-90,scale=0.39]{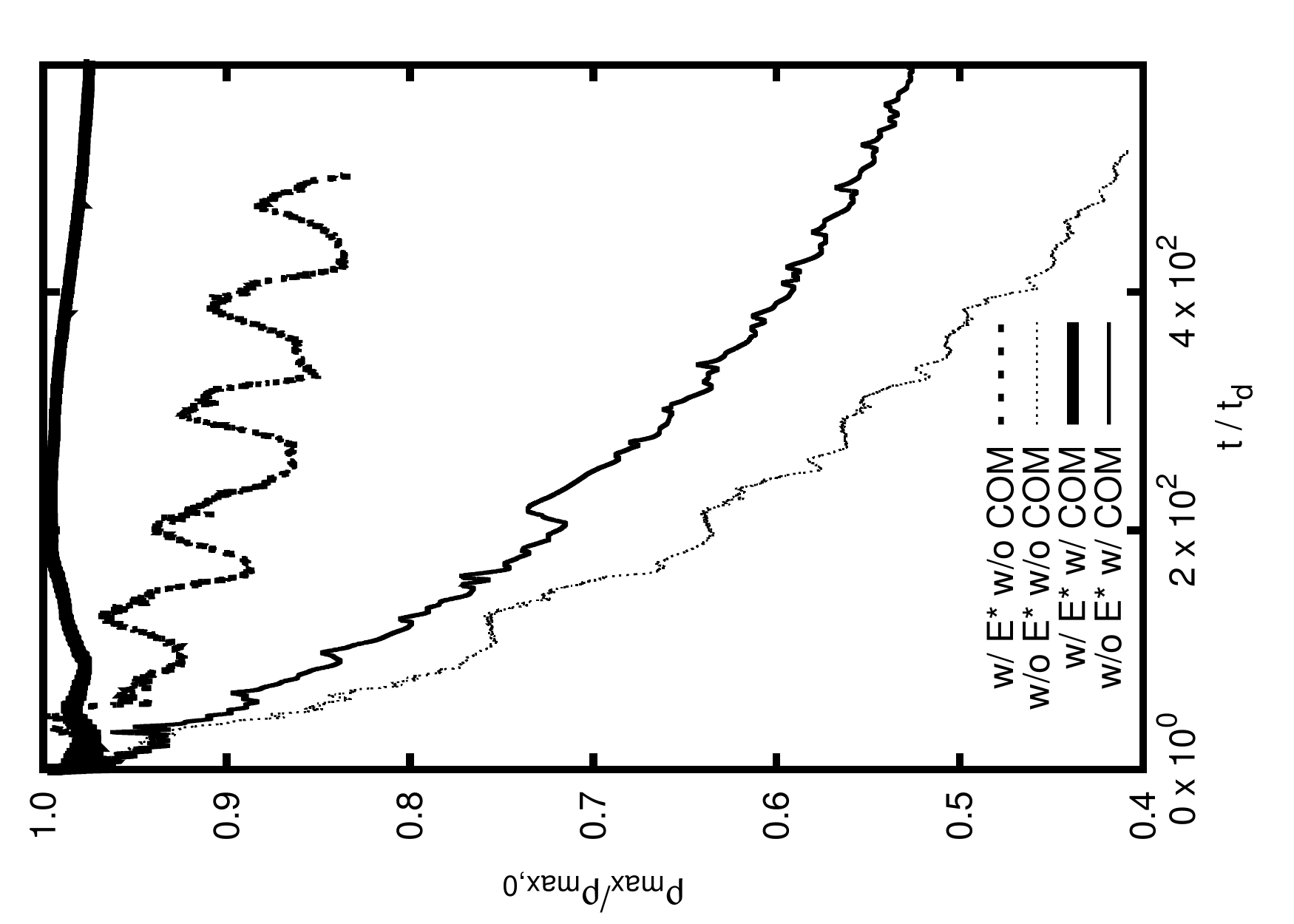} &
\includegraphics[angle=-90,scale=0.39]{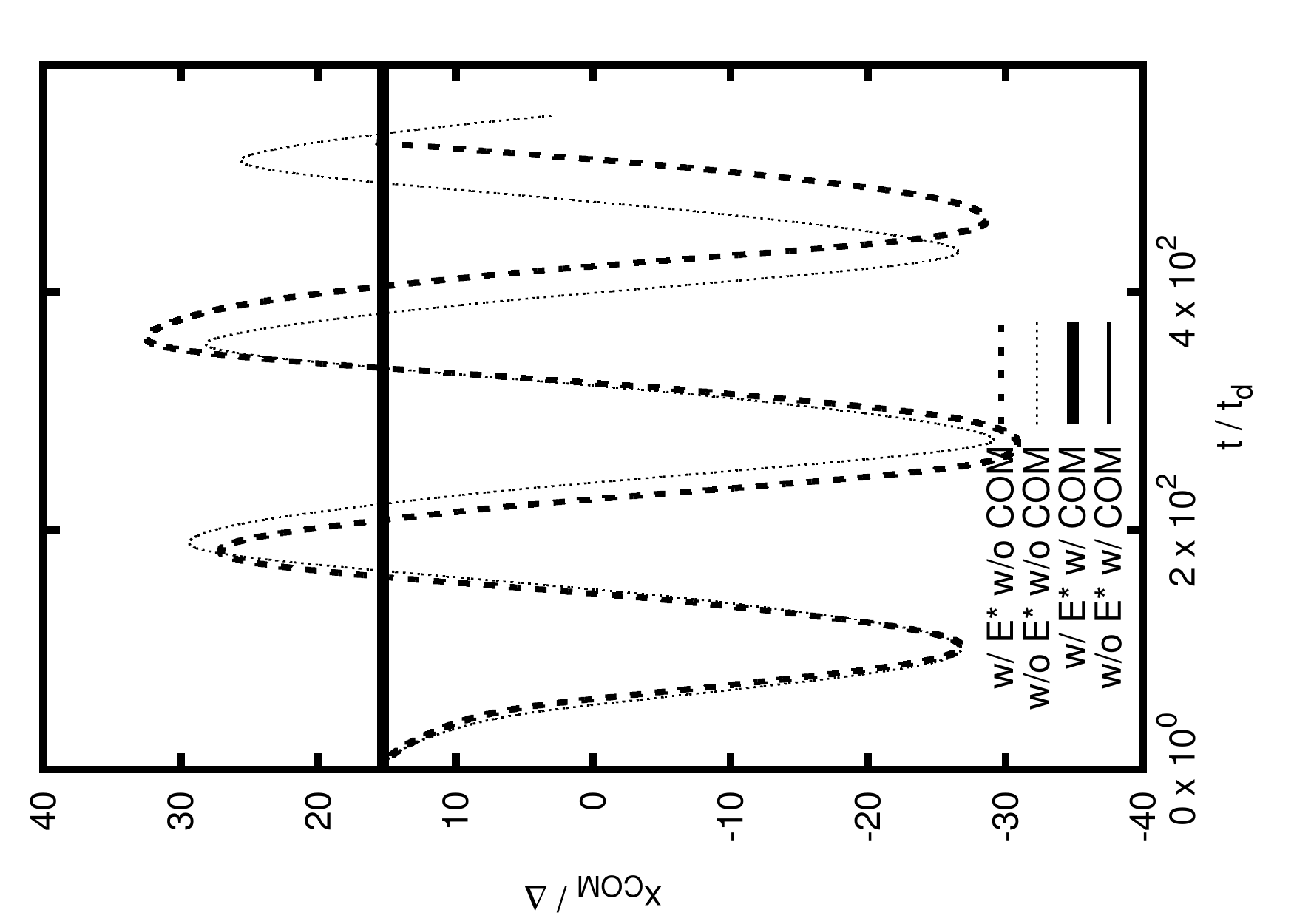} \\
\hline
\end{tabular}
\caption[Off-center polytrope]{  Off-center polytrope.
{\it Left:} Relative change in total energy.
{\it Middle:} Relative change in central density.
{\it Right:} The x-location of the center of mass.
}
\label{poly2}
\end{center}
\end{figure}

We have also run four simulations with the spherical polytrope placed off of the coordinate center. Because we have chosen to evolve the radial, azimuthal, and vertical momenta instead of three Cartesian momenta, we cannot expect that the $x$ and $y$ Cartesian momenta will be conserved.  Cartesian momenta are globally conserved for the centered polytrope due to symmetry. The off-center polytropes do not have this symmetry, and as a result, they act as if they are pushed by an outside force. As shown in the right panel of Figure \ref{poly2},
the polytrope is pushed toward the coordinate center, passes it, and eventually changes directions, resulting in a roughly sinusoidal pattern. A similar effect  was noted by \cite{M2001}, except  the direction of the net force was away from the coordinate origin. 

In Appendix \ref{com_correct} we detail a method we have used to correct for unphysical center of mass motion.  For the simulations detailed in this paper, we have applied the center of mass correction to the above coordinate centered polytrope runs and to two of the off center runs. For the off-center polytrope runs, this correction cancels out the net force pushing on the polytrope. As seen in Figure \ref{poly2}, both the center of mass correction and the E* corrections result in better conservation of total energy (left panel) and a more stable equilibrium configuration. Unlike the coordinate centered runs, however, even with both corrections in use, there is a noticeable increase in the total energy over many dynamical timescales.

\section {Binary Simulations}
\label{binary_section}

\begin{figure}
\includegraphics[scale=0.40]{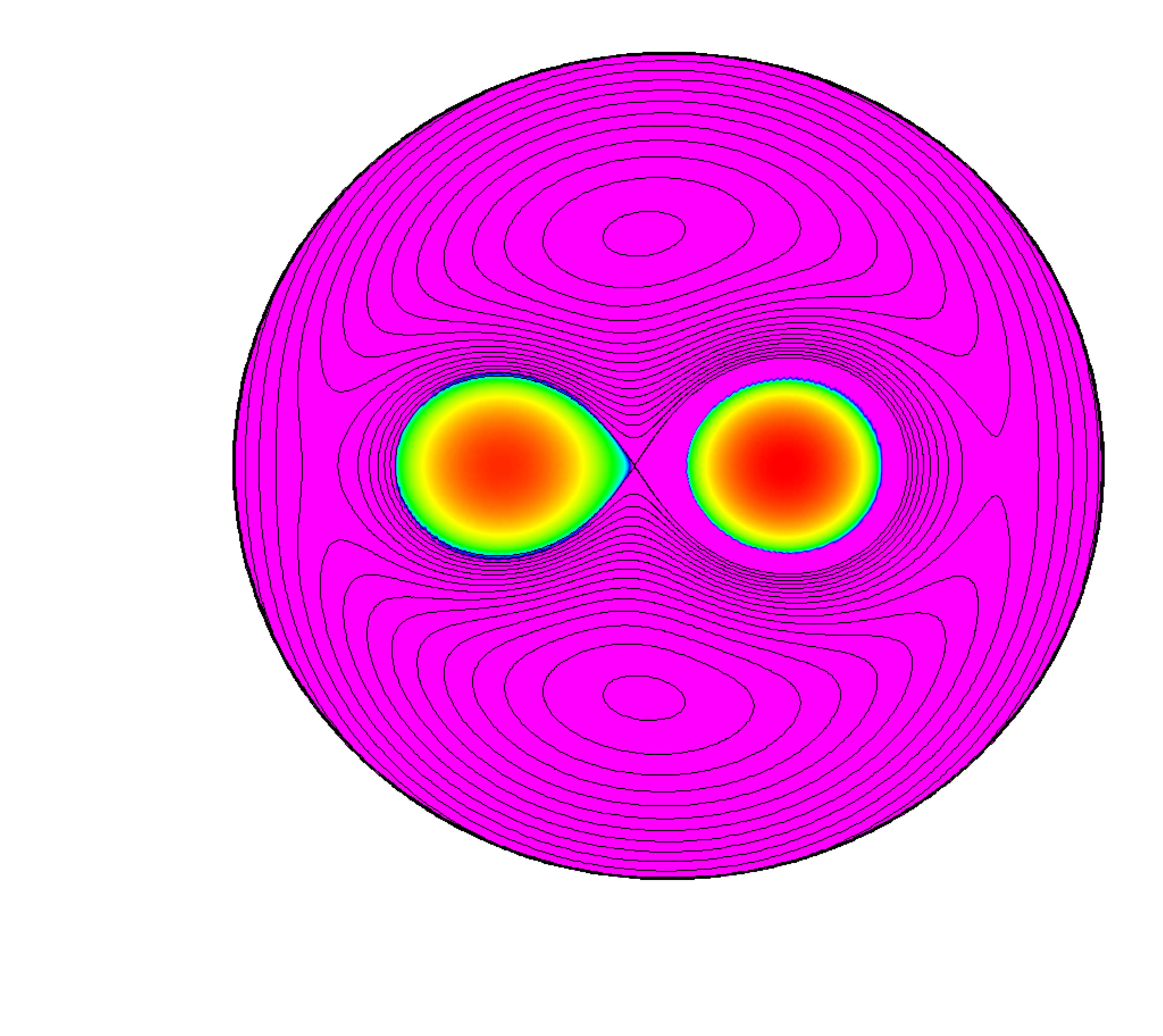}
\caption[Initial density profile in equatorial plane for $q=0.7$ runs]{ Equatorial plane mass distribution for the $q=0.7a$ and $q=0.7b$ runs at $t=0$. The logarithmic color scale runs from $10^{-6}$ to $10^0$ in code units. The black lines are contours of effective potential for $\Phi_\mathrm{eff} \ge \Phi_\mathrm{L1}$. $\Phi_\mathrm{L1}$ is the effective potential at the stationary point between donor and accretor (the ``L1" point).}   
\label{roche_pot}
\end{figure}

Here we present the results of two binary simulations. Both begin with the same initial condition of a polytropic binary with mass ratio $q=0.7$ with the less massive star filling its Roche lobe.
In one of the runs, we have disabled the radiation feature. We call this run ``$q=0.7b$''. We refer to the run with the radiation feature enabled as ``$q=0.7a$". We evolved each on a grid with $159$ radial, $256$ azimuthal, and $49$ vertical interior grid zones. We used reflective symmetry about the equatorial plane, so the effective size in the vertical direction is $98$ zones. The initial equatorial plane mass density distribution and Roche configuration is shown in Figure \ref{roche_pot}.
 The logarithmic color scale runs from $10^{-6}$ to $10^0$ in code units and the contour lines are contours of the effective potential. 

\subsection {Initial Conditions}
\label{initialq7}

The initial conditions were generated using a self-consistent field (SCF) technique similar to that of \cite{H1986}. We have used this technique for several of our previous simulations (e.g. \cite{NT1997}, \cite{MTF2002}, \cite{DMTF2006}), and recently we have extended it to include a cold white dwarf equation of state (\cite{ET2009}). For the present simulations we use the SCF technique with a polytropic equation of state. The polytropic index is set to $n = \frac{3}{2}$. The SCF code generates an initial density configuration and a polytropic constant, $K$, for each component, and determines the orbital period and separation. The parameters of the SCF model used for these simulations are given in Table \ref{scf_params}.
\begin{table}
\caption{ SCF Binary Parameters \tablenotemark{*}  \label{scf_params}}
\begin{tabular}{p{0.45\textwidth} p{0.15\textwidth} p{0.1\textwidth} }
  \tableline
  \tableline
        & Donor & Accretor \\
  \tableline
   Mass & 0.282 & 0.403  \\
   Effective Radius & 0.887 & 0.840  \\
   Central Mass Density & 0.608 & 1.000 \\
   Polytropic Constant ($K$) & 0.236 & 0.257 \\
  Period & \multicolumn{2}{ c }{ 31.19 } \\
  Separation & \multicolumn{2}{ c }{ 2.58 } \\
  Grid Spacing & \multicolumn{2}{ c }{ $\frac{\pi}{128} \approx 0.0245$ } \\
	\tableline
\end{tabular}
\tablenotetext{*}{These values are in ``code" units.}
\end{table}

Given the density and polytropic constants, $p$ in equation (\ref{pressure}) is set equal to $P_\mathrm{poly}$ in equation (\ref{polytropiceos}), to obtain the initial internal energy density for the $q=0.7b$ run. The $q=0.7a$ run requires that we also compute an initial value for the radiation energy density. For the interiors of the stars, where the radiation diffusion approximation applies, the radiation and gas temperatures are equal and the pressure due to radiation is isotropic. In this limit the radiation pressure is 
\begin{equation}
\label{radpress}
P_\mathrm{rad} = \frac{1}{3} E_R = \frac{4}{3}\frac{\sigma}{c} T^4.
\end{equation}
We set the sum of the radiation and gas pressures equal to the polytropic pressure, 
\begin{equation}
\label{radinit}
P_\mathrm{poly} = K \rho^{1+\frac{1}{n}} = \frac{\mathcal{R}}{\mu} \rho T  + \frac{4}{3} \frac{\sigma}{c} T^4,
\end{equation}
and numerically solve for $T$. Then, using Equations (6) and (8) and Equation (125), we set the initial values for internal and radiation energy densities, respectively.

The results of our previous simulations without radiation and with a polytropic equation of state had the benefit of scalability. They were evolved using equations which contain three fundamental units of measure (length, time, and mass), but only one physical constant, Newton's gravitational constant. For a given value of that constant as represented in the code, one is free to choose two out of three scaling constants for the length, mass, and time. With the introduction of radiation transport to the simulation, there are now four independent physical constants in the equation set: (1) the speed of light, (2) the Stefan-Boltzmann constant, (3) the gas constant, and (4) Newton's gravitational constant. Setting these constants fixes the ratio of code units to physical units for length, time, mass, and temperature to only one possible value for each. Therefore the simulation results correspond to a unique physical system. In Table \ref{physcons}
\begin{table}
\caption{ $q=0.7$  Physical Constants in Code Units \label{physcons}}
\begin{tabular}{ p{0.4\textwidth} p{0.25\textwidth}  p{0.25\textwidth} }
  \tableline
  \tableline
  Newton's gravitational constant ($G$) & $1.00 \times 10^{0}$ & $l_\mathrm{code}^3 / m_\mathrm{code} / t_\mathrm{code}^2$ \\
  speed of light ($c$) & $1.98 \times 10^{2}$ & $l_\mathrm{code}^3 / t_\mathrm{code}$ \\
  gas constant  ($\frac{\mathcal{R}}{\mu}$) \tablenotemark{a} & $4.40 \times 10^{-1}$ & $l_\mathrm{code}^2 / t_\mathrm{code}^2 / K_\mathrm{code} $ \\
  Stefan-Boltzmann constant ($\sigma$) &  $2.18 \times 10^{0}$ & $m_\mathrm{code} / t_\mathrm{code}^3 / K_\mathrm{code}^4 $ \\
  \tableline
\end{tabular}
\tablenotetext{a}{The gas constant only appears in the equations divided by the mean molecular weight.}
\end{table}
we list our choice of physical constants for the binary runs, in code units. Note that only $G$ has to be specified in the $q=0.7b$ run. For the $q=0.7a$ run, this choice of constants fixes the ratio of code units to cgs units. These are shown in Table \ref{codetoreal}.
\begin{table}
\caption{ $q=0.7$ Real Units per Code Unit \label{codetoreal}}
\begin{tabular}{ p{0.45\textwidth} p{0.45\textwidth} }
  \tableline
  \tableline
  $l_\mathrm{code}$ & $8.18 \times 10^{9} \ \mathrm{cm}$\\
  $m_\mathrm{code}$ & $2.81 \times 10^{33} \ \mathrm{g}$\\
  $t_\mathrm{code}$ & $5.40 \times 10^{1} \ \mathrm{s}$\\
  $K_\mathrm{code}$ & $1.62 \times 10^{8} \ ^{\circ}\mathrm{K}$\\
  \tableline
\end{tabular}
\end{table}
The binary presented here has a period of $1685 \ \mathrm{s}$ and components with masses $0.57 M_\odot$ and $0.40 M_\odot$ and respective radii $0.098 R_\odot$ and $0.104 R_\odot$. Physically the radii are about ten times too large for fully degenerate white dwarfs (there do exist, however, semi-degenerate helium stars with radii of the same order). The main purpose of these two simulations was to test the method described above rather than to simulate a particular realistic system.

We must also be careful in our choice of $\gamma$. If the entropy gradient is opposite the pressure gradient, convective instability results. To avoid instability without radiation, setting $\gamma \ge 1 + \frac{1}{n}$ is sufficient. Since $\gamma=\frac{5}{3}$ for a monatomic ideal gas, and $1+\frac{1}{n} = \frac{5}{3}$ for an $n = \frac{3}{2}$ polytrope, usually we would set $\gamma=\frac{5}{3}$. This results in an isentropic entropy profile. The addition of radiation, however, changes the entropy profile. Since the temperature is now set by equation (\ref{radinit}) instead of equation (\ref{idealgas}), the entropy profile runs with the pressure gradient. 
One way to get the entropy gradient to point in the right direction is to set $\gamma$ to a higher value. We have found empirically that setting $\gamma=1.671$ results
in stability against convection for the accretor and donor.

Outside of the two stars we simply set the internal gas and radiation energy densities to their floor values. For the present simulations, theses values are $1.49 \times 10^{-23}$ and $1.174 \times 10^{-24}$, respectively, in code units. They were chosen such that the radiation coupling terms (i.e. the RHS of equation (\ref{internal_gas_energy})) will sum to zero.

\subsection{Quality of Results}
\label{qorsection}
\begin{figure}
\begin{center}
\begin{tabular}{|c|c|c}
\hline
\includegraphics[angle=-90,scale=0.39]{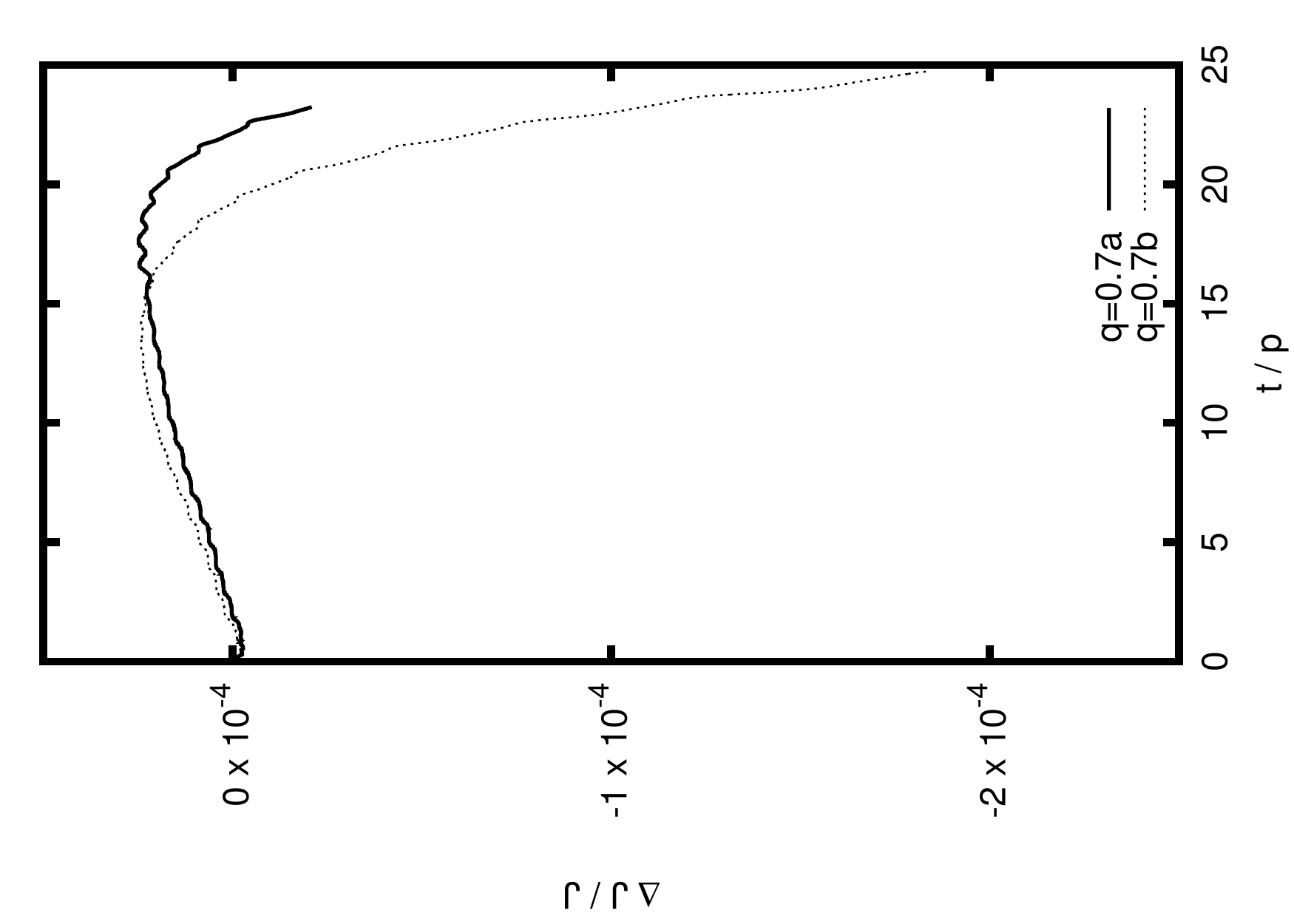} &
\includegraphics[angle=-90,scale=0.39]{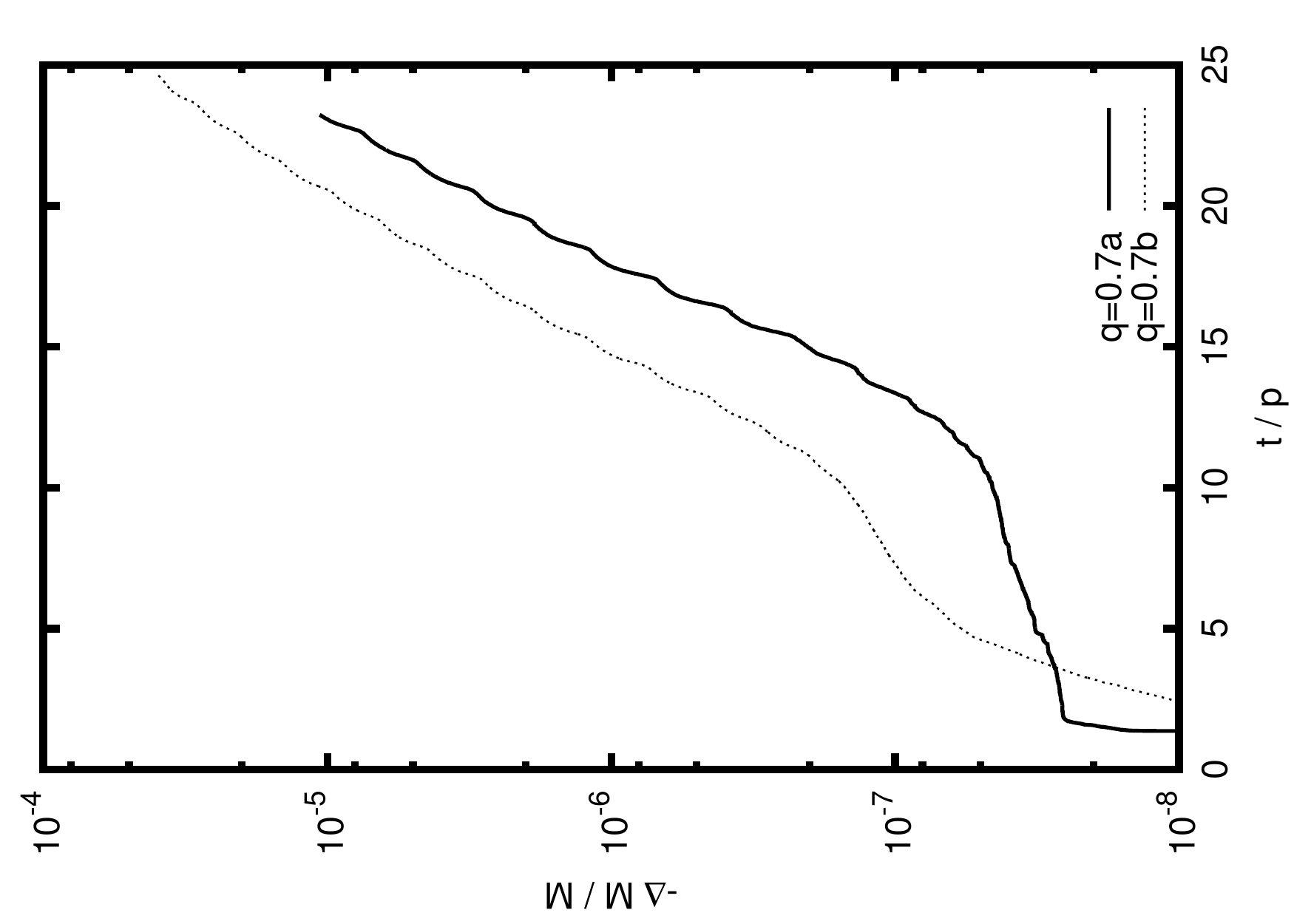} &
\includegraphics[angle=-90,scale=0.39]{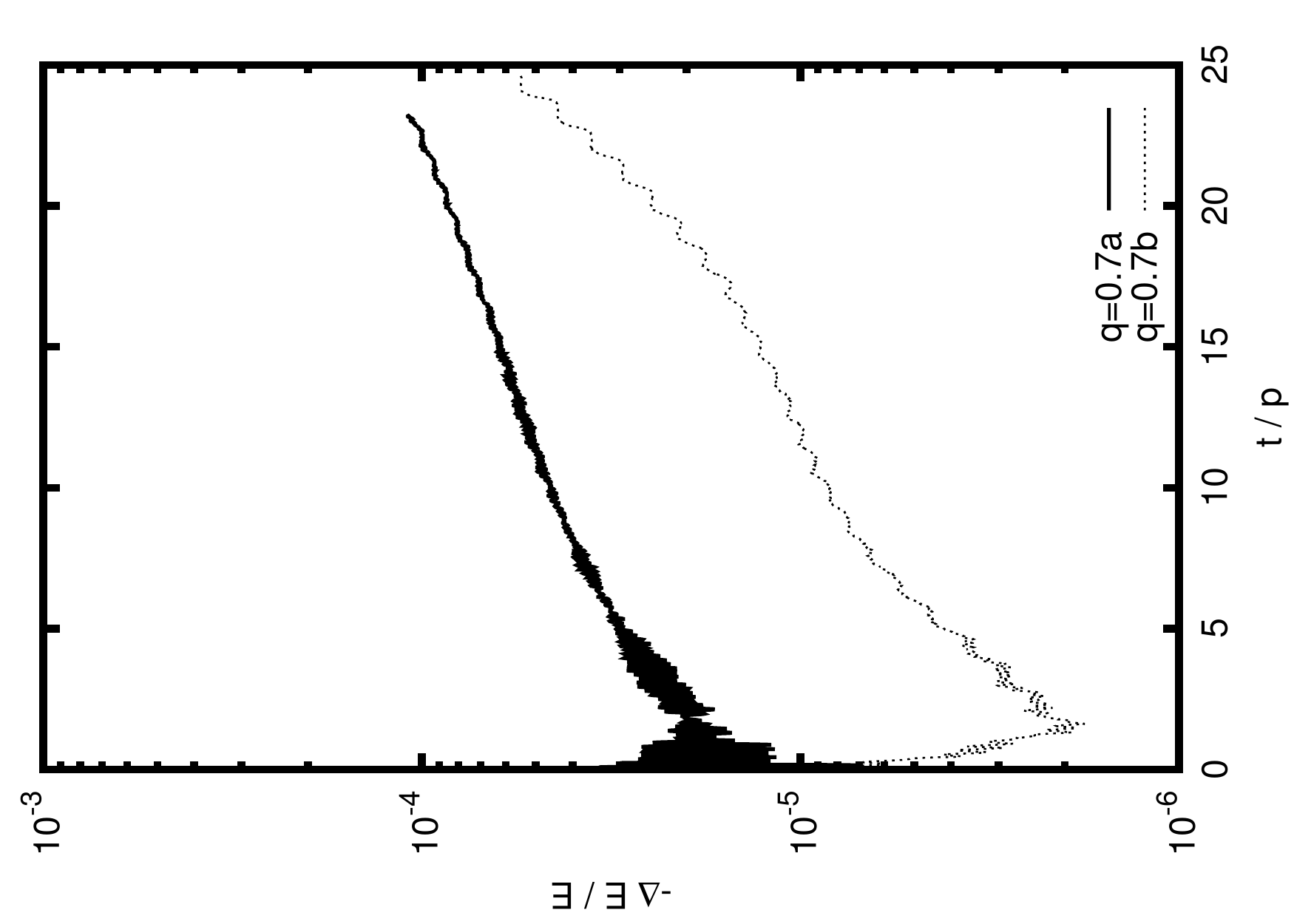} \\
\hline
\end{tabular}
\caption[Total angular momentum, mass, and energy for the $q=0.7$ runs]{Binary runs $q=0.7a$ (solid curve) and $q=0.7b$ (dotted curve). 
{\it Left:} The relative change in the total z-angular momentum on the grid from its initial value.
{\it Middle:} The negative of relative change in the total mass on the grid from its initial value.
{\it Right:} The negative of relative change in the total energy on the grid from its initial value.} 
\label{q7com}
\end{center}
\end{figure}

\begin{figure}
\begin{center}
\begin{tabular}{|c|c|c|}
\hline
\includegraphics[angle=-90,scale=0.39]{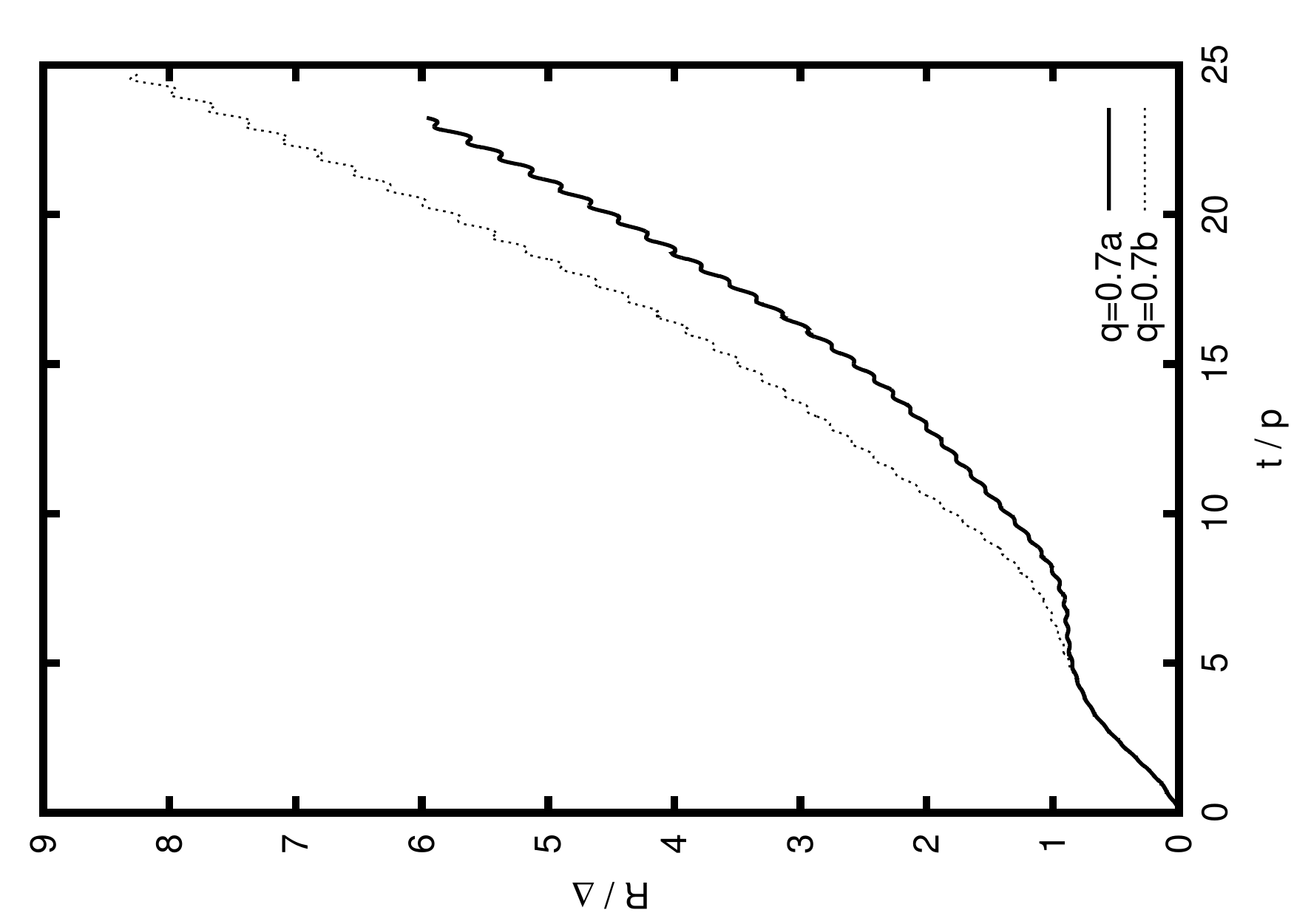} & 
\includegraphics[angle=-90,scale=0.39]{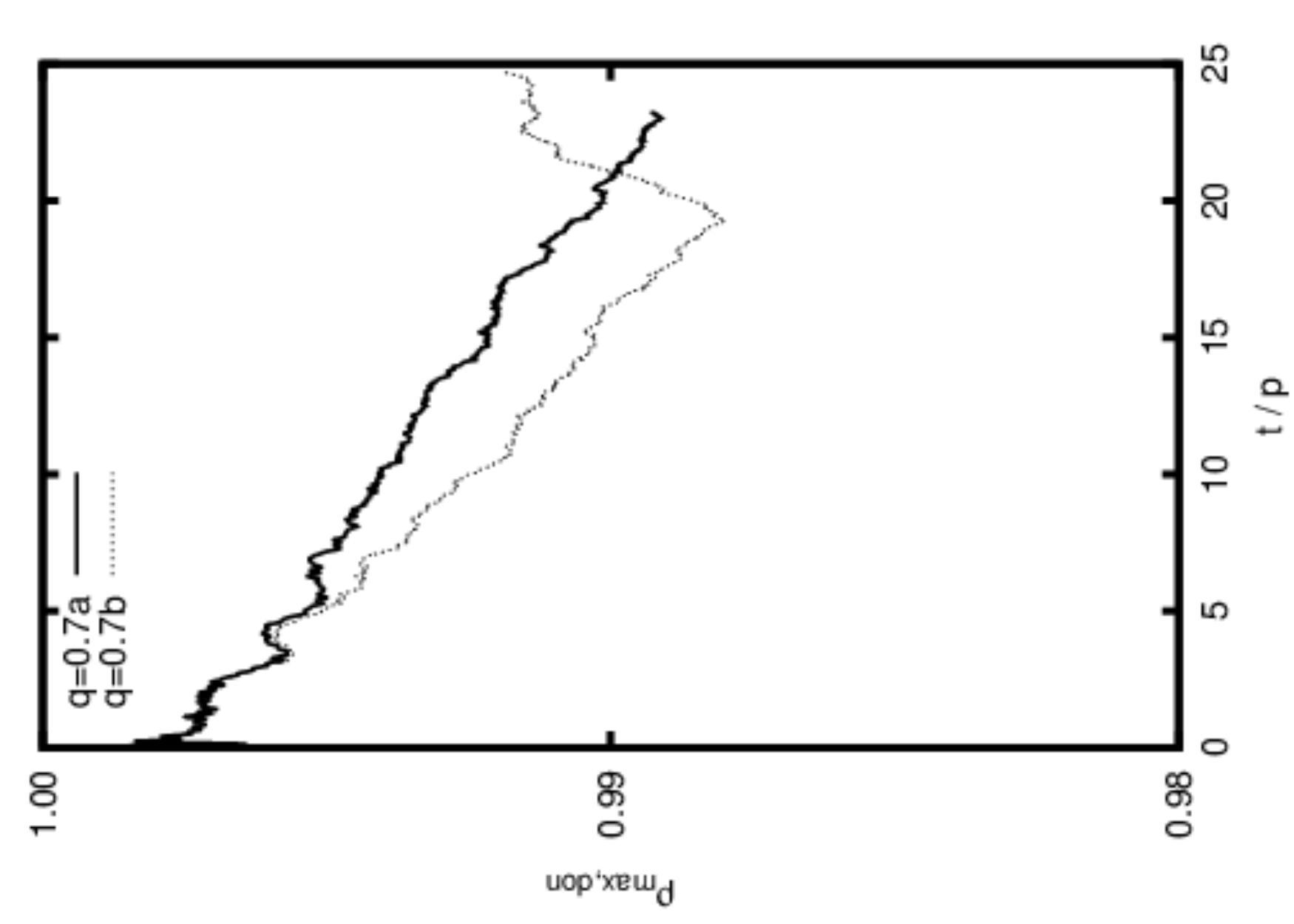} & 
\includegraphics[angle=-90,scale=0.39]{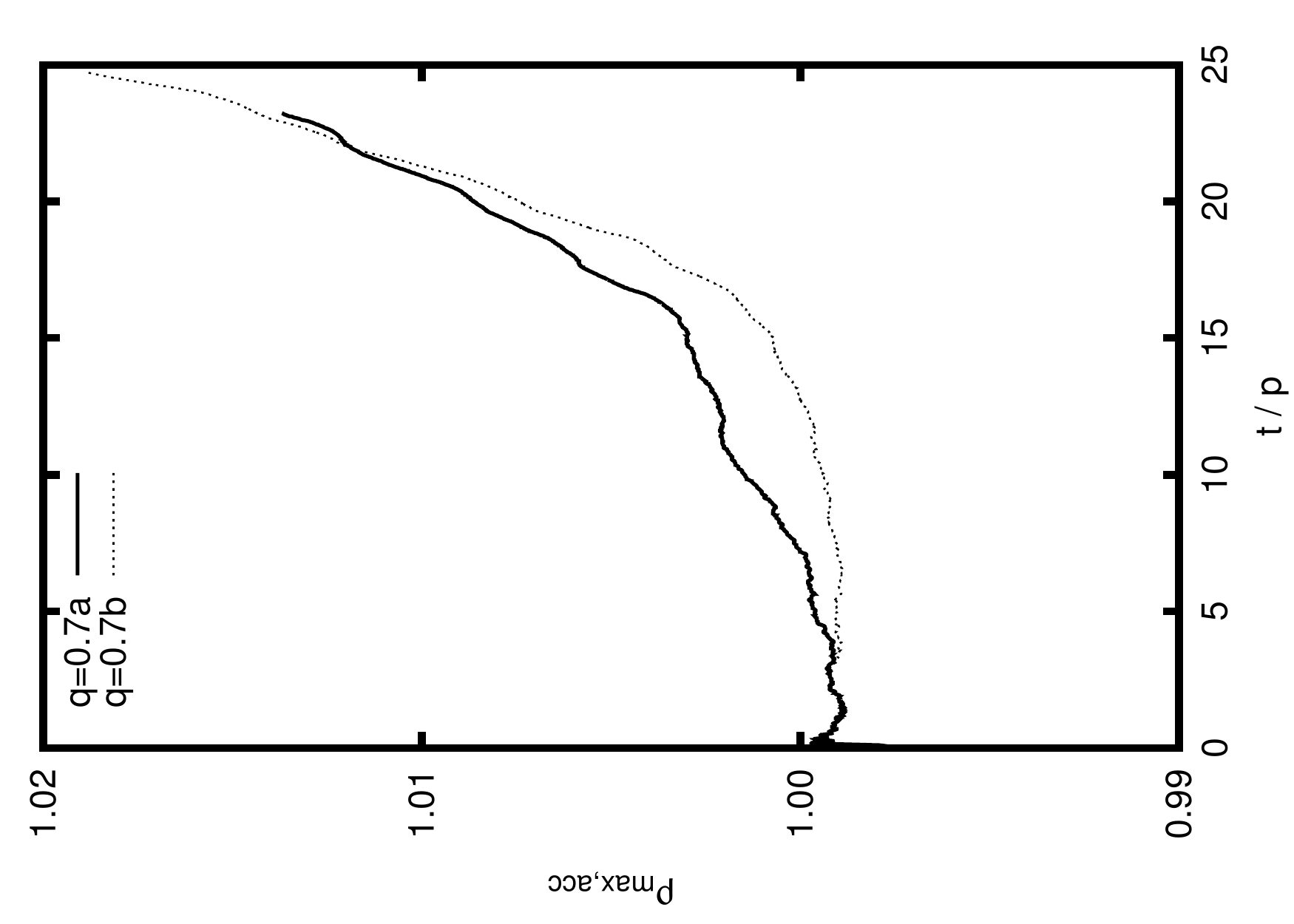} \\
\hline
\end{tabular}
\caption[Center of mass location and maximum densities for $q=0.7$ runs]{ Binary runs $q=0.7a$ (solid curve) and $q=0.7b$ (dotted curve). 
{\it Left:} The orbit averaged radial location of the center of mass of the entire grid.
{\it Middle:} The orbit averaged maximum density of the donor, normalized to its initial value.
{\it Right:} The orbit averaged maximum density of the accretor, normalized to its initial value.}
\label{q7rho_max}
\end{center}
\end{figure}

\begin{figure}
\begin{center}
\begin{tabular}{|c|c|}
\hline
\includegraphics[angle=-90,scale=0.39]{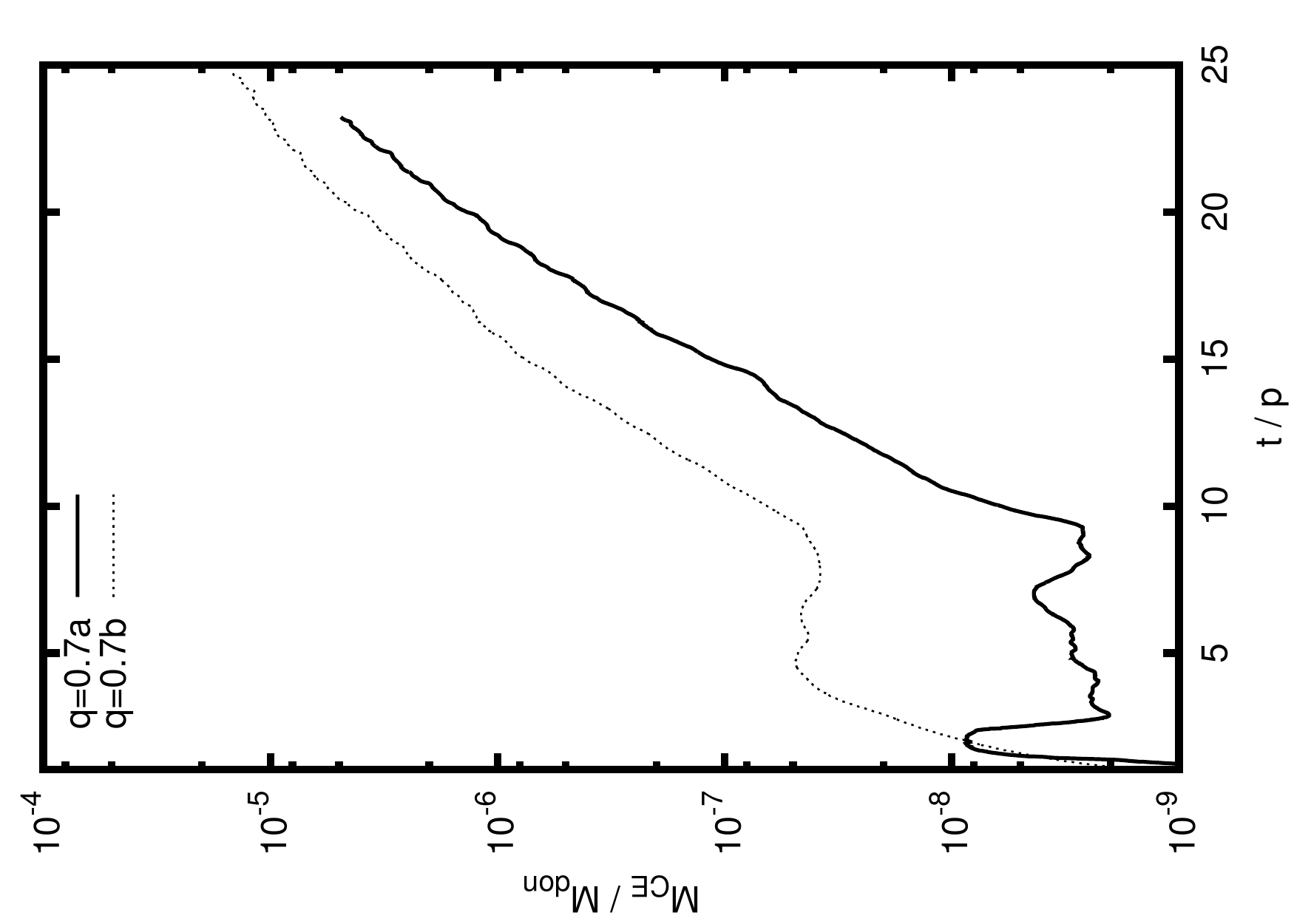} & 
\includegraphics[angle=-90,scale=0.39]{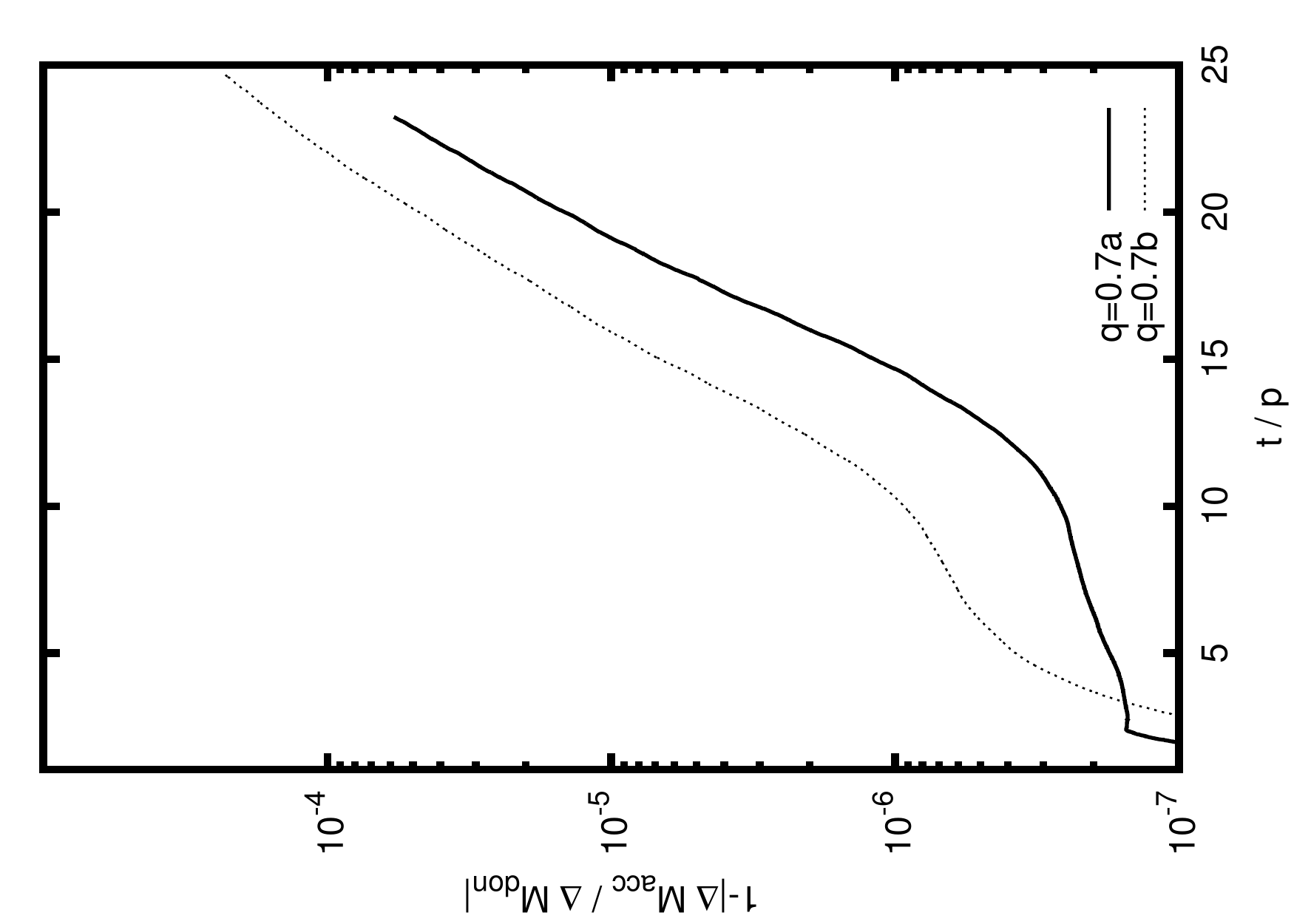} \\
\hline
\end{tabular}
\caption[Common envelope mass and fraction of mass lost for $q=0.7$ runs]{Binary runs $q=0.7a$ (solid curve) and $q=0.7b$ (dotted curve).  {\it Left:} The orbit averaged mass of the common envelope, normalized to the mass of the donor.
{\it Right:} The running total of the fraction of mass lost from the donor that is not captured by the accretor.
}
\label{q7mdot}
\end{center}
\end{figure}

The $q=0.7a$ and $q=0.7b$ runs were evolved for approximately $23$ and $25$ orbits, respectively. The relative change in the total angular momentum (left panel), mass (middle panel), and energy (right panel) are plotted in Figure \ref{q7com}.
 The runs were terminated because the center of mass moved off the coordinate origin by several grid zones (see left most panel in Figure \ref{q7rho_max}).
\cite{DMTF2006} used a center of mass correction.  We did not use such a correction for the $q=0.7a$ and $q=0.7b$ runs. There is significant epicyclic variation evident in many of the Figures towards the end of each run, and we believe this is primarily due to the wandering center of mass. Our chosen grid size also turned out to be too small. The expanded atmospheres of the stars were beginning to come into contact with the upper vertical boundary shortly before terminating each run. This is the reason for the accelerated rate of mass
loss seen in the middle panel of Figure \ref{q7mdot}.
Since the $q=0.7a$ and $q=0.7b$ runs we have added a  correction to the present code (see Appendix \ref{com_correct}). This correction is used in some of the single polytrope runs described in \S \ref{singlesection}.
As seen in Figure \ref{q7com}, relative mass loss through the grid boundaries of the same order as the relative changes in total angular momentum and energy does not begin until just after the $15^\mathrm{th}$ period. We may therefore assume that most all of the changes in angular momentum and total energy at early times are due to error in the numerical scheme. During this period of the evolution, the total z-angular momentum is conserved to within a relative error of approximately $1.7 \times 10^{-6}$ per orbit. This is a marked improvement over our previous code, which conserves angular momentum to within an error of about $1 \times 10^{-4}$ per orbit (\cite{MTF2002}), and it is on par with recent SPH codes (\cite{YPR2007}, \cite{GGI2004}). Because of the E* scheme, our code is able to conserve total energy to even better accuracy. Over the first $15$ orbits, total energy is conserved to within a relative error of about $8 \times 10^{-7}$ per orbit for the $q=0.7b$ run and $4 \times 10^{-6}$ per orbit for the $q=0.7a$ run. Although we should expect the $q=0.7a$ run to lose total energy by radiation leaving the grid, as will be shown below the amount of radiative luminosity is not enough to account for the higher error in the $q=0.7a$ run relative to the $q=0.7b$. Each of the runs conserve total energy better than the SPH codes mentioned above.

For the purposes of producing the figures, we have defined the ``common envelope" to be any point on the grid for which $\Phi_\mathrm{eff} + \frac{1}{2} \mathbf{u}^2 < \Phi_{L2}$, where $\Phi_{L2}$ is the effective potential at the stationary point opposite the donor from the accretor. If its gravitational binding energy is below this threshold, a grid cell belongs to either the accretor or donor depending on which of the two exerts more gravitational acceleration at that point. 
 
\begin{figure}
\begin{center}
\begin{tabular}{|c|c|c|c|}
\hline
\includegraphics[scale=0.155]{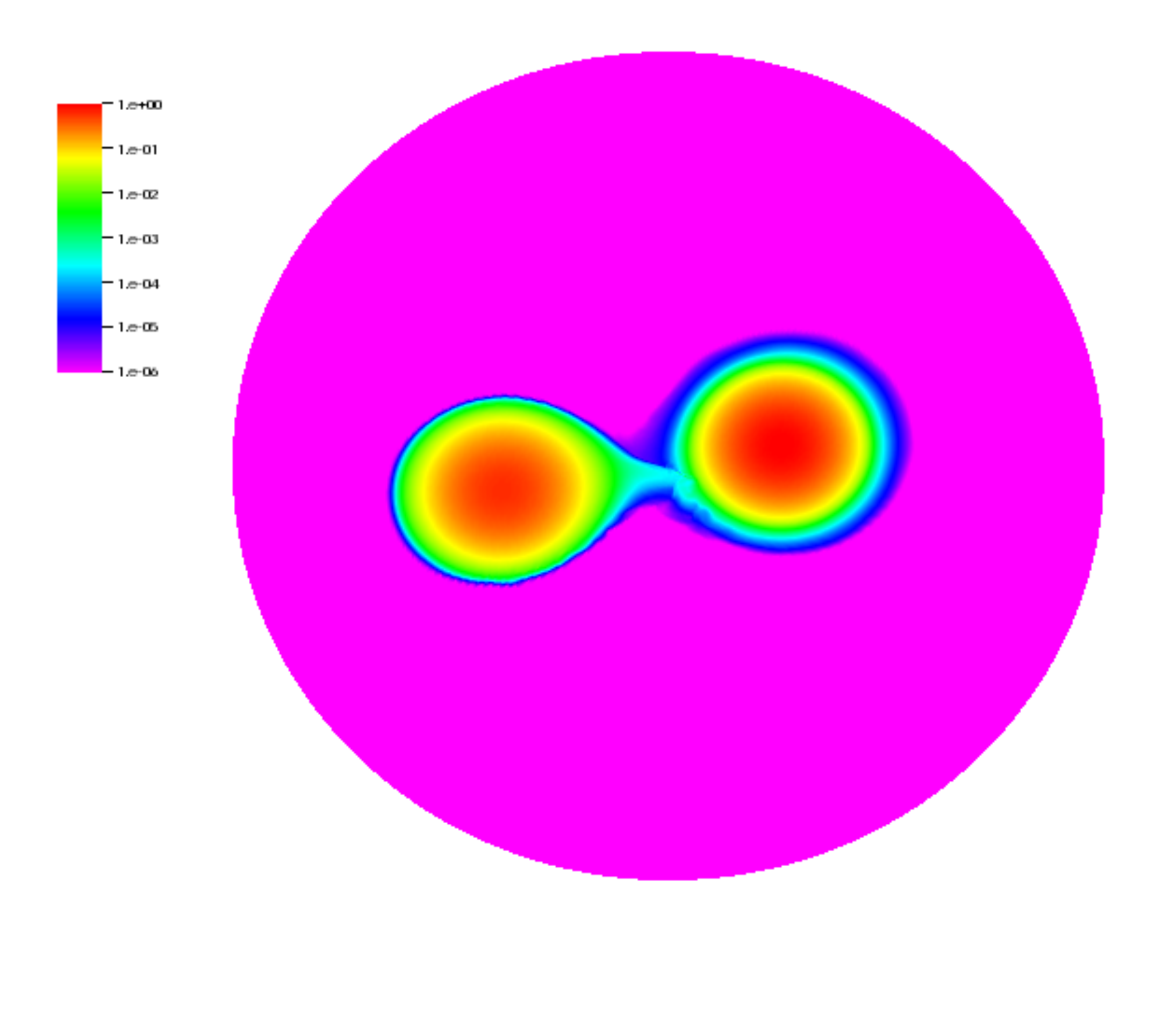} & 
\includegraphics[scale=0.155]{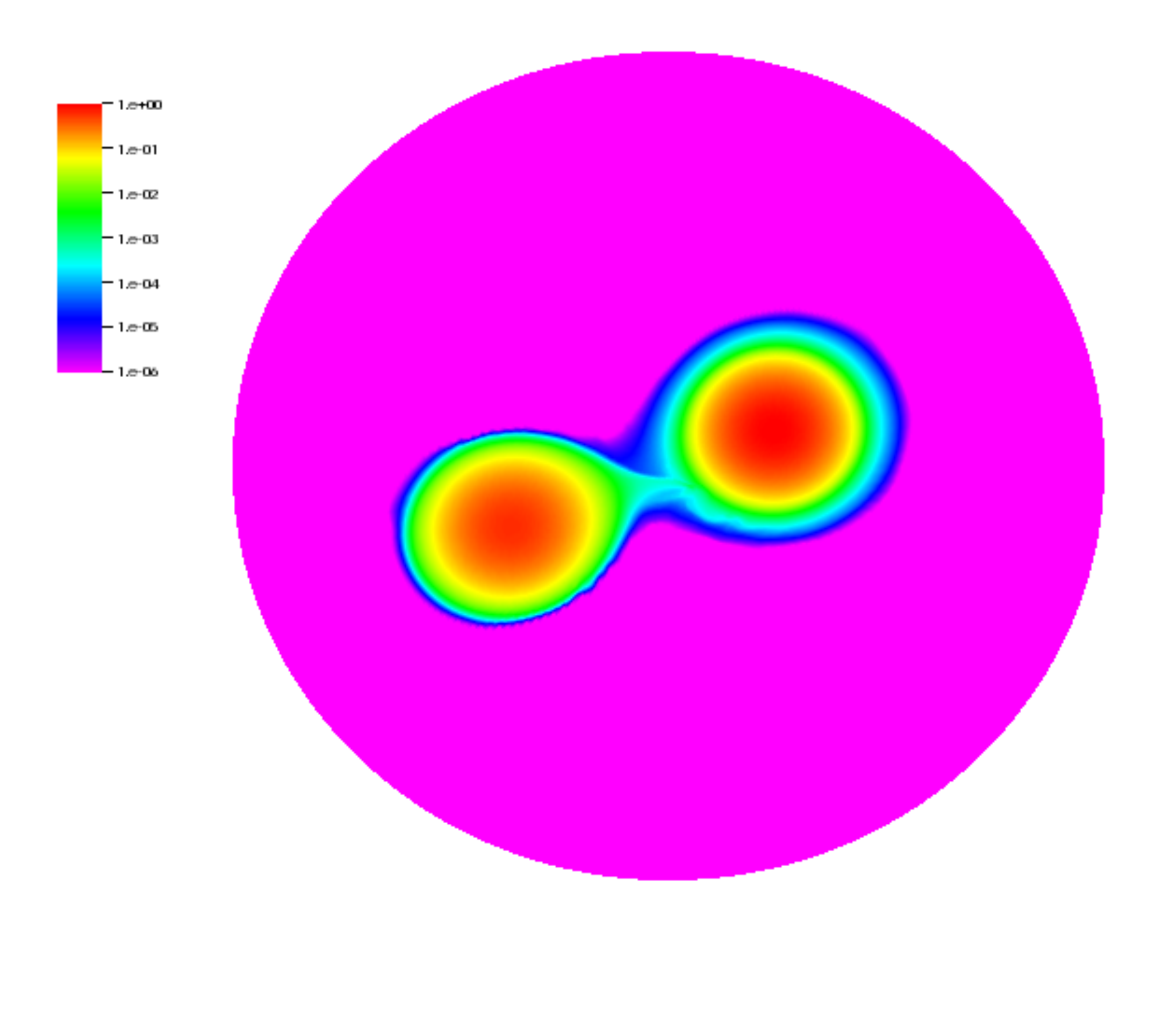} &
\includegraphics[scale=0.155]{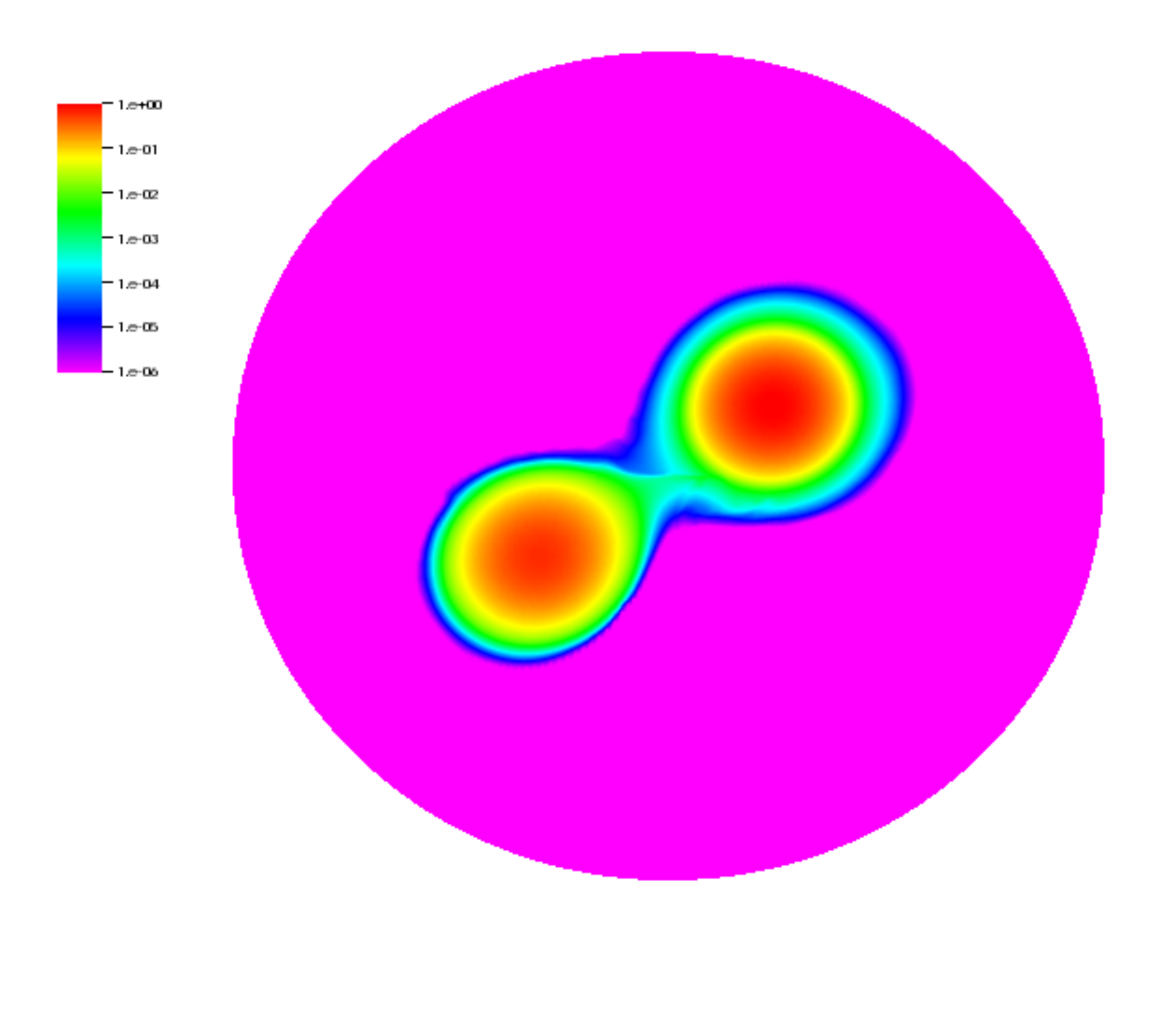} &
\includegraphics[scale=0.155]{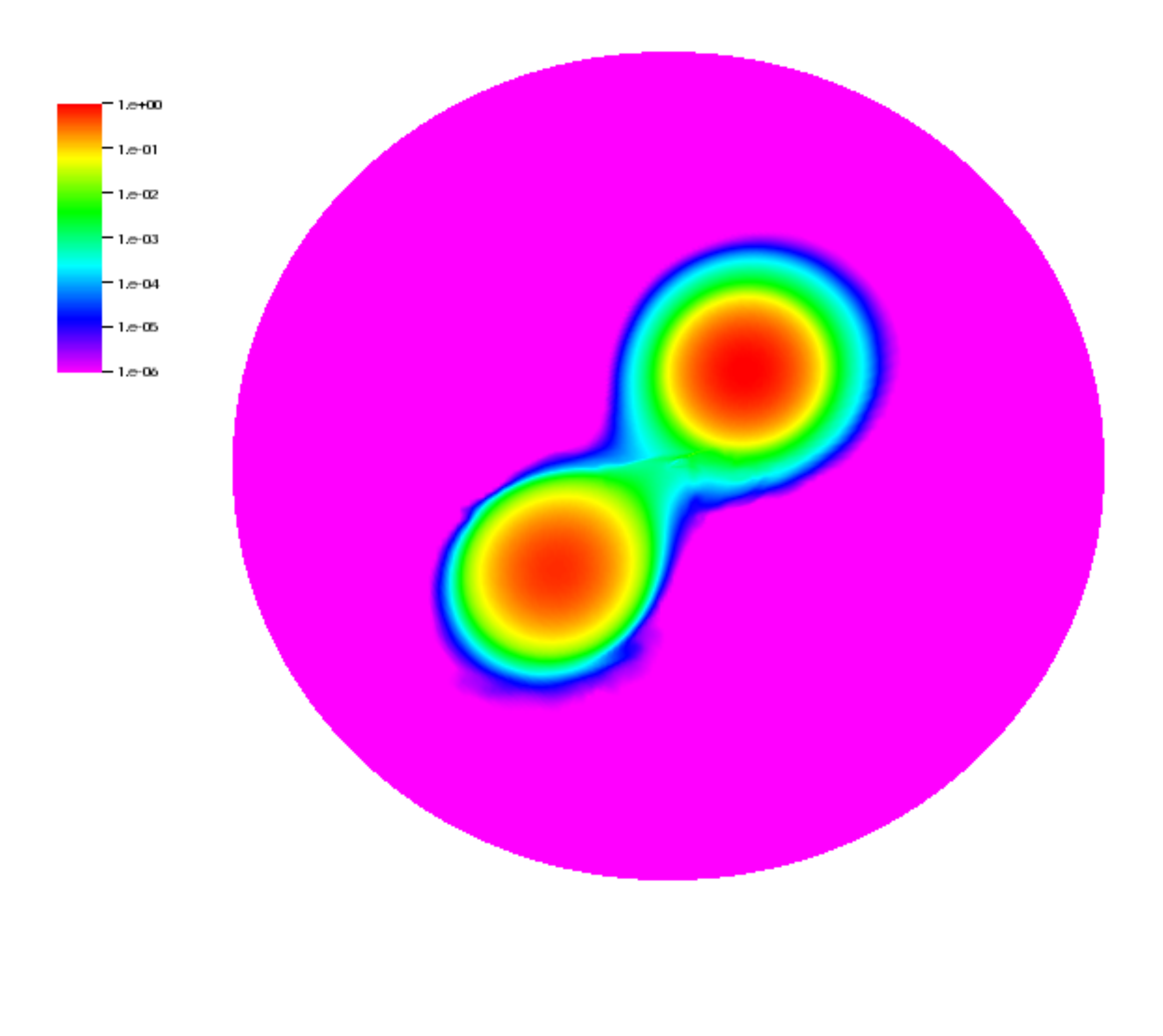} \\
\hline
\includegraphics[scale=0.155]{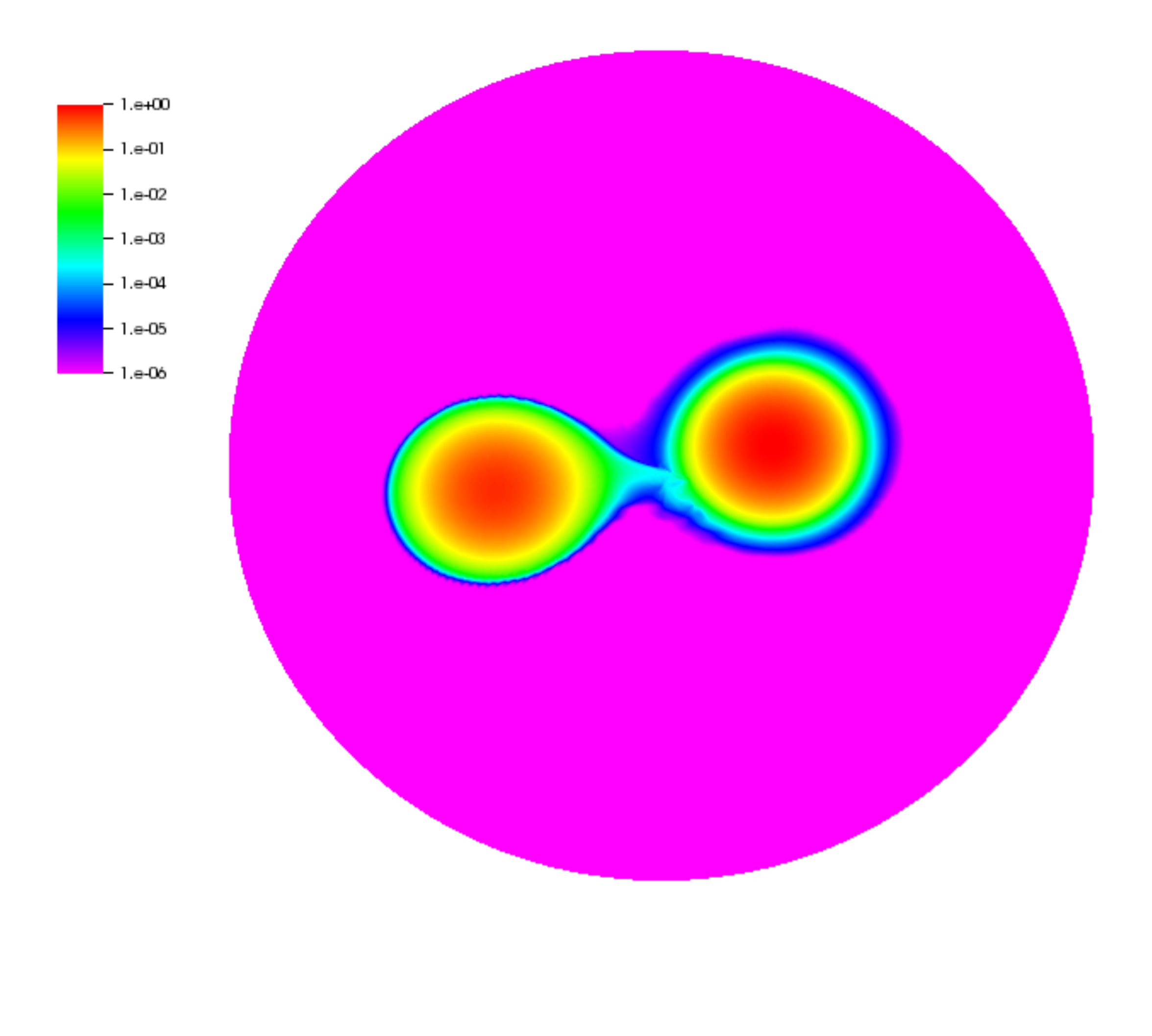} &
\includegraphics[scale=0.155]{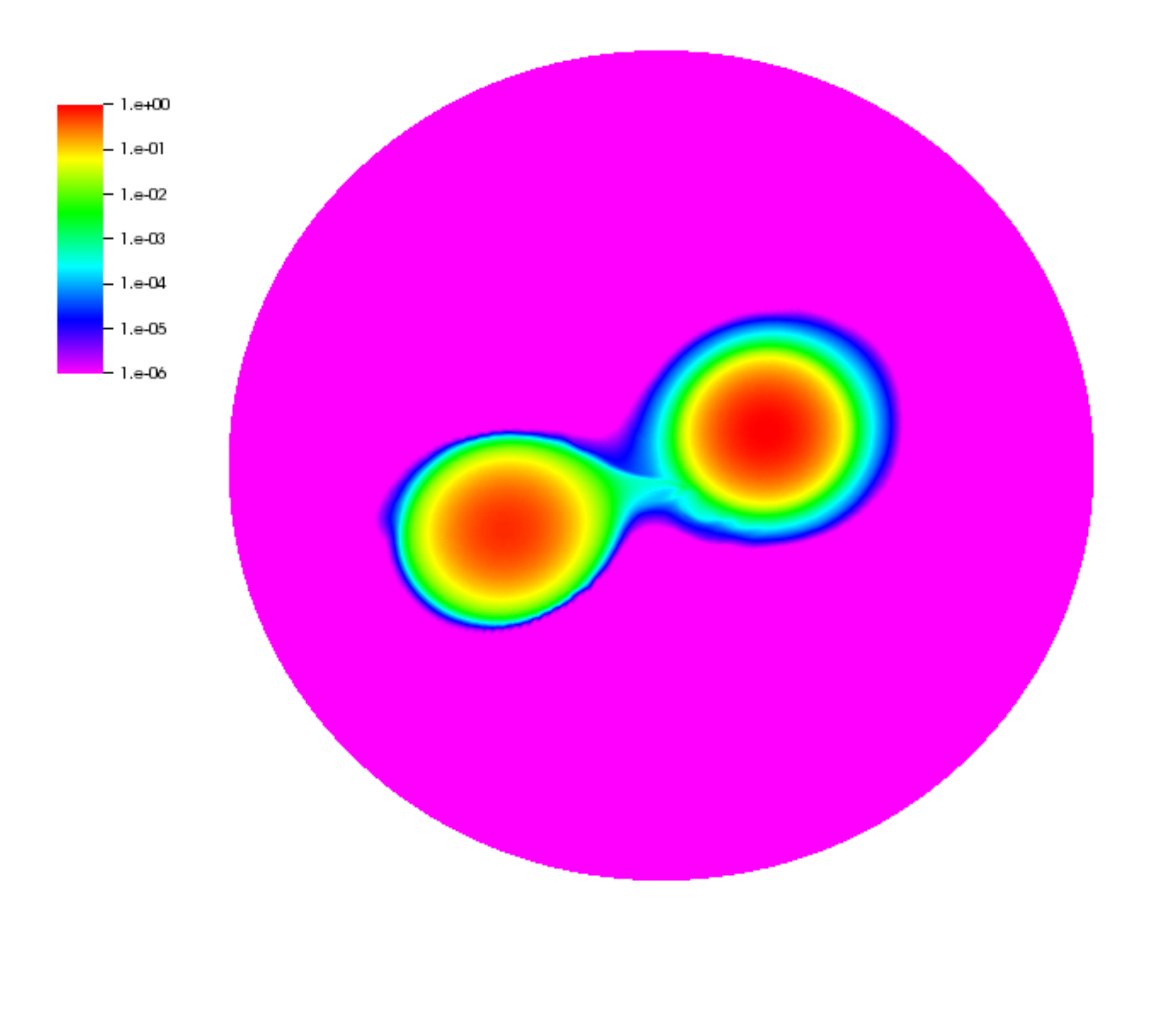} &
\includegraphics[scale=0.155]{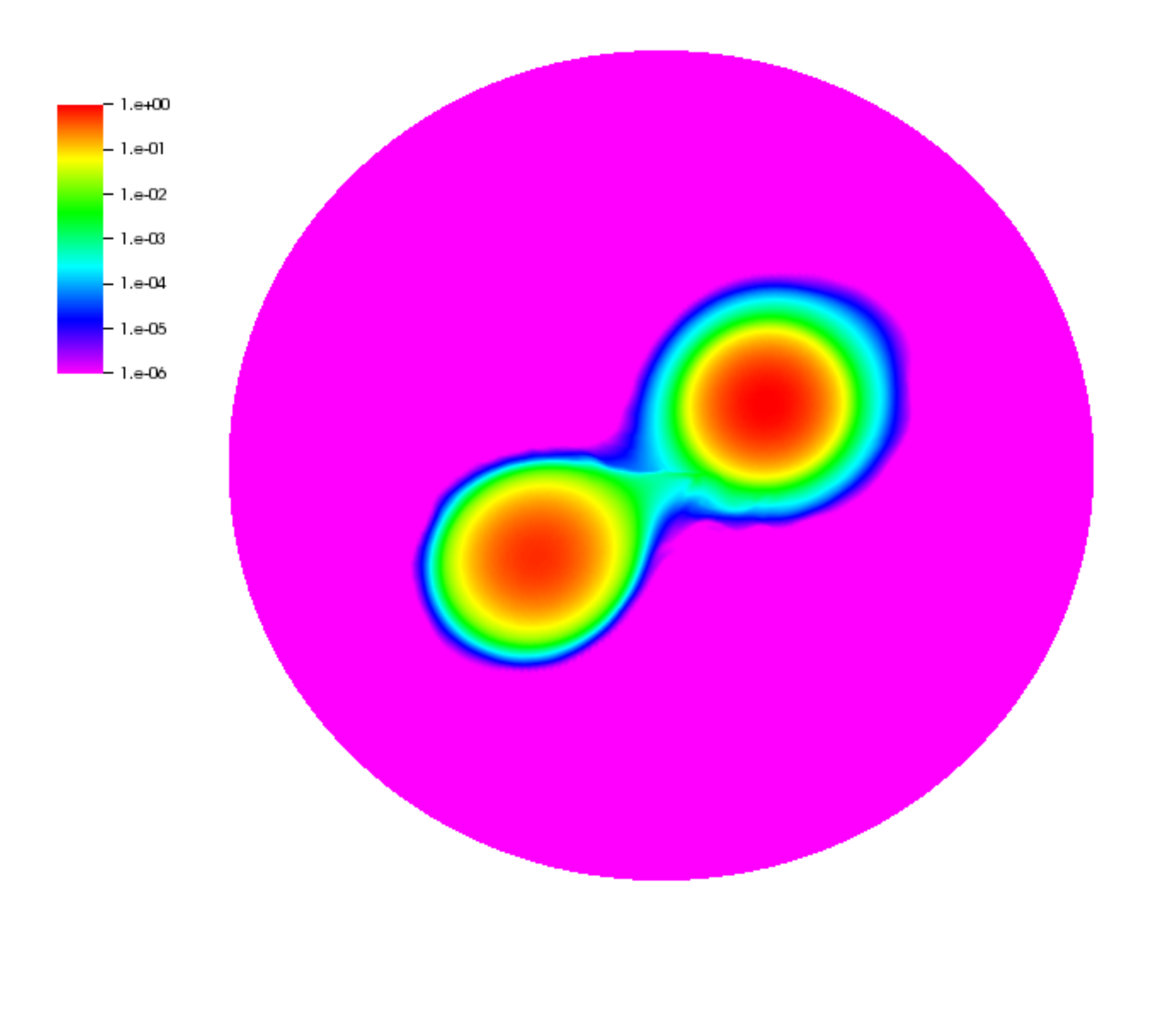} &
\includegraphics[scale=0.155]{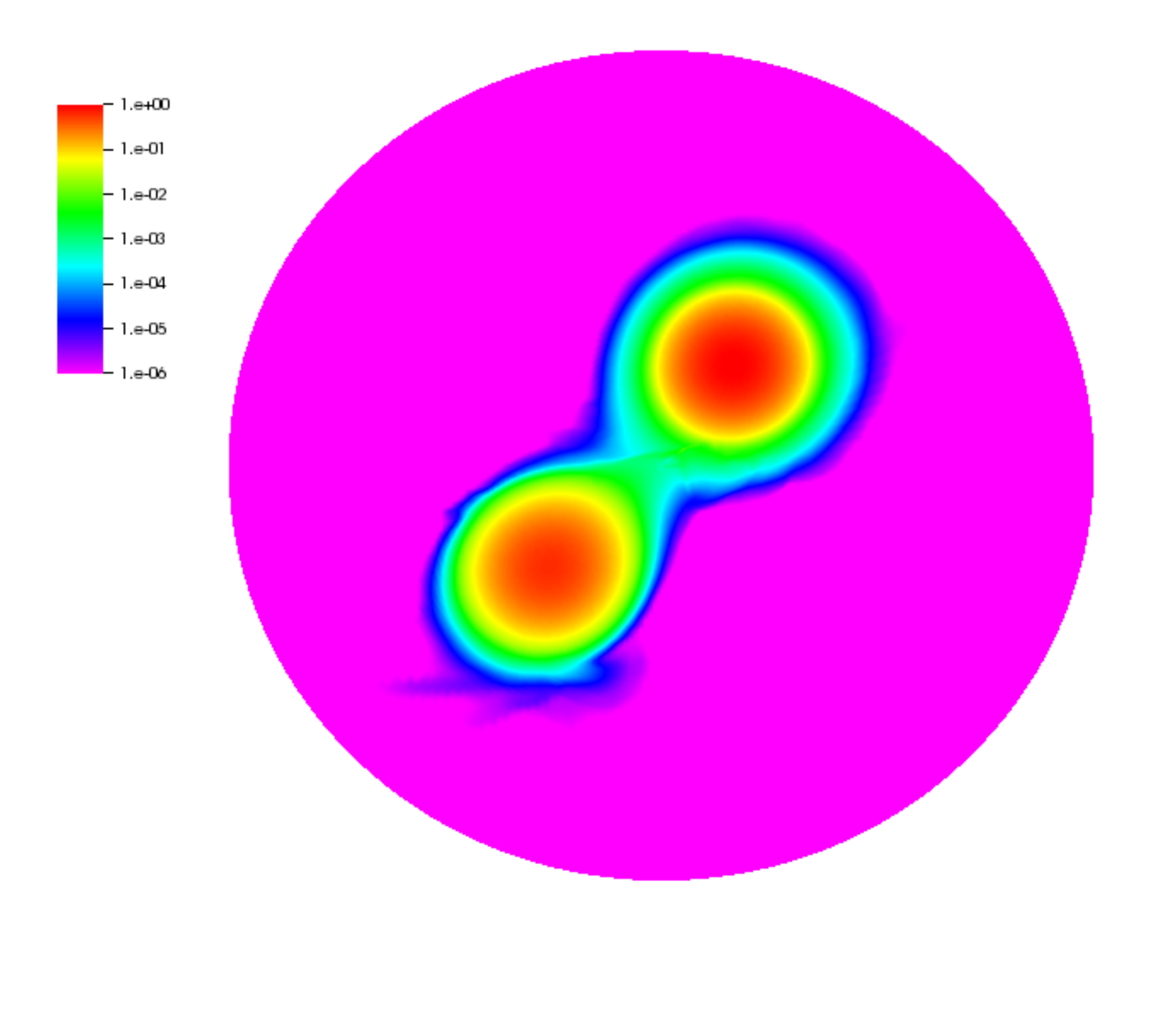} \\
\hline 
\end{tabular}
\caption[High density range equatorial slices for $q=0.7$ runs]{ These are density plots of a slice through the equatorial plane for the $q=0.7$ runs. The top row is the $q=0.7a$ run and the bottom row the $q=0.7b$ run. From left to right, the columns correspond to $t = 5$ orbits, $10$ orbits, $15$ orbits, and $20$ orbits. The color density scale runs from $10^{-6}$ to $1$ in code units.} 
\label{panel1}
\end{center}
\end{figure}

\begin{figure}
\begin{center}
\includegraphics[scale=0.324]{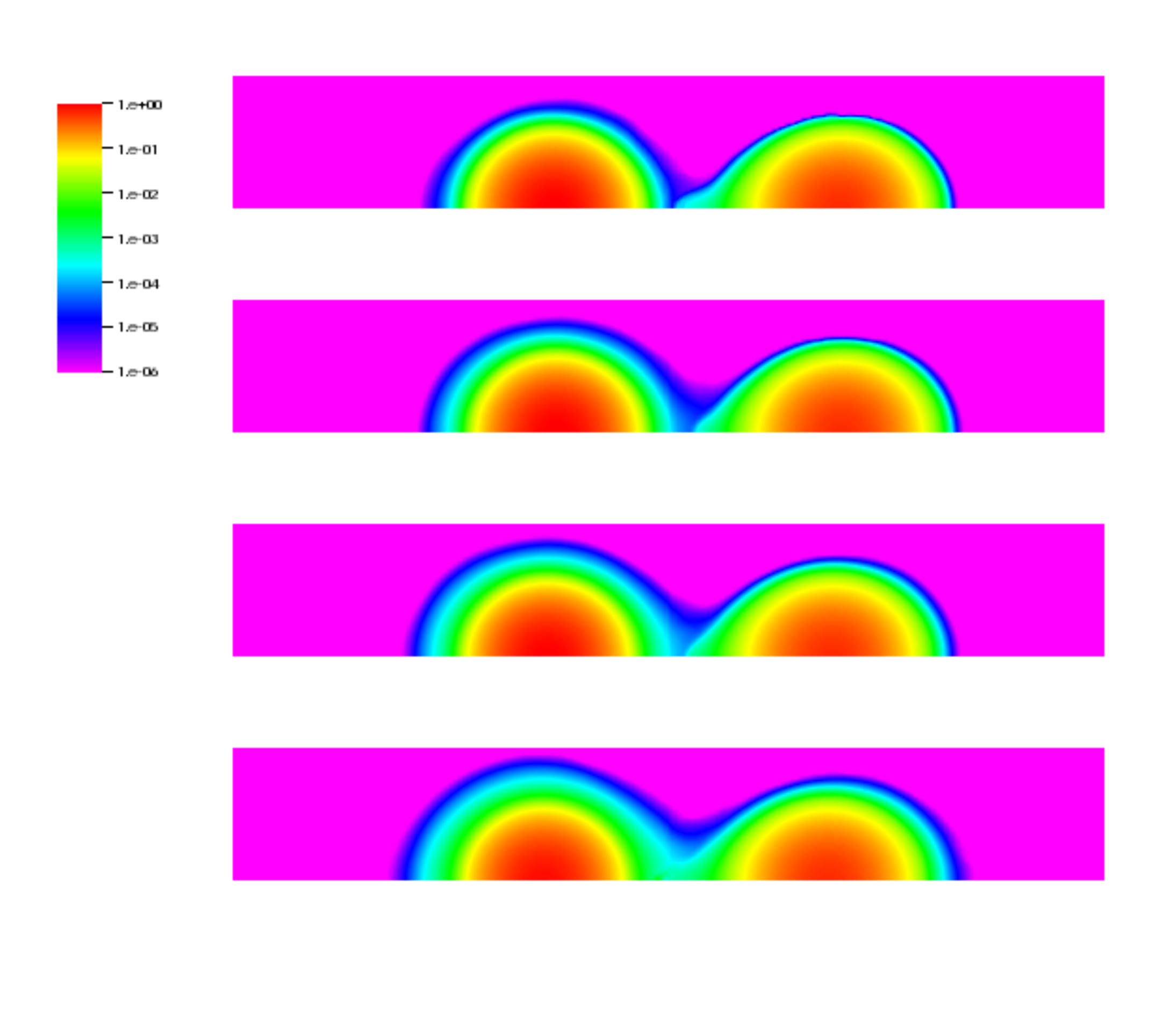}
\includegraphics[scale=0.20]{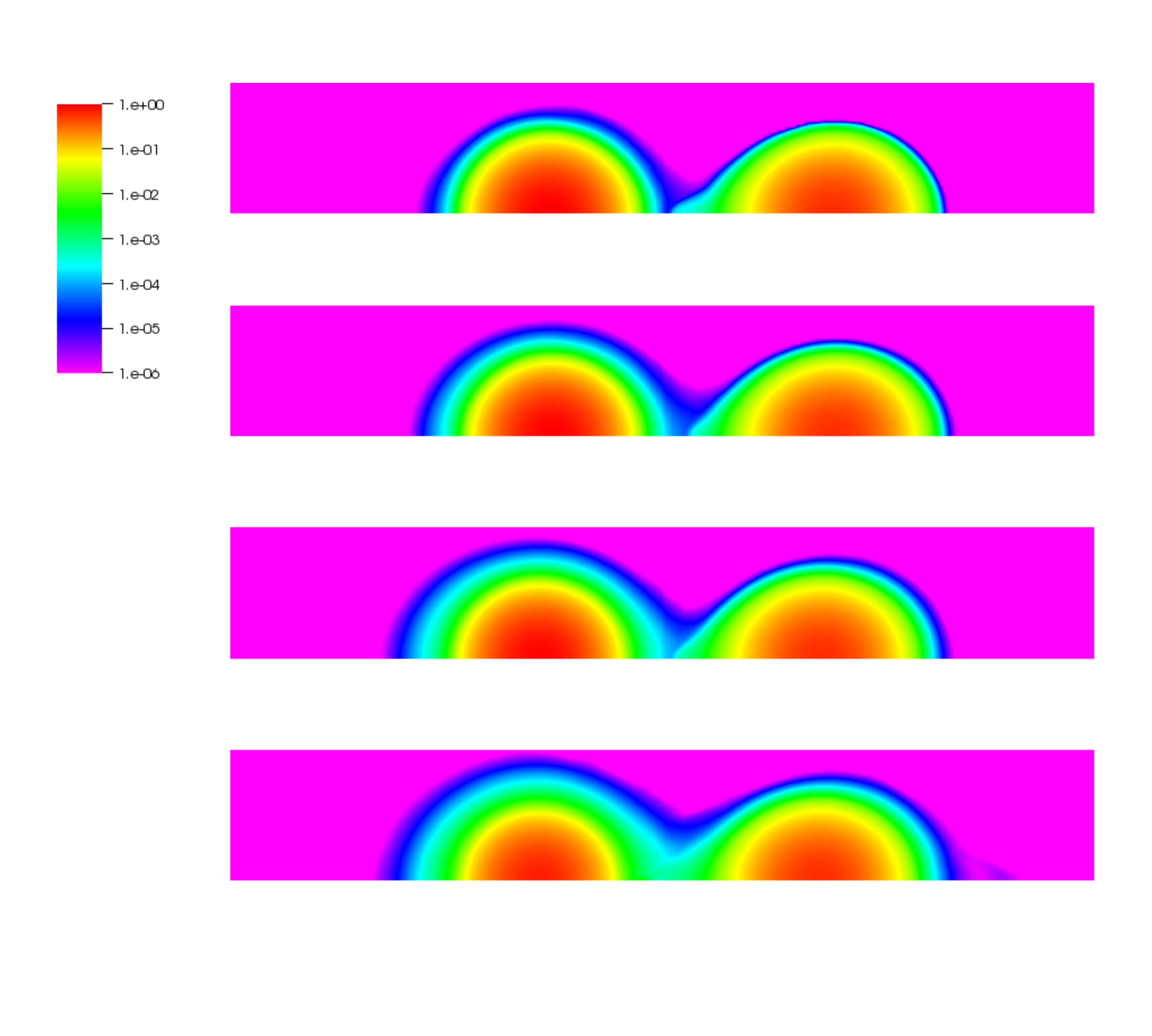}
\caption[Vertical density slice for $q=0.7$ runs]{ These are density plots, for the $q=0.7$ runs, of a vertical slice perpendicular to the equatorial plane and coinciding with the line running from one star's center of mass to the other's. The left column is the $q=0.7a$ run and the right column the $q=0.7b$ run. From top to bottom, the rows correspond to $t = 5$ orbits, $10$ orbits, $15$ orbits, and $20$ orbits. The color density scale runs from $10^{-6}$ to $1$ in code units. Note that we use symmetry across the equatorial plane in these simulations, so there is only a ``top" half of the grid. } 
\label{panel0}
\end{center}
\end{figure}

\begin{figure}
\begin{center}
\begin{tabular}{|c|c|c|c|}
\hline
\includegraphics[scale=0.155]{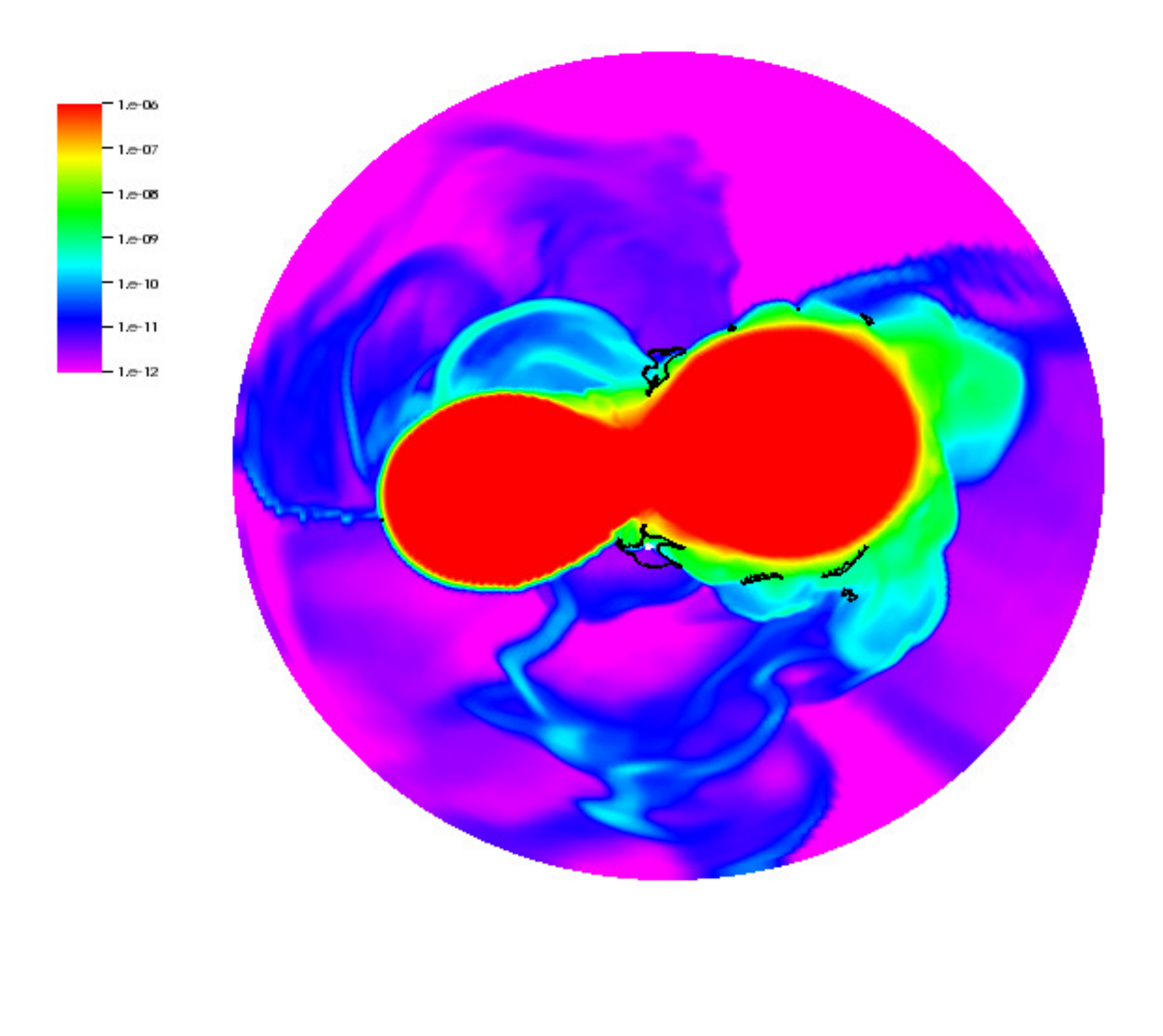} & 
\includegraphics[scale=0.155]{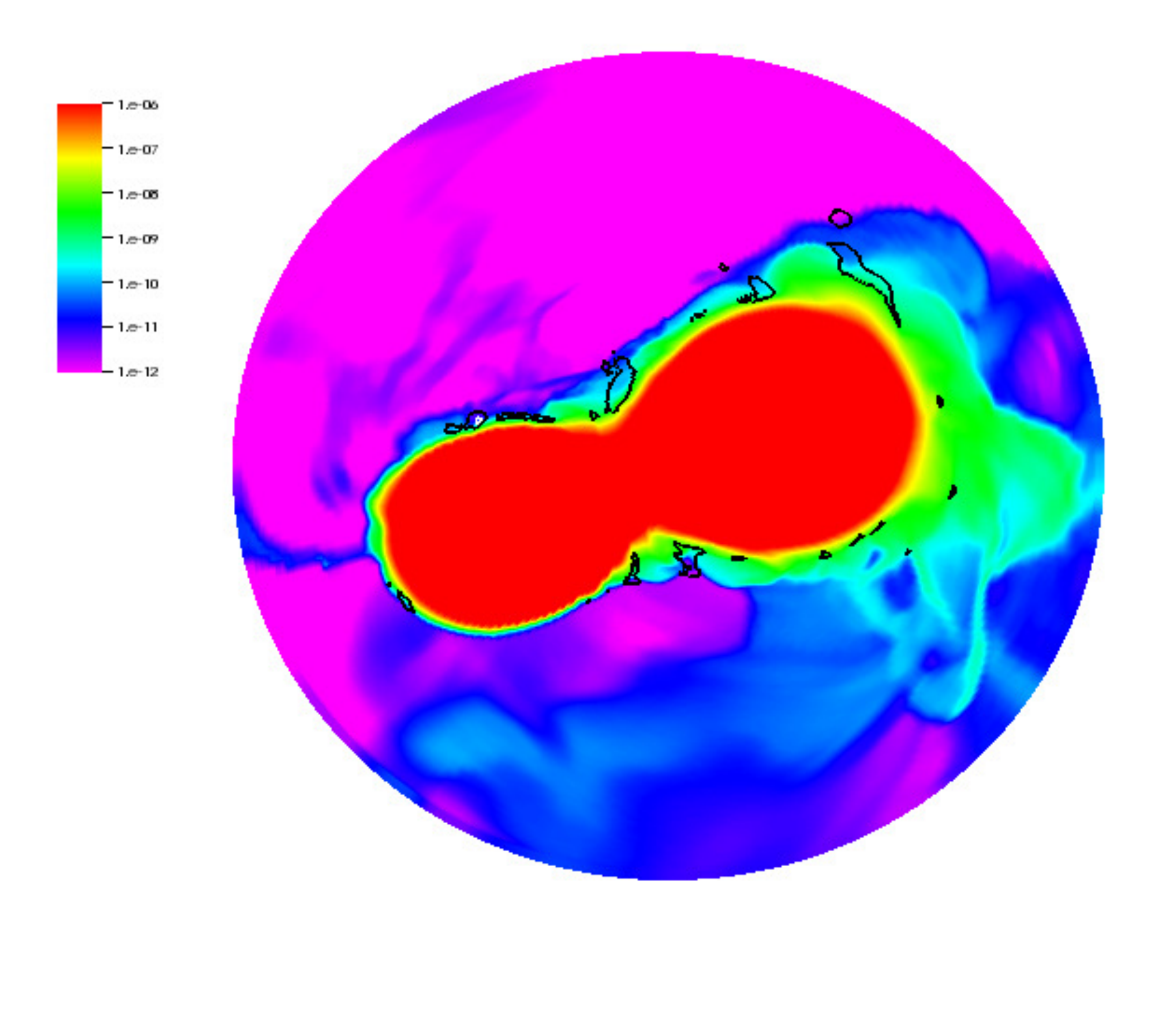} &
\includegraphics[scale=0.155]{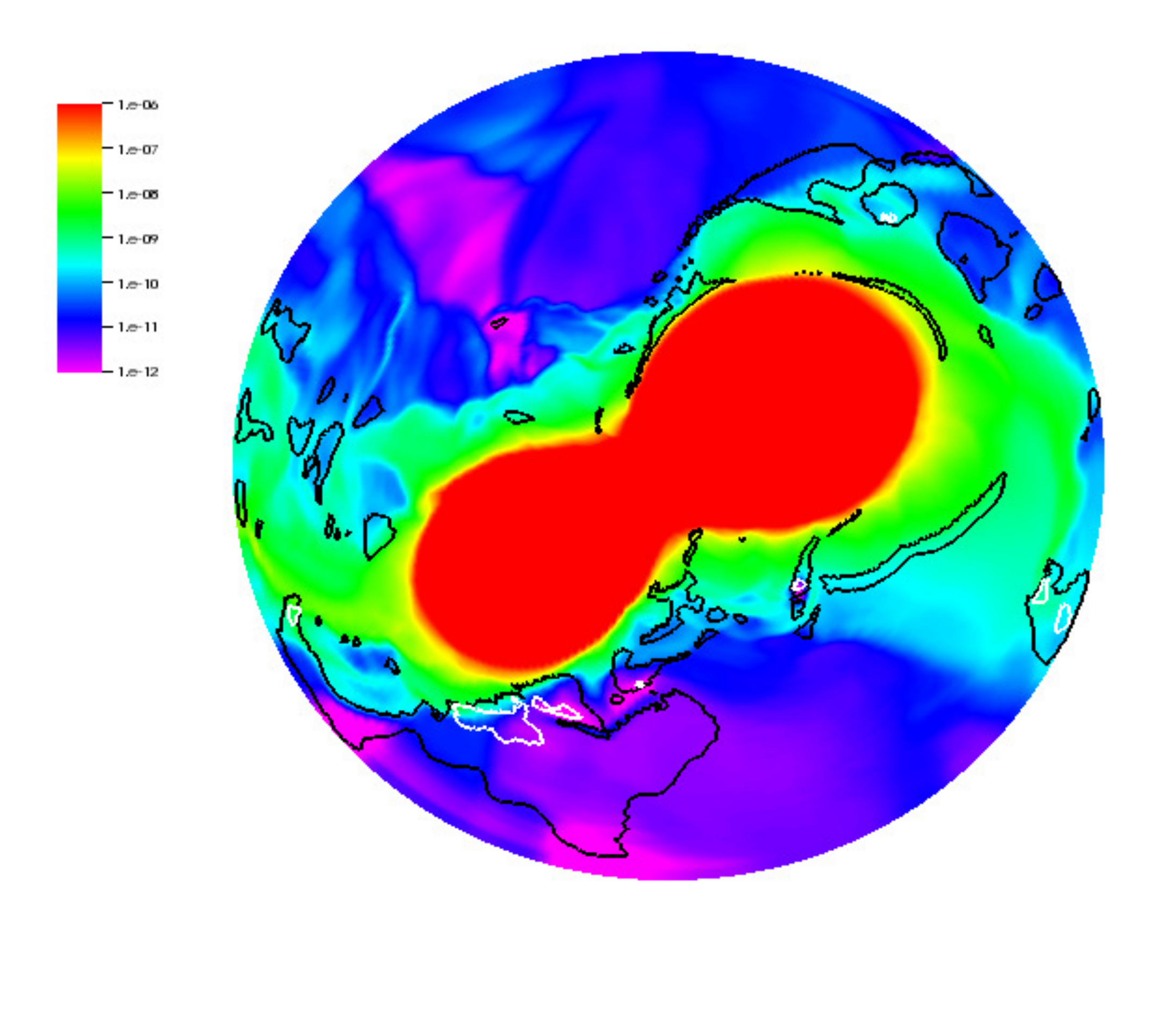} &
\includegraphics[scale=0.155]{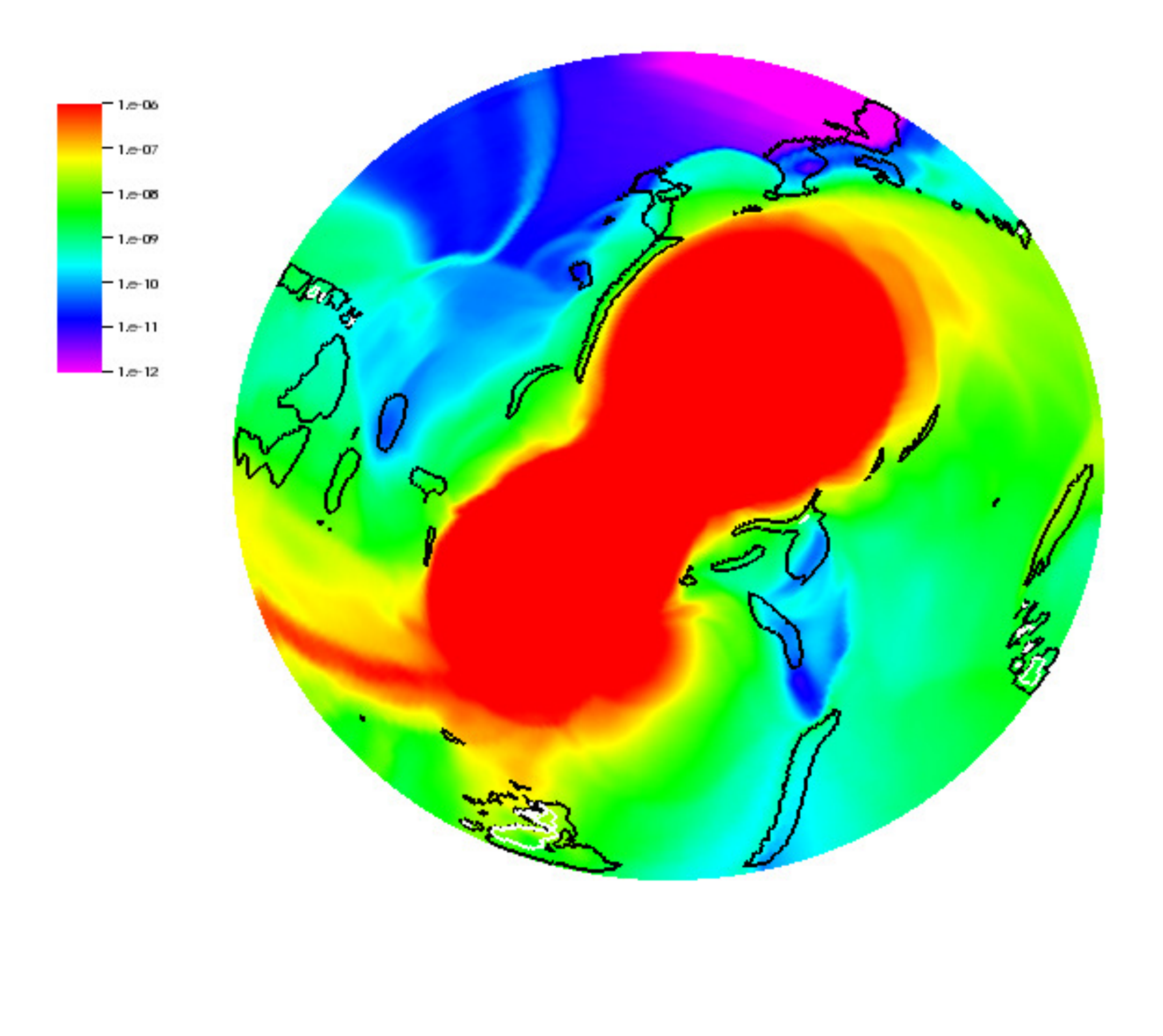} \\
\hline
\includegraphics[scale=0.155]{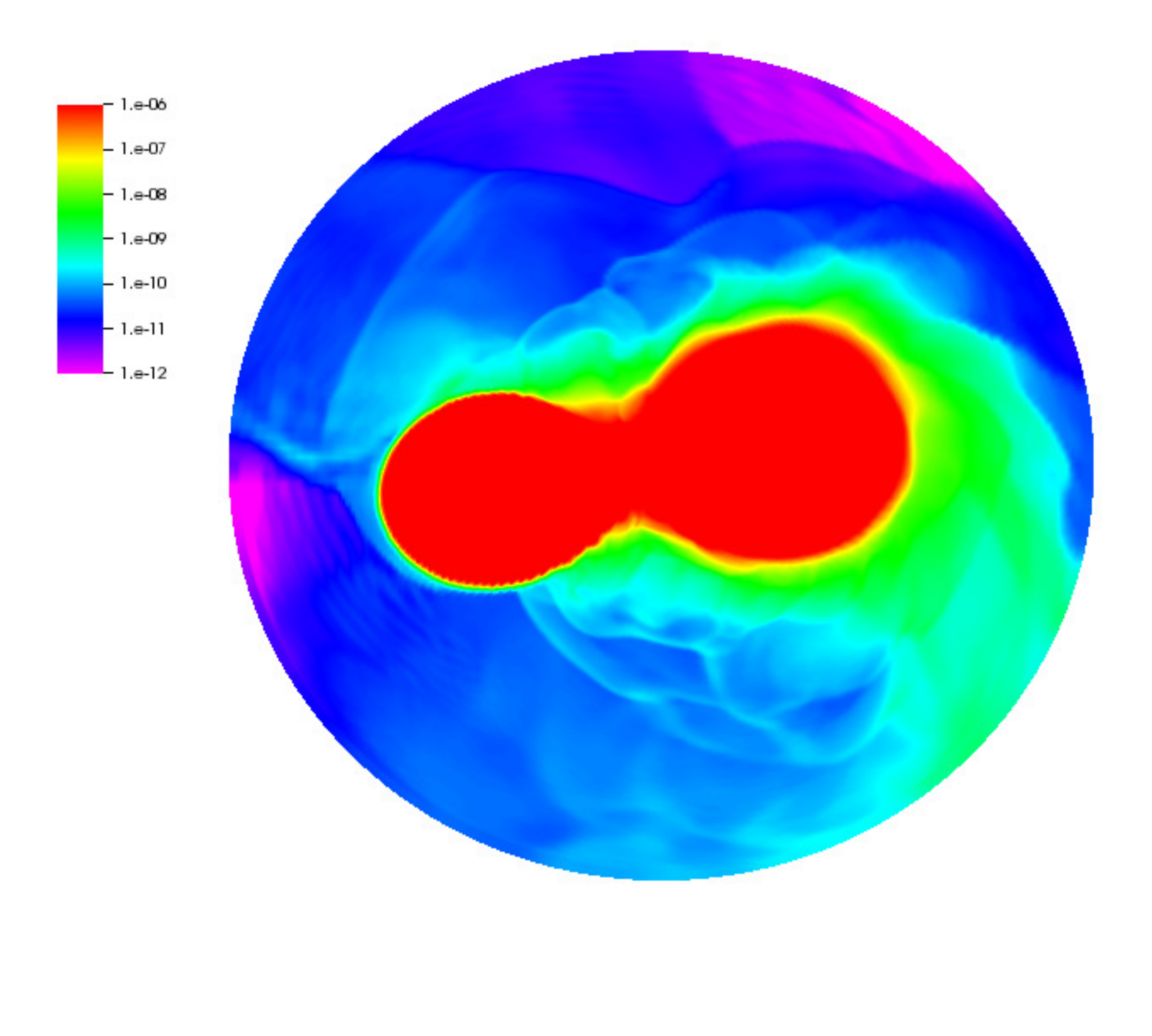} &
\includegraphics[scale=0.155]{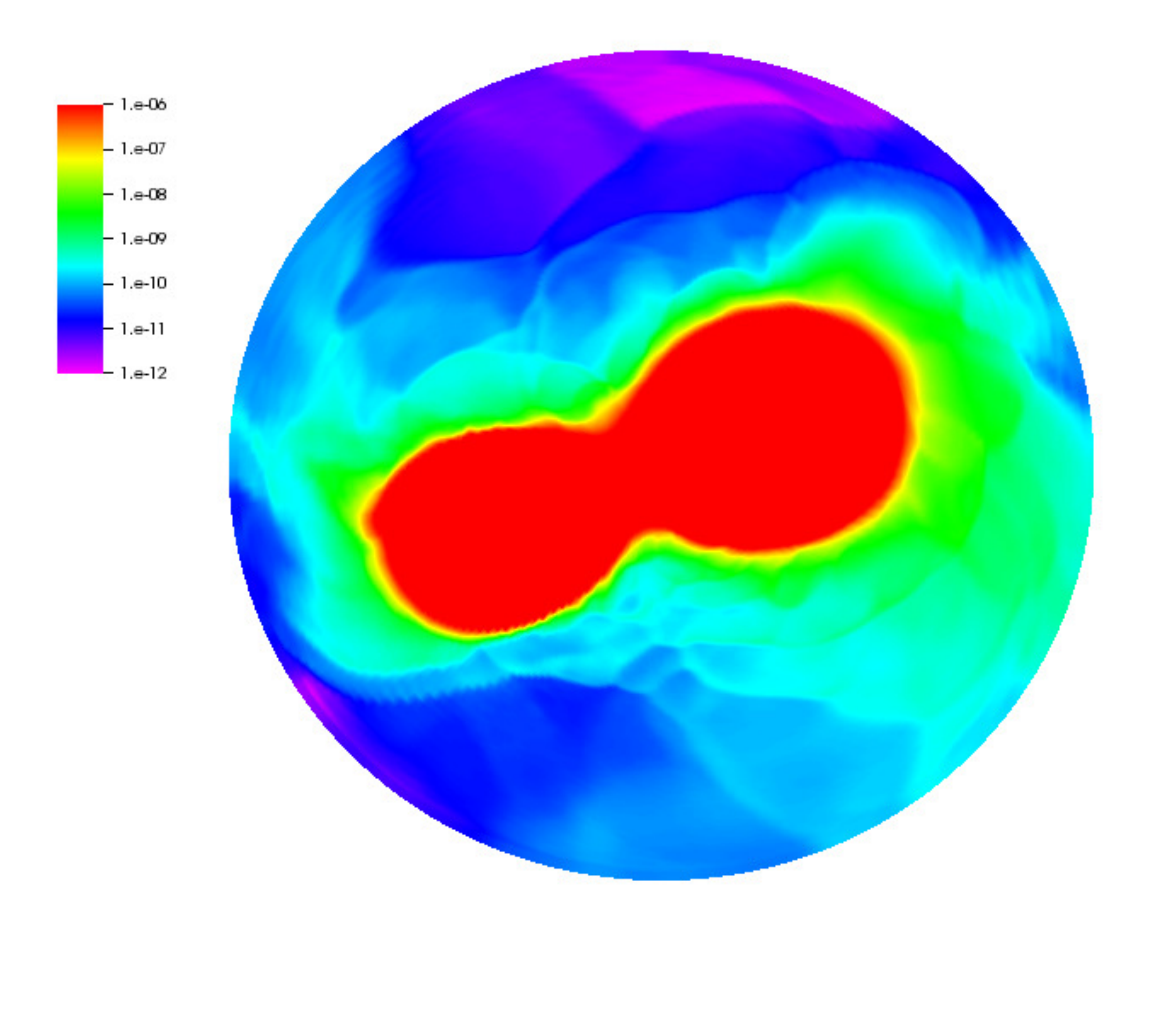} &
\includegraphics[scale=0.155]{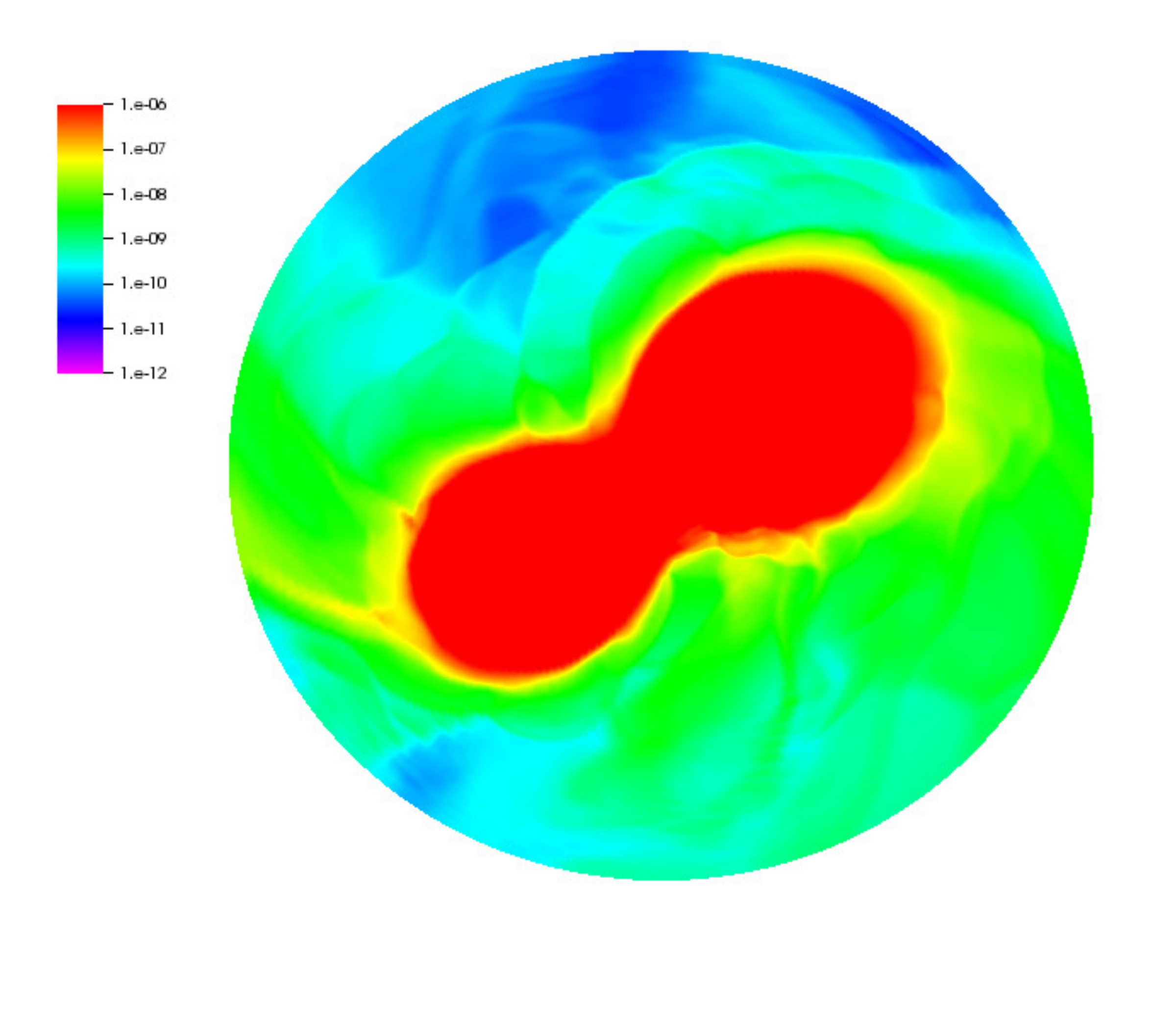} &
\includegraphics[scale=0.155]{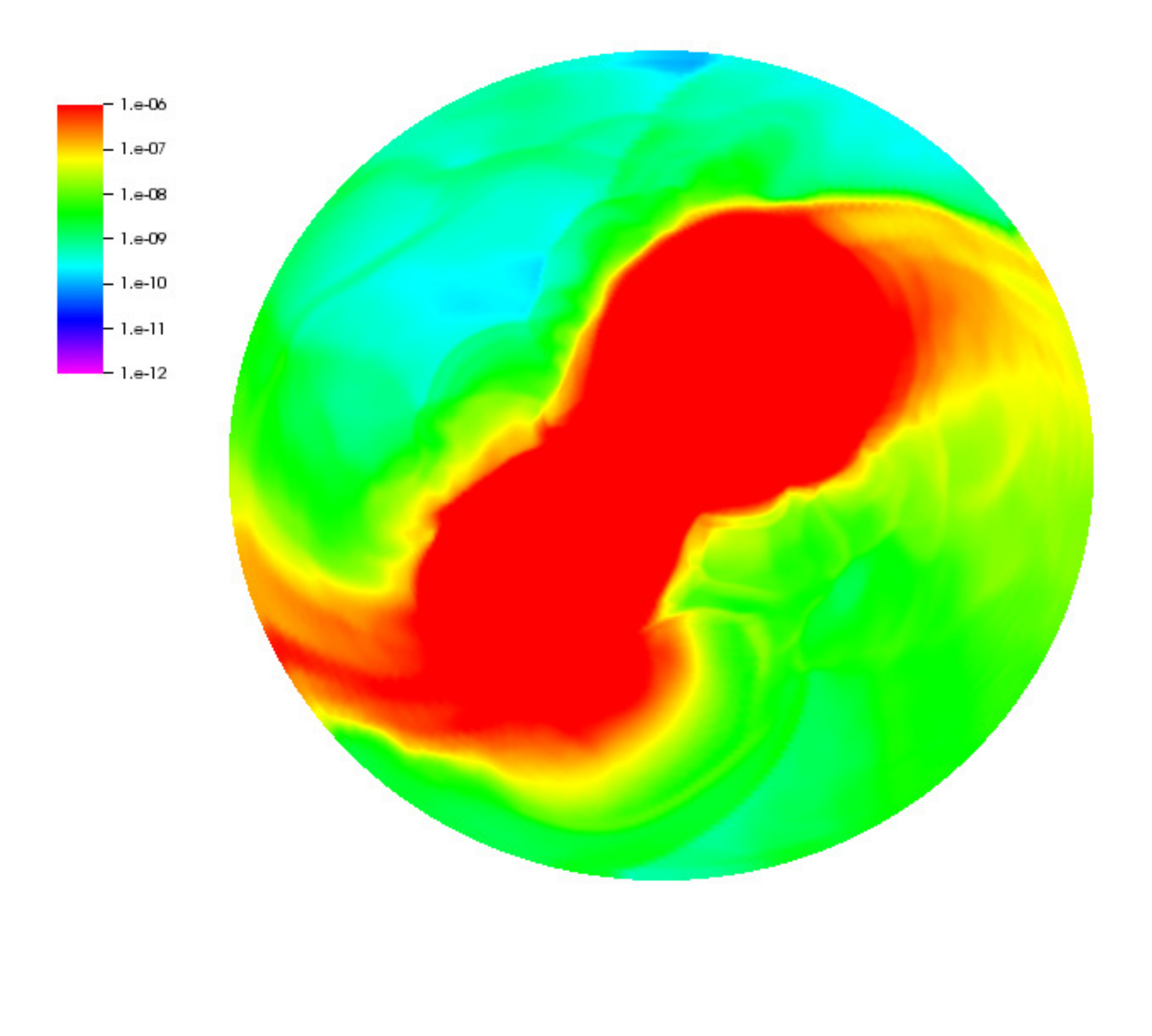} \\
\hline 
\end{tabular}
\caption[Low density range equatorial slices for $q=0.7$ runs]{ These are density plots of a slice through the equatorial plane, with the scale altered to highlight low density regions. The top row is the $q=0.7a$ run and the bottom row the $q=0.7b$ run. From left to right, the columns correspond to $t = 5$ orbits, $10$ orbits, $15$ orbits, and $20$ orbits. The color density scale runs from $10^{-12}$ to $10^{-6}$ in code units. For the $q=0.7a$ run we have also plotted black contours around regions in which the flow is super-Eddington. The white contours are regions where the radiation is ten times or more super-Eddington. The definition of these contours is provided in the last paragraph of \S \ref{qorsection}}.
\label{panel2}
\end{center}
\end{figure}

\begin{figure}
\begin{center}
\includegraphics[scale=0.75]{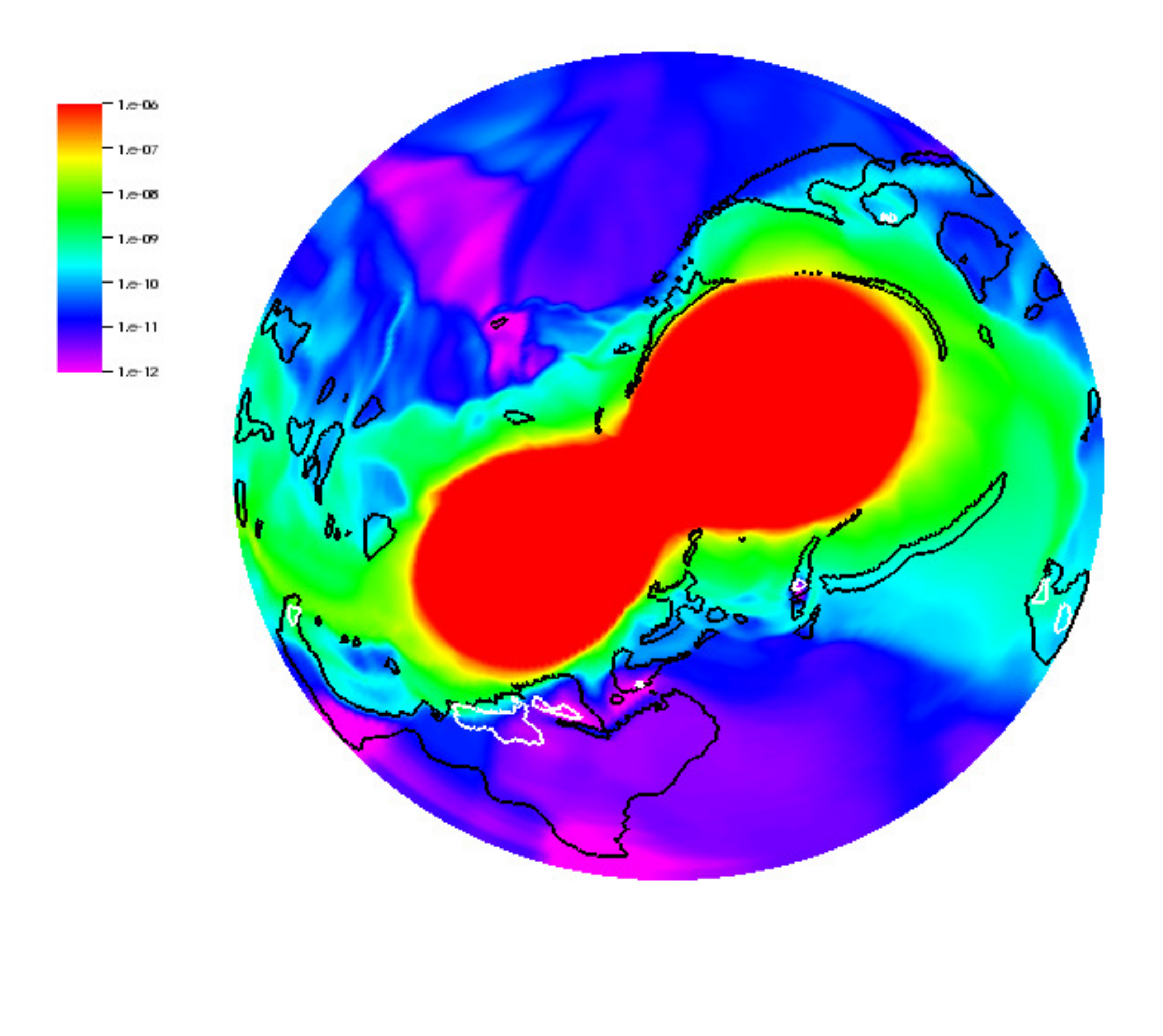} 
\caption[Close up of low density plot for $q=0.7a$ at $15$ orbits.]{A close up of the $q=0.7a$ low density plot at $15$ orbits, seen in the upper row of Figure \ref{panel2}, third from the left.}
\label{panel2_close}
\end{center}
\end{figure}

Four frames from the $5^\mathrm{th}$, $10^\mathrm{th}$, $15^\mathrm{th}$, $20^\mathrm{th}$ orbit for both runs are shown in Figures \ref{panel1}
through \ref{panel0}. 
Figures \ref{panel1} and \ref{panel0} show density with a logarithmic color scale, in code units, running from $10^{-6}$ to $10^0$. To highlight the low density regions, the logarithmic density scale in Figure \ref{panel2} 
runs from $10^{-12}$ to $10^{-6}$. Figures \ref{panel1} and \ref{panel2} depict slices through the equatorial plane, while Figure \ref{panel0} is a slice through the plane perpendicular to the equatorial plane and coincident with the center of mass of both stars. The $q=0.7a$ run in Figure \ref{panel2} also contains contour lines. For this plot, we have defined regions of super-Eddington accretion to be any region for which the condition
\begin{equation}
-\mathbf{f}_\mathrm{rad} \cdot \mathbf{f}_\mathrm{grav} \le  \mathbf{f}_\mathrm{grav} \cdot \mathbf{f}_\mathrm{grav},
\label{eddcond}
\end{equation}
is satisfied. Here, $\mathbf{f}_\mathrm{rad} := -\Lambda_E \gradient{E_R}$ and $\mathbf{f}_\mathrm{grav} := -\rho \gradient{\Phi}$ are the forces of radiation and gravity, respectively. This condition is satisfied inside the black contours. Inside the white contours, the force of radiation is ten times more than what is needed to satisfy equation (\ref{eddcond}). Figure \ref{panel2_close}
 is a close-up of the $15^{\mathrm{th}}$ orbit for the $q=0.7a$ run in Figure \ref{panel2}.

\subsection{Discussion}

\begin{figure}
\begin{center}
\begin{tabular}{|c|c|c|}
\hline
\includegraphics[angle=-90,scale=0.39]{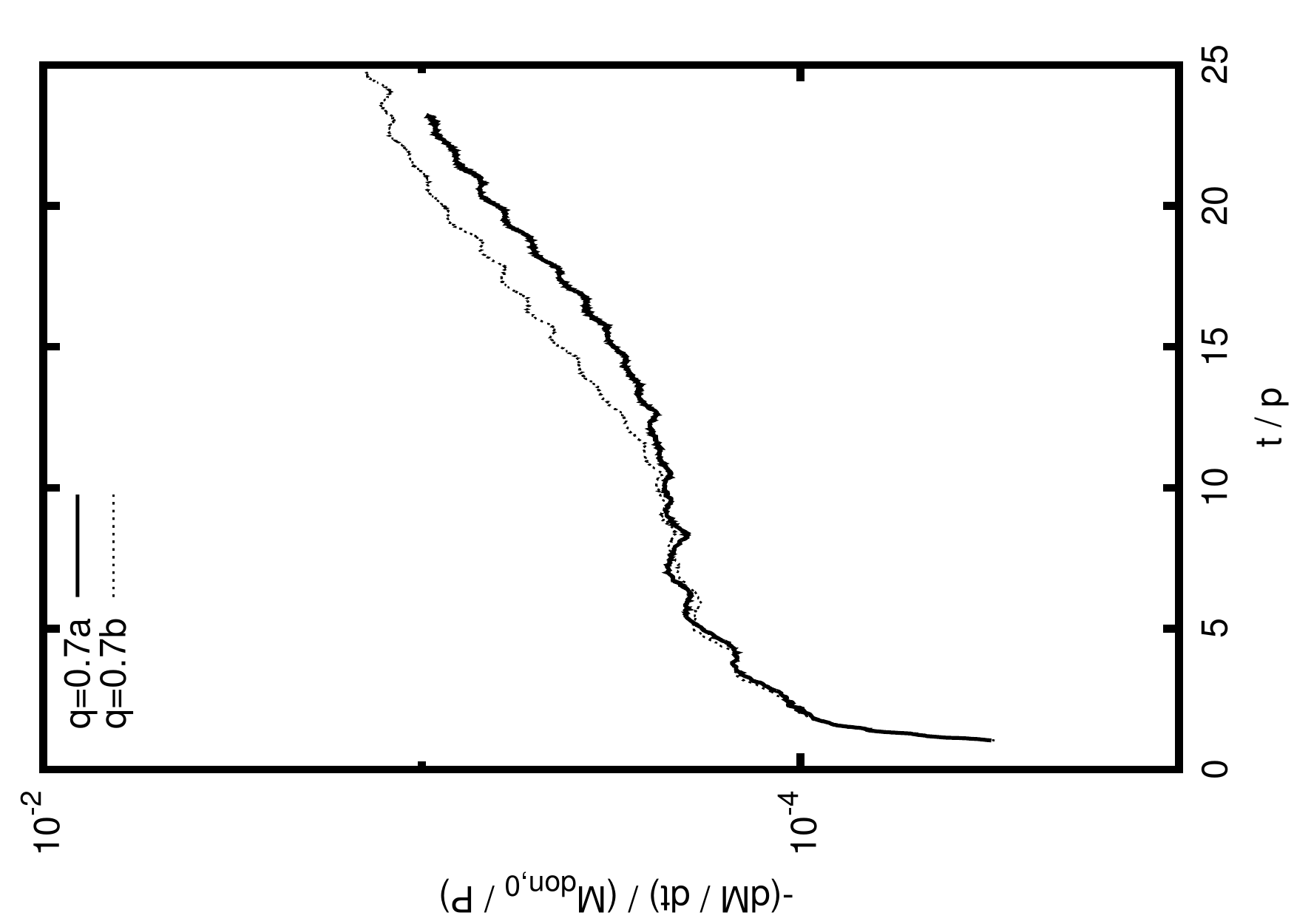} & 
\includegraphics[angle=-90,scale=0.39]{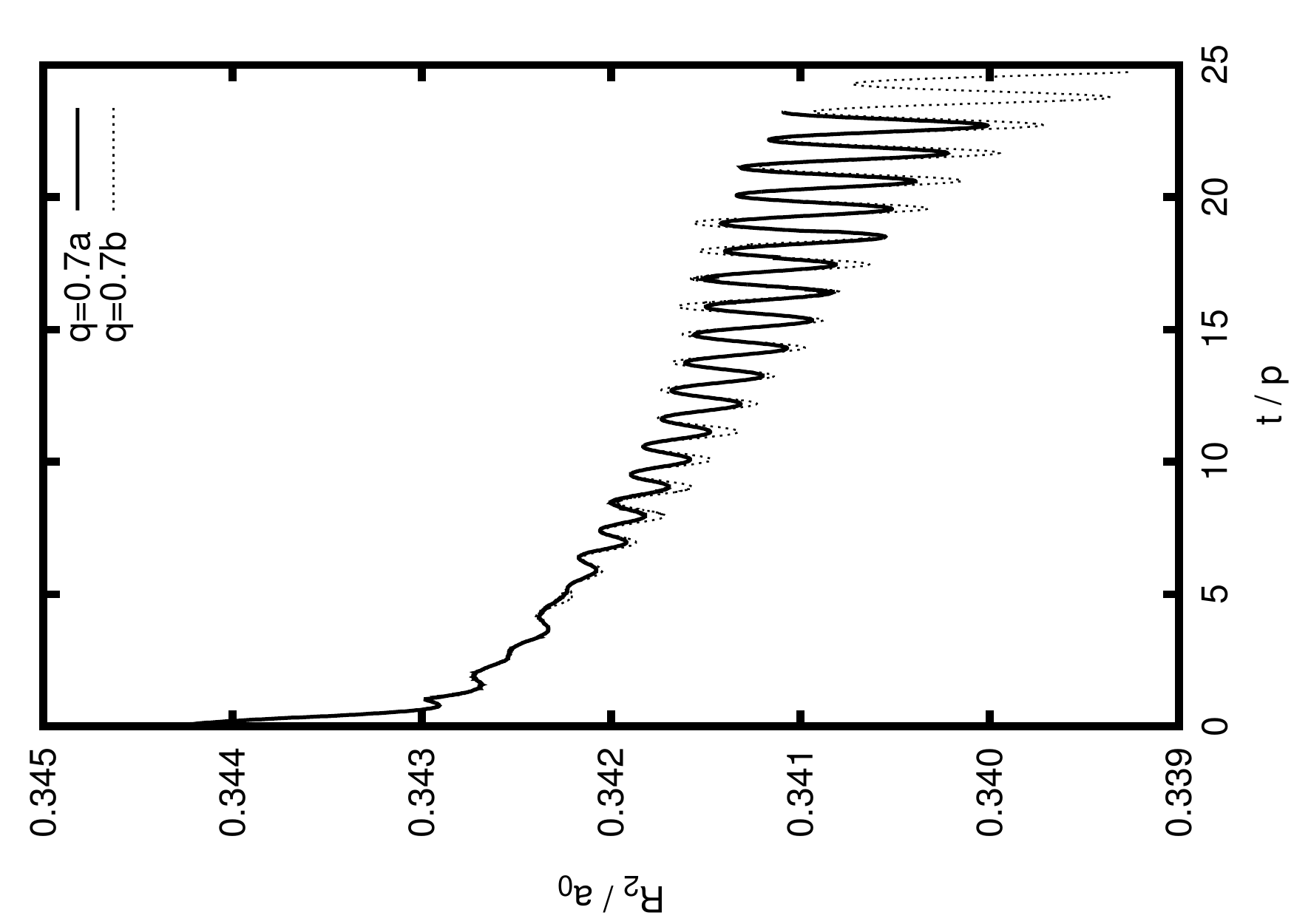} &
\includegraphics[angle=-90,scale=0.39]{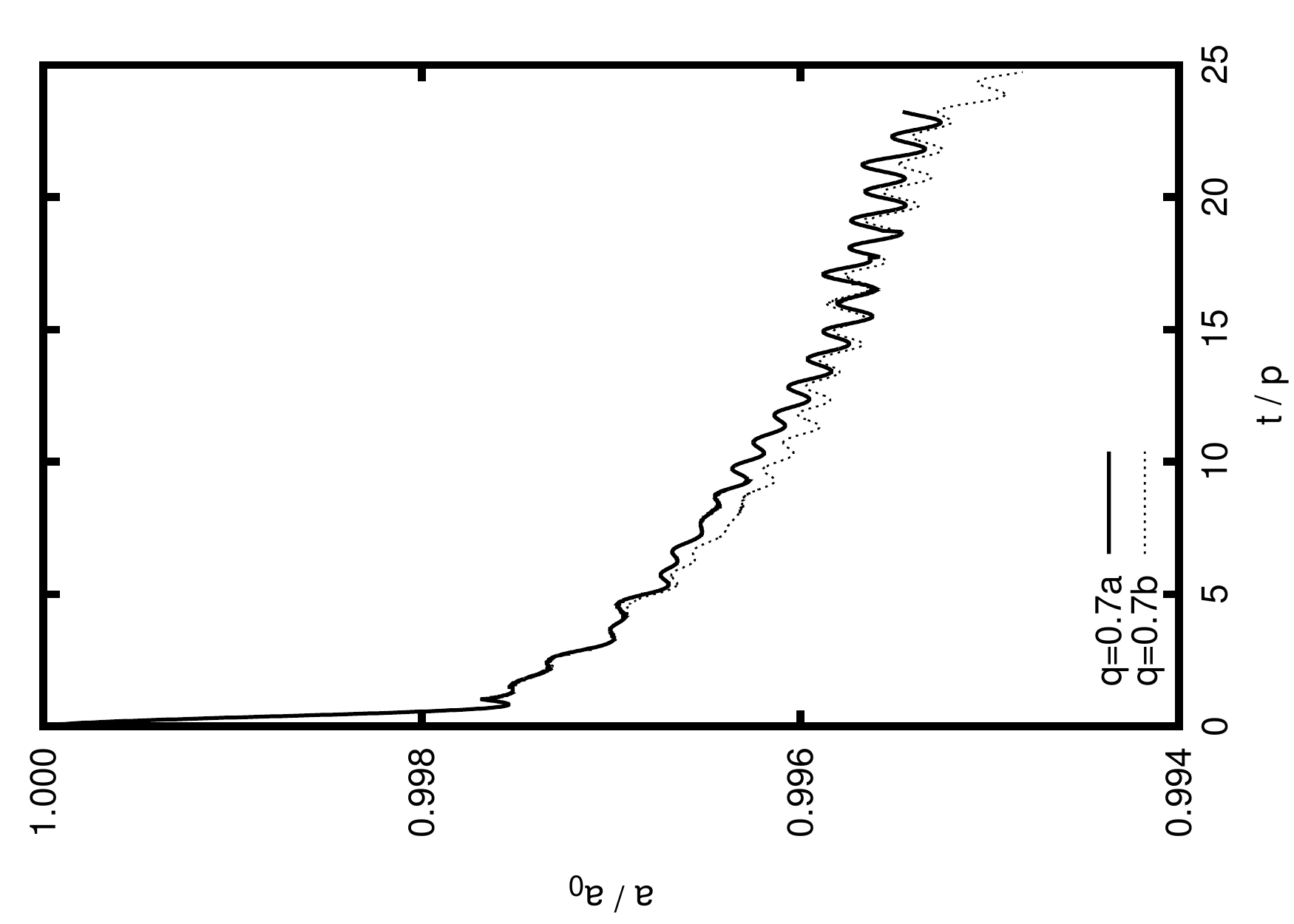} \\
\hline
\includegraphics[angle=-90,scale=0.39]{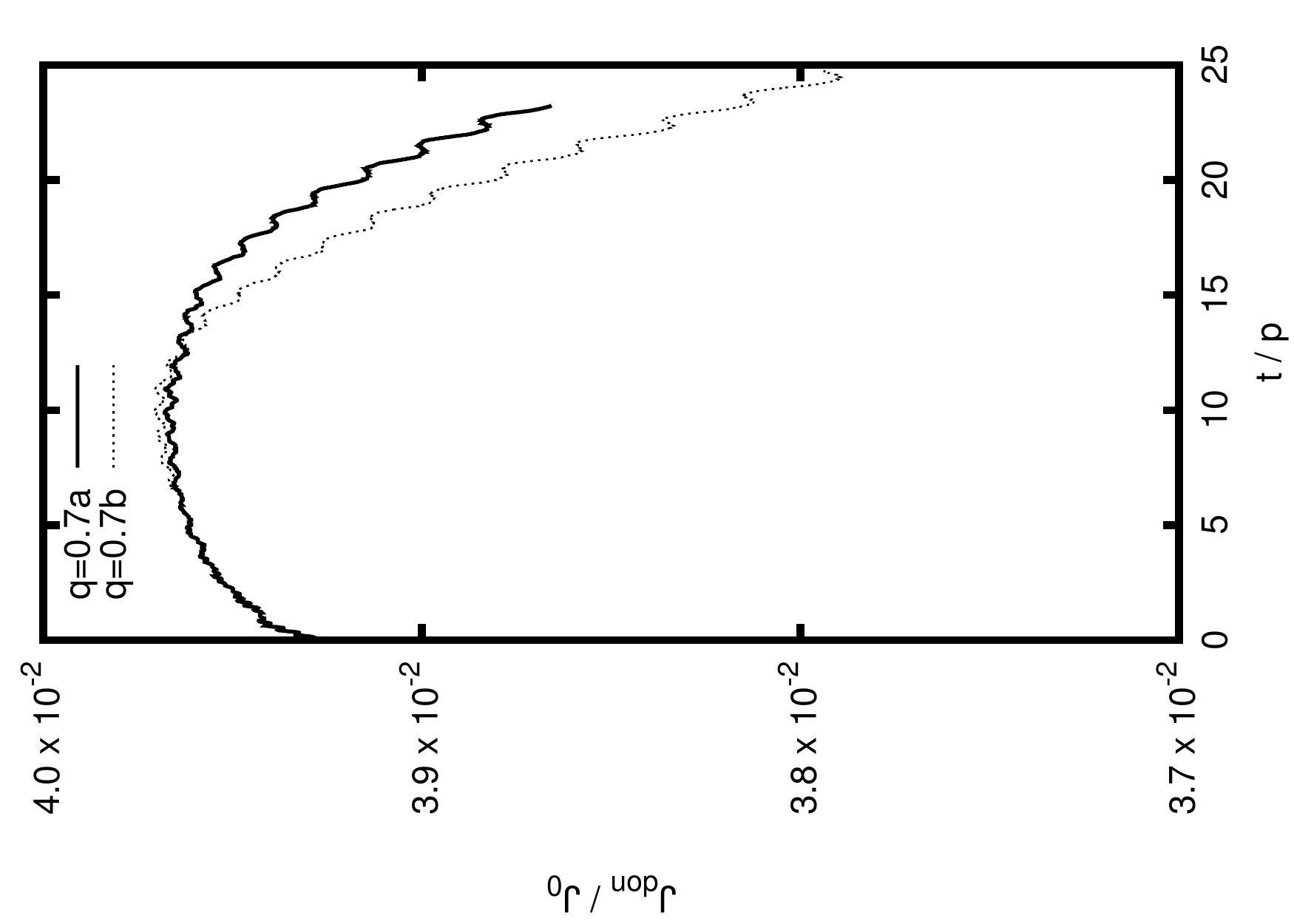} & 
\includegraphics[angle=-90,scale=0.39]{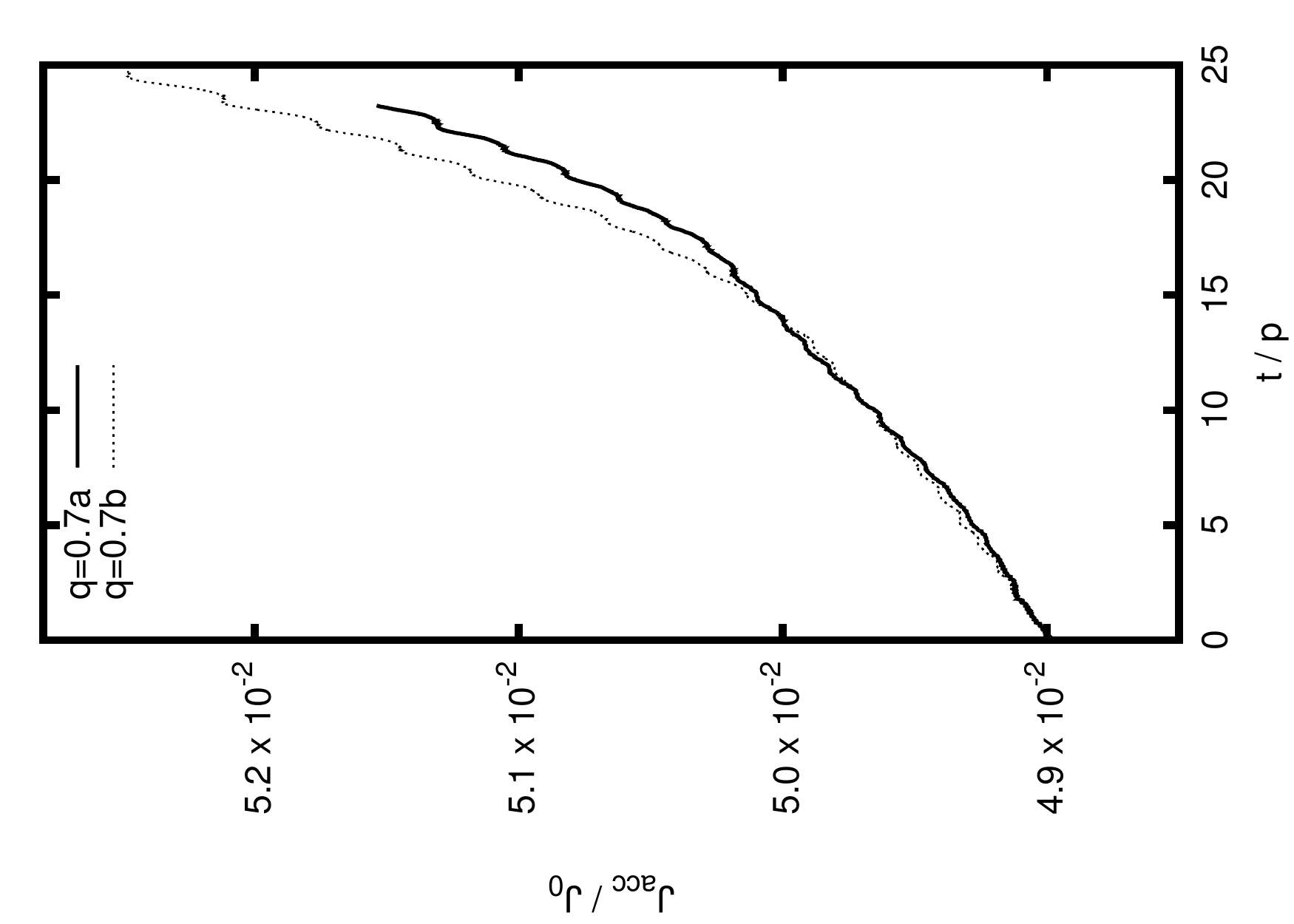} &
\includegraphics[angle=-90,scale=0.39]{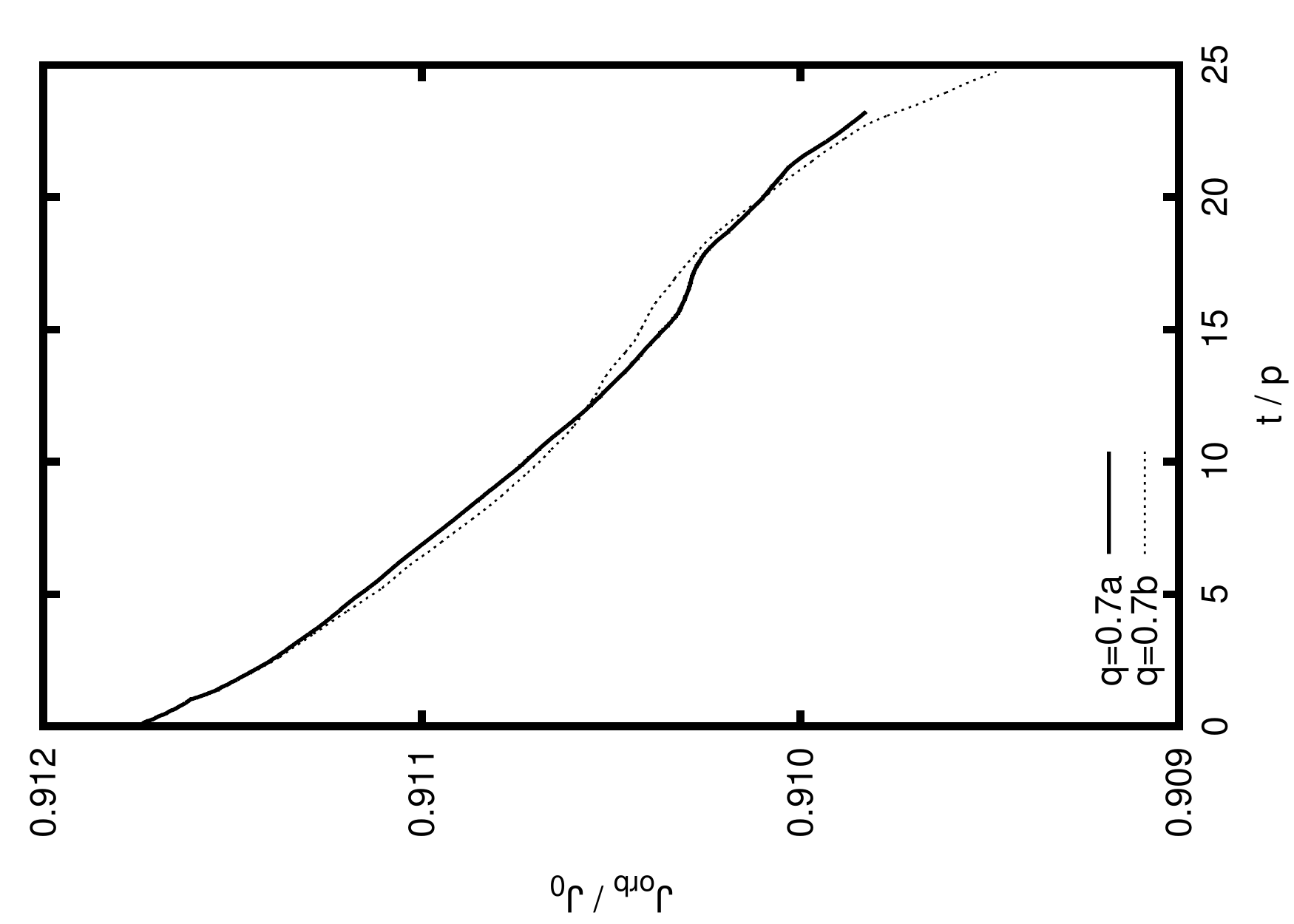} \\
\hline
\end{tabular}
\caption[Mass transfer rate, Roche lobe radius, orbital separation, and angular momenta of the $q=0.7$ runs]{ Binary runs $q=0.7a$ (solid curve) and $q=0.7b$ (dotted curve). 
{\it Top Left:} The orbit averaged rate of mass transfer from the donor, normalized to donor masses per orbital period.
{\it Top Middle: } The orbit averaged effective Roche lobe radius of the donor, in units of the initial orbital separation.
{\it Top Right: } The orbit averaged orbital separation normalized to its initial value;
{\it Bottom Left: } The spin angular momentum of the donor, in units of initial total angular momentum.
{\it Bottom Middle: } The spin angular momentum of the accretor, in units of initial total angular momentum.
{\it Bottom Right: } The orbit averaged orbital angular momentum, in units of initial total angular momentum.} 
\label{q7}
\end{center}
\end{figure}

Immediately after the evolution begins, the donor in both runs overflows its Roche lobe. Although the initial configuration is in equilibrium according to the algebraic system of equations used by the SCF code, it is slightly out of equilibrium once placed in the hydrodynamics code. As seen in the upper left panel of Figure  \ref{q7},
mass transfer proceeds at a steadily increasing rate until about the $15^\mathrm{th}$ orbit.  At this point the transfer rate begins to grow suddenly. It is generally expected that runaway mass transfer will occur for binaries of this mass ratio. The mass within the common envelope is relatively very small, and hence the rate of increase of the accretor's mass is nearly the same as the rate of decrease of the donor's. In the upper middle and upper right panel, we see that the Roche lobe effective radius and orbital separation shrink
throughout the evolution, consistent with dynamically unstable mass transfer. We define the Roche lobe effective radius as the radius of the sphere with the same volume as the Roche lobe. As more mass is removed from the donor and piled onto the accretor, the central density of the donor decreases while increasing for the accretor (see the middle and right panels of Figure \ref{q7rho_max}, respectively). 
As seen in the bottom left panel of Figure \ref{q7}, initially the donor's spin angular momentum increases very slightly, but after a few orbits it decreases monotonically for the remainder of the evolution. The accretor's spin angular momentum increases throughout the evolution (see bottom middle panel of Figure \ref{q7}). There are a variety of forces in play here: the transport of angular momentum from the donor to accretor by advection; the gravitational tidal interaction between donor and accretor; and the torque created by the accretion stream impacting the accretor off-center.

\begin{figure}
\begin{center}
\begin{tabular}{|c|c|c|}
\hline 
\includegraphics[angle=-90,scale=0.39]{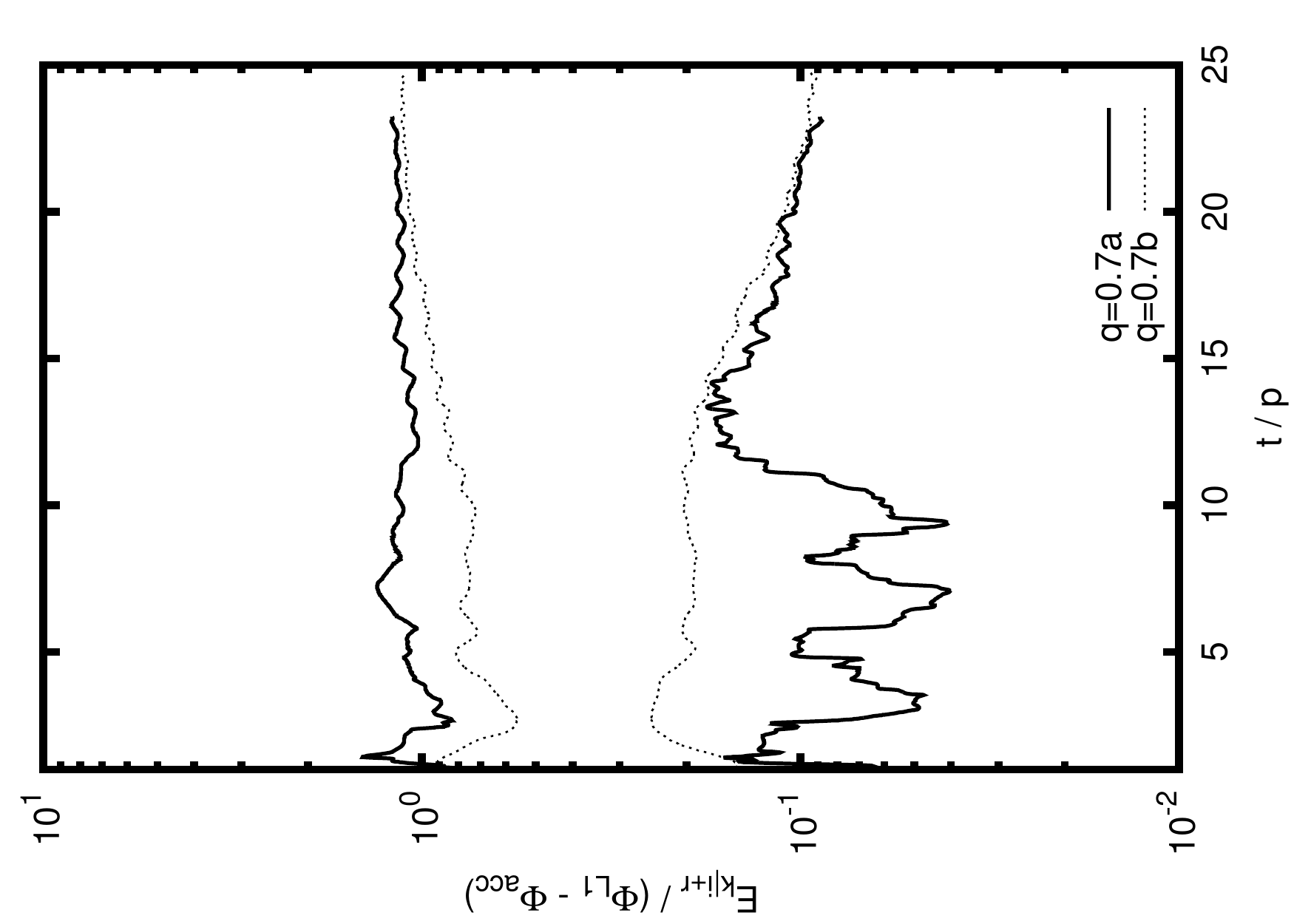} & 
\includegraphics[angle=-90,scale=0.39]{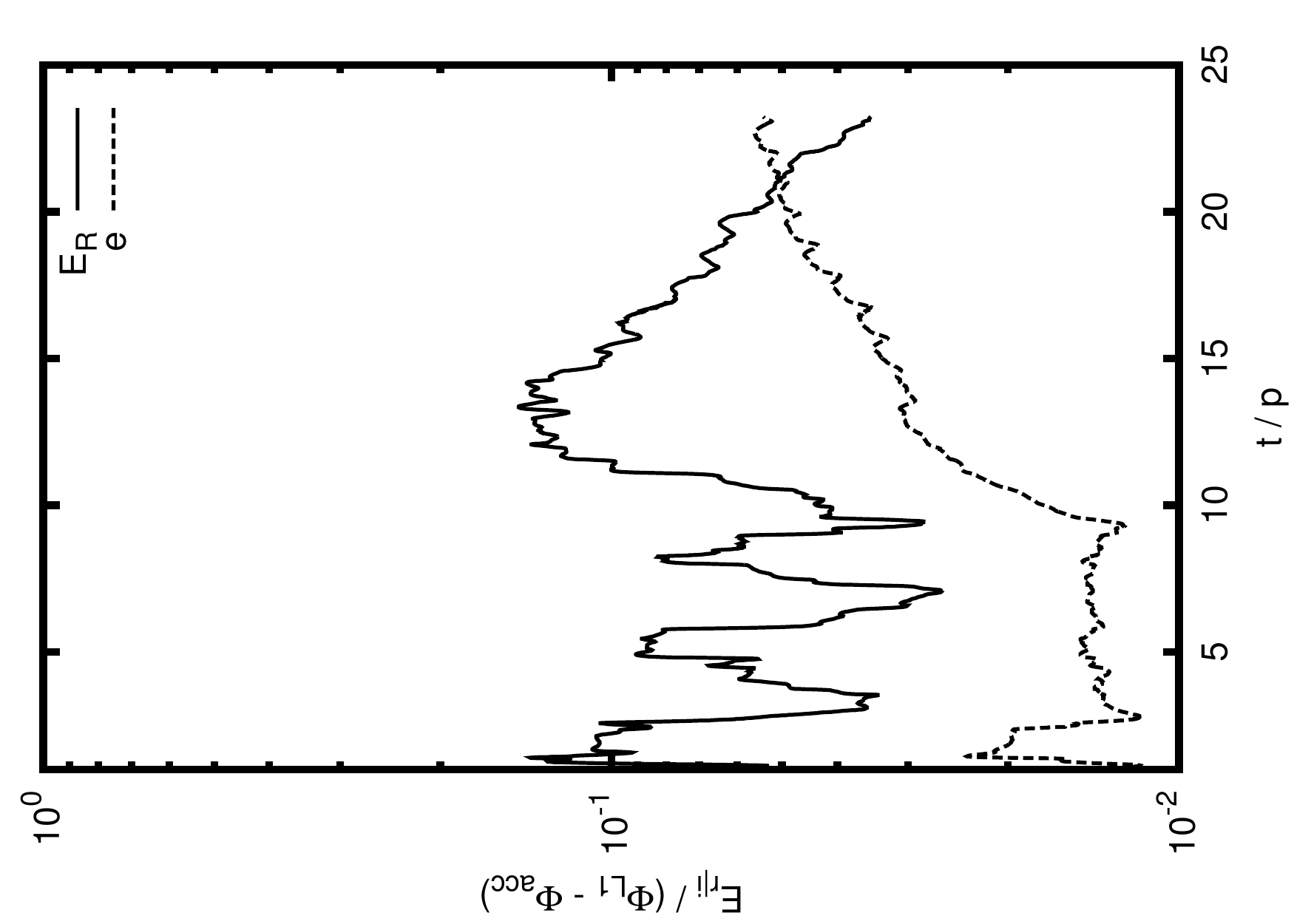} &
\includegraphics[angle=-90,scale=0.39]{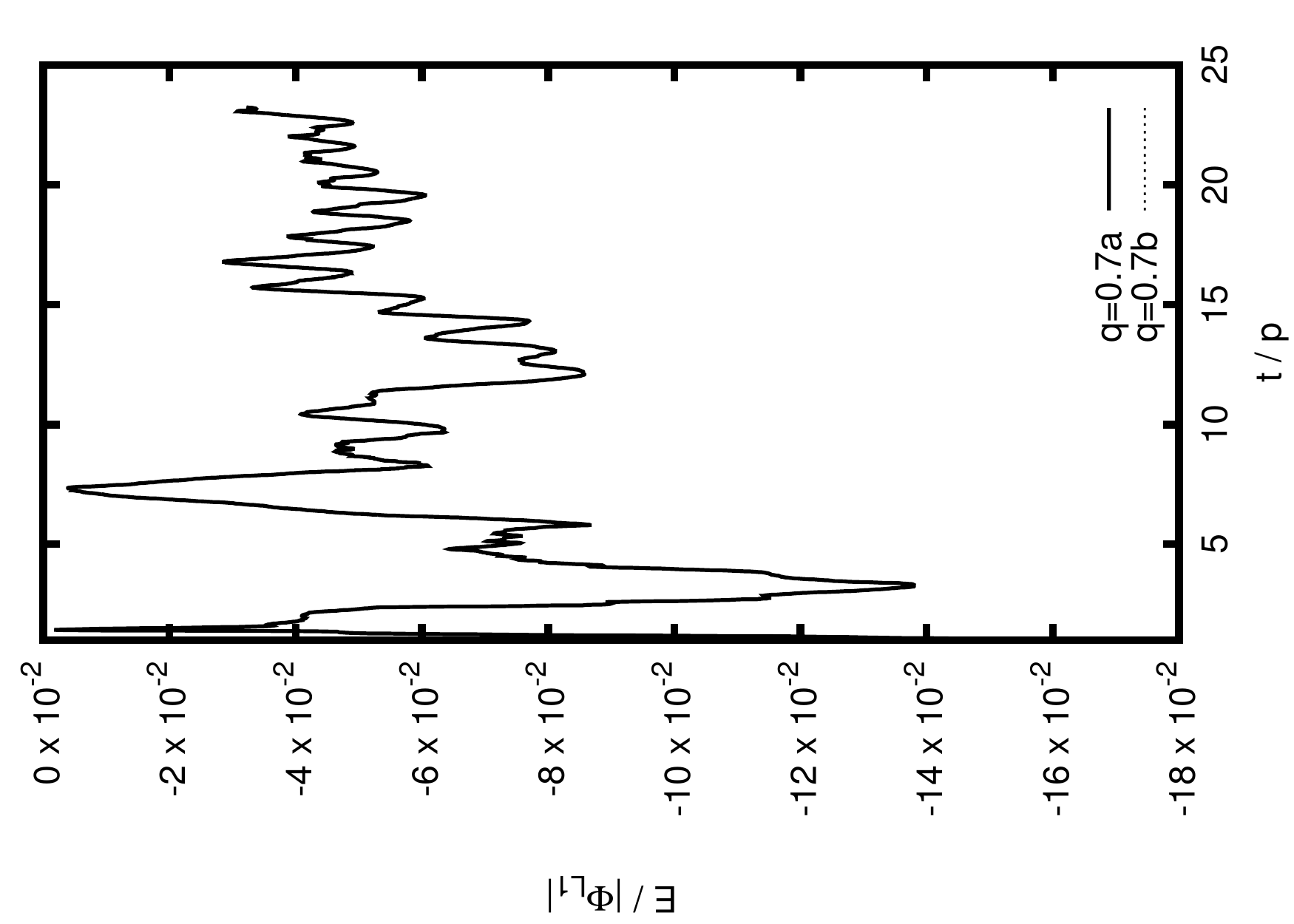} \\
\hline
\end{tabular}
\caption[Common envelope specific energies for the $q=0.7$ runs]{ Common envelope of the $q=0.7$ binary runs. 
{\it Left:}  The top two curves are the orbit averaged specific kinetic energies. For the $q=0.7b$ run (dotted curve), the bottom curve is the orbit averaged  specific internal energy. For $q=0.7a$ run (solid curve), the bottom curve is the sum of the orbit averaged  specific internal and specific radiation energies. All are plotted in units of $\Phi_{L1} - \Phi_{acc}$. 
{\it Middle:}  The orbit averaged  specific radiation energy (solid curve) and the orbit averaged specific internal energy (dotted curve) for the $q=0.7a$ run, in units of $\Phi_{L1} - \Phi_{acc}$. 
{\it Right:} The orbit averaged specific gravitational binding energy of the common envelope, in units of $\left|\Phi_\mathrm{L1}\right|$, taken in the inertial frame for the $q=0.7a$ (solid curve) and $q=0.7b$ (dotted curve) runs. } 
\label{q7ce}
\end{center}
\end{figure}

\begin{figure}
\begin{center}
\begin{tabular}{|c|c|}
\hline
\includegraphics[angle=-90,scale=0.39]{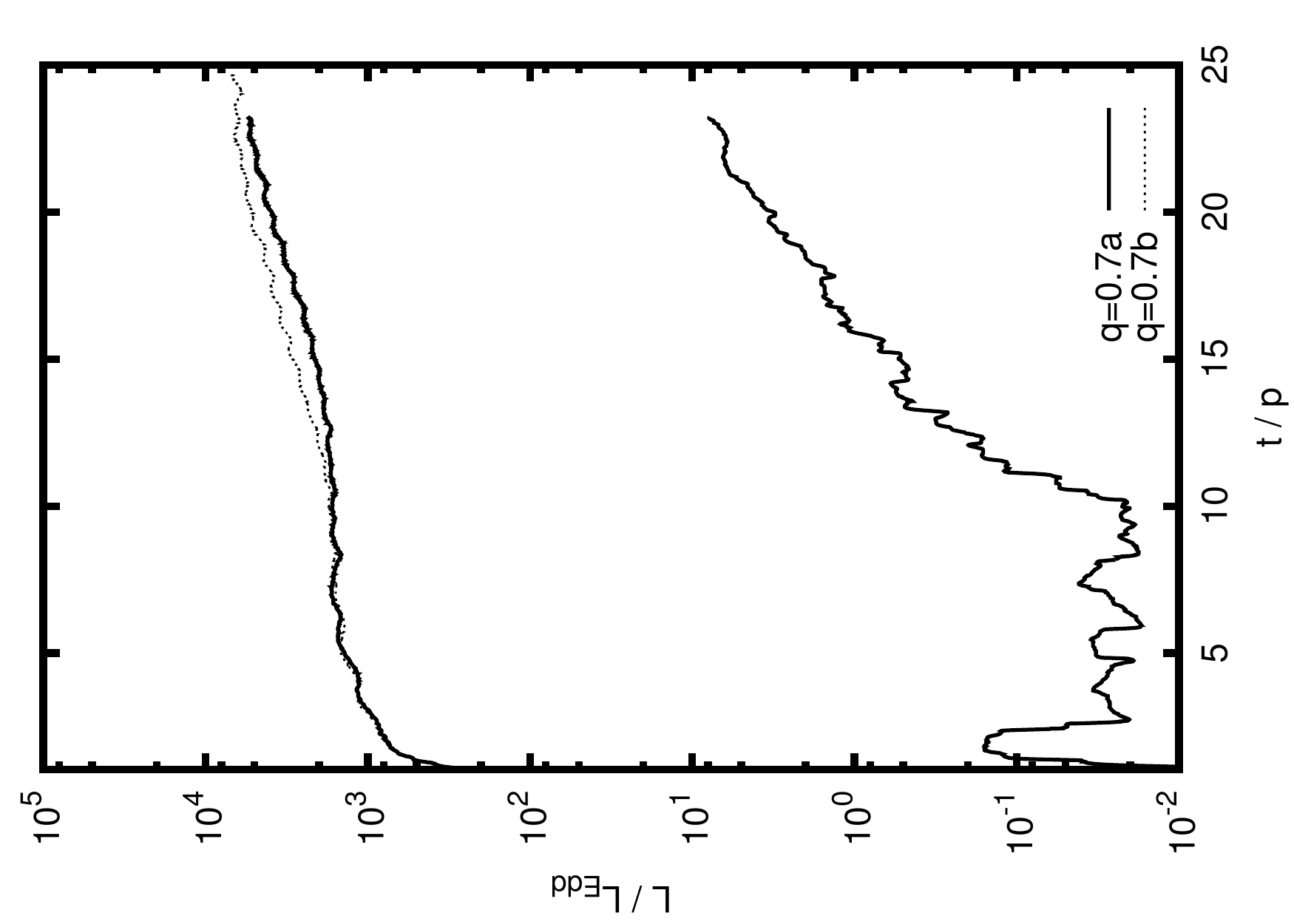} & 
\includegraphics[angle=-90,scale=0.39]{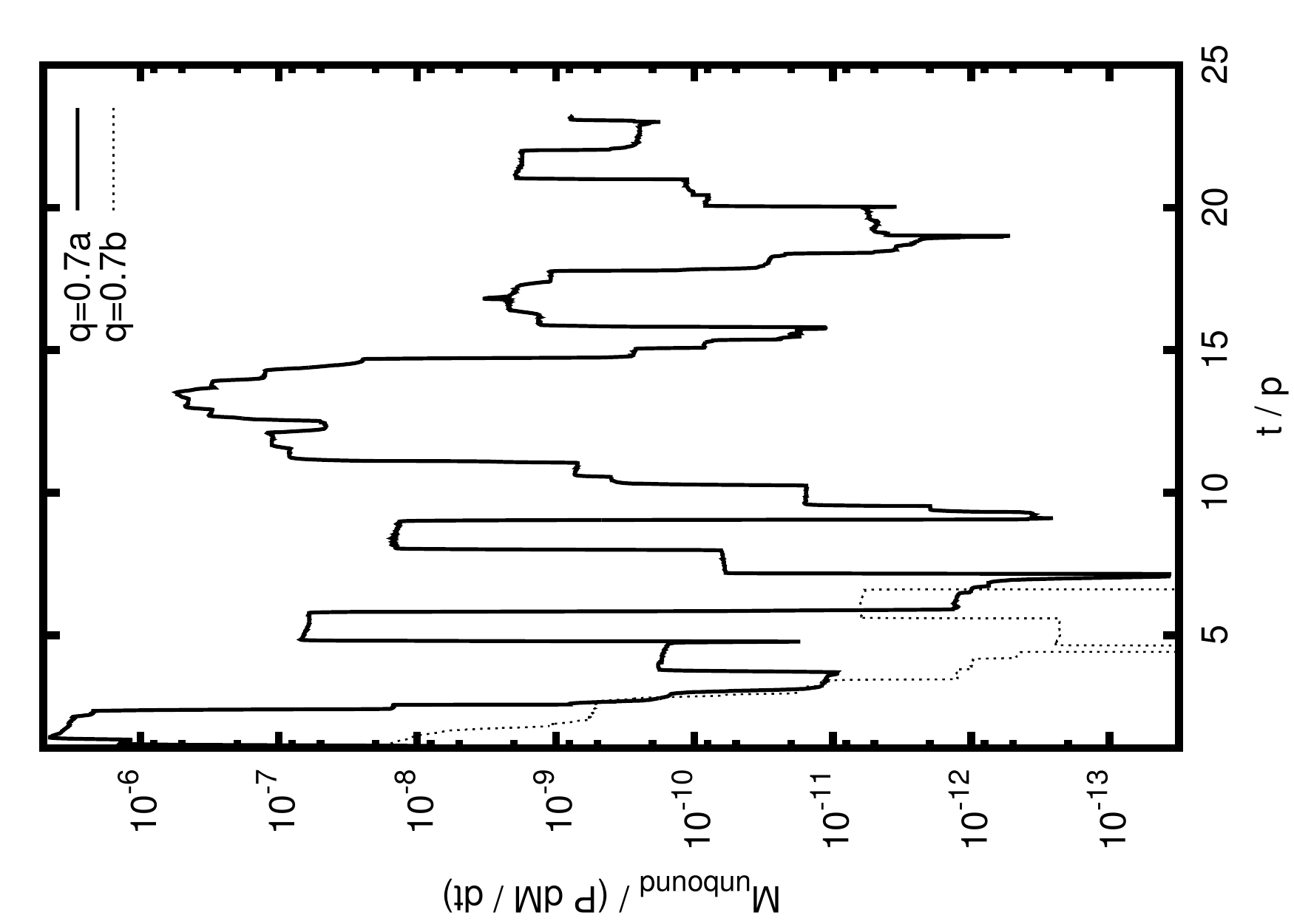} \\
\hline
\end{tabular}
\caption[Accretion luminosity, radiative luminosity, and above zero-binding energy mass fraction for $q=0.7$ runs]{ Binary runs $q=0.7a$ (solid curve) and $q=0.7b$ (dotted curve). 
{\it Left:} The top two curves are the orbit averaged accretion luminosities, in units of Eddington luminosity, for the $q=0.7a$ (solid curve) and $q=0.7b$ (dotted curve) runs. The bottom curve is the radiative luminosity for the $q=0.7a$ run. 
{\it Right:} The orbit averaged fraction of mass on the grid with a gravitational binding energy above zero  for the $q=0.7a$ (solid curve) and $q=0.7b$ (dotted curve) runs, normalized to the orbit-averaged mass transfer rate.
} 
\label{eddlum}
\end{center}
\end{figure}

	The stationary point in $\Phi_\mathrm{eff}$ which lies between the two stars is the L1 point. As seen in the left panel of Figure \ref{q7ce},
almost all of the common envelope has roughly the same specific kinetic energy as the difference in effective potential between the L1 point and the surface of the accretor, independent of time.  This is consistent with physical expectations, as in order for a piece of the fluid to reach a potential high enough to escape into the common envelope, it must have sufficient kinetic energy. In the middle panel, we see that initially the specific radiation energy exceeds specific internal energy in the common envelope for the $q=0.7a$ run. This is reversed as the envelope cools by radiative transport and as a higher rate of mass transfer produces more internal gas energy. In the right panel, we show the inertial frame specific gravitational binding energy of the envelope in units of $\left|\Phi_\mathrm{L1}\right|$. The envelope remains above the energy of the L1 point throughout the evolution, yet very little of the mass on the grid is above zero binding energy. However, as seen in the right panel of Figure \ref{eddlum},
relative to the $q=0.7b$ run, far more grid material is gravitationally unbound in the $q=0.7a$ run. After approximately the $6^\mathrm{th}$ orbit, none of the material in the $q=0.7b$ run is unbound, while a fraction (albeit tiny) of the material in the $q=0.7a$ run is unbound throughout the simulation. 

The accretion luminosity of both runs is shown in the left panel of Figure \ref{eddlum}. 
For the $q=0.7a$ run, the radiation luminosity that escapes through the grid boundaries is also shown. These luminosities are normalized to the nominal Eddington luminosity for spherical accretion, 
\begin{equation}
L_\mathrm{Edd} := \frac{4 \pi G m_p M_\mathrm{acc} c }{\sigma_T},
\end{equation} 
where $M_\mathrm{acc}$ is the mass of the accretor, $\sigma_T$ is the Thomson scattering cross section, and $m_p$ the mass of a proton. The accretion luminosity of both runs exceeds the radiative luminosity of the $q=0.7a$ run by many orders of magnitude. The radiative luminosity itself is roughly on the order of $L_\mathrm{Edd}$. This is consistent with the prediction by \cite{HW1999} that most of the radiation in a highly super-Eddington mass transfer will be swept up by the accretion flow, leaving approximately the Eddington luminosity to escape. Between the $q=0.7a$ and $q=0.7b$ runs, however, we see very little substantial differences between the state of the donor and accretor. The mass transfer rates and central densities begin to diverge from one another at about the $11^{th}$ orbit, and the transfer rate of the $q=0.7a$ run is less than the $q=0.7b$ run. Although this is consistent with super-Eddington accretion, other evidence suggests that the force of radiation cannot be the cause of these differences.  As seen in Figure \ref{panel2}, very little of the matter is actually in a region of space where the force of radiation is sufficient to cancel the force of gravity. The lowest density depicted in Figure \ref{panel2}, represented by the color purple, is on the order of the lowest optically thick density. As the run progresses the grid fills with optically thick material, and the radiation field is mostly in the diffusion limit. Since the flow is dominated by advection, the radiation produced where the stream impacts the accretor simply moves with the flow of the material, rather than escaping the material and exerting a force on the stream further up. If anything, it would appear the presence of radiation {\it reduces} the flow of material into the common envelope. In the color figures of the evolution (Figures \ref{panel1}, \ref{panel2}, and \ref{panel0}), the envelope seems to be  less extended for the $q=0.7a$ run. Because the envelope begins to flow off the grid, however, we cannot make any firm conclusions about its evolution past the first few orbits using these simulations. 

\section{Conclusions}

In this paper we have presented an Eulerian based grid method for evolving an astrophysical fluid on a  rotating cylindrical mesh. The method simulates the physical processes of inviscid hydrodynamical fluid flow, Newtonian self-gravity, and radiation transport in the FLD approximation.  We have tailored the method for the study of close binary systems at the onset of dynamically unstable mass transfer. These systems are initialized with a state of near equilibrium and remain in such a state throughout most of their simulated evolutions. Therefore, careful attention has been paid to constructing accurate equilibrium models as initial configurations and to the globally conserved quantities of mass, momentum, and energy. In particular, we have incorporated the potential energy terms into the K-T method in such a way that conserves total energy to a very high precision. Over the first $15$ orbits, total energy is conserved to within a relative error of about $8 \times 10^{-7}$ per orbit for the $q=0.7b$ run and $4 \times 10^{-6}$ per orbit for the $q=0.7a$ run.

In addition to a set of verification tests, we have demonstrated our method by running two test cases of a binary system of mass ratio $q=0.7$; one with the radiation feature enabled ($q=0.7a$), and one with it disabled ($q=0.7b$).  The radiation energy tends to be swept up by the accretion flow and, consistent with the arguments made by \cite{HW1999}, the radiative luminosity is of roughly the same order as the Eddington luminosity. Over the time frame of the simulations, radiation transport did not appear to significantly effect the accretion flow itself. The main difference between the runs was in the low density common envelope surrounding the stars.  The envelope of the run with radiation enabled ($q=0.7a$) is less massive and more condensed. No significant fraction of the envelope achieves positive gravitational binding energy in either run. The results tend to suggest that radiation transport plays no significant role at the onset of dynamically unstable mass transfer for systems of mass ratio $q=0.7$, even when the accretion luminosity is significantly super-Eddington. It is of course quite possible that radiation transport plays a role when such an object merges, however, due to numerical difficulties with the center of mass of the system, we terminated the simulations before merger. With the center of mass correction enabled for future runs, we intend to re-run the $q=0.7$ simulation to merger with radiation enabled.

 As seen in Figure \ref{eddlum}, the accretion rate is over $10^2$ times the critical Eddington rate as soon as the simulation starts, and quickly rises to over $10^3$ times Eddington before reaching the $5^\mathrm{th}$ orbit.  Because of this, we are not able to accurately model the trans-Eddington phase, where the accretion luminosity begins to just barely cross over the Eddington luminosity. Although we have shown that the radiation is swamped by the hydrodynamic flow of optically thick material when the flow rate is highly super-Eddington, it is possible that there exists a regime where the flow of accreting material is unable to carry the entirety of the radiated accretion luminosity with it and the radiation is significant enough to affect the flow. Ultimately, the question of stability would depend on the opposing effects of mass loss, which favors stability, and consequent angular momentum loss carried by the outflowing mass, which favors merger. The mass transfer instability may grow so quickly in DWDs with $q \gtrsim 0.7$ that effects due to radiation would not have time to affect the orbital dynamics and prevent merger. DWDs of lower mass ratio, however, take longer to merge. There is evidence that suggests such DWDs may in fact never merge and instead experience long term stable mass transfer. \cite{MFTD2007} were able to run a $q=0.4$ polytropic model for over $40$ orbits, and the mass transfer appeared to be stable when the run terminated. \cite{DMTF2006} obtained similar results for a $q=0.5$ model.  Although earlier SPH simulations suggest DWD systems of this mass ratio  fall prey to dynamical, and in some cases, secular instability, and merge within a few orbital periods (\cite{RS1995}), recently an SPH model of a $q=0.5$ DWD took over $60$ orbital periods to merge (\cite{DRB2009}). If such a system exists in a trans-Eddington regime, mass loss may cause it to  never transition to super-Eddington. Furthermore,  when significant mass is lost over a sufficient number of orbital periods, it is possible  merger of  an otherwise unstable system might be prevented.  While it has not been possible for us to answer this question via the simulations presented here, it may be possible to move our models into such a regime by appropriate scaling of the physical constants in Table \ref{physcons}. We plan to carry out such simulations in the future.
 
	The long-term stability of such systems will also likely depend on the interaction between the common envelope and the binary components. If mass transfer occurs over many orbits, the frictional forces of the common envelope will tend to favor merger. Evidence from our $q=0.7$ simulations suggests this effect is insignificant over the time it takes for such a system to merge. It is possible, however, for the cumulative effect of friction to become significant over many orbits, and the common envelope may cause otherwise stable systems of lower mass ratio to eventually merge. If, on the other hand, the envelope is able to attain sufficient energy to escape the system, merger may be avoided. For the $q=0.7$ models, as seen in the left panel of Figure \ref{q7ce}, most of the common envelope's energy is kinetic. However, as seen in the right panel of Figure \ref{eddlum}, only a tiny fraction of the envelope exceeds the gravitational binding energy. Excluding other effects, based on this alone we would expect that very little of the mass is dynamically ejected.  As shown in Figure \ref{panel2}, however, the common envelope quickly grows to extend past the computational grid. To realistically model interactions with the common envelope requires a larger spatial domain. With the present code, we cannot simultaneously model the binary components with a similar resolution to the models presented here and model a large, extended common envelope. In the future, incorporation of AMR into our models may permit us to accomplish this.

\section{Acknowledgements}

We acknowledge valuable interactions that we have had with J. Frank, G. Clayton, P.M. Motl, and W. Even over the course of this project.  We also acknowledge the referee for his thorough analysis of our work and his insightful comments. This work has been supported, in part, by grants PHY-0803629, AST-0708551, EPS-1006891, OIA-0963375, and DGE-0504507 from the U.S. National Science Foundation and, in part, by NASA/ATP grants NNX10AC72G, and NNX07AG84G.  This research also has been made possible by grants of high-performance computing time on the TeraGrid (TG-AST090104), at LSU, and across LONI (Louisiana Optical Network Initiative), especially awards loni\_astro08 and loni\_astro09.

\begin{appendices}

\section{Opacities}
\label{opac_app}
The three opacities in our equation set, $\chi$, $\kappa_P$, and $\kappa_E$, represent the frequency integrated opacity weighted by the frequency dependent radiative flux, $\mathbf{F}_\nu$, the frequency dependent Planck function, $B_\nu$, and the frequency dependent radiative energy density, $E_{R,\nu}$, respectively. Their definitions are given by equations (8) through (10) in \cite{HNF2006}. Physically, $\kappa_P$ and $\kappa_E$ should contain absorption terms only, while $\chi$ includes absorption as well as scattering. If the opacity is frequency dependent we have to make assumptions about the spectrum in order to integrate the opacity over frequency. In the diffusion limit we may assume a blackbody spectrum. When the radiation is free-streaming, however, this assumption does not generally hold. Additionally, when Thompson scattering and free-free absorption terms are both present in $\chi$, we cannot obtain an analytic expression even in the diffusion limit. Due to these limitations, we adopt a simplistic expression for the opacities. For the $q=0.7a$ run, we have set 
\begin{equation}
\chi := \sigma_T \rho + a_\mathrm{ff,s} \rho^2 T^{-3.5}, 
\label{chifudge}
\end{equation}
and
\begin{equation}
\kappa_E := \kappa_P := a_\mathrm{ff,a} \rho^2 T^{-3.5}, 
\end{equation}
where 
\begin{equation}
\sigma_T := 8.4 \times 10^{12} \left(l_\mathrm{code}^2 / m_\mathrm{code}\right), 
\end{equation}
\begin{equation}
a_\mathrm{ff,s} := 2.12 \times 10^{11} \left(l_\mathrm{code}^5 K_\mathrm{code}^{3.5} / m_\mathrm{code}^2\right), 
\end{equation}
and 
\begin{equation}
a_\mathrm{ff,a} := 6.50 \times 10^{12} \left(l_\mathrm{code}^5 K_\mathrm{code}^{3.5} / m_\mathrm{code}^2\right).
\end{equation}
 In practice, for the $q=0.7a$ run, we find that the first term of equation (\ref{chifudge}) (the scattering term) is generally several orders of magnitude higher than the second term.

\section{Three Different Gas Energy Schemes}
\label{energy_schemes}
We refer to the gas energy scheme of equation (\ref{estar_}) as the E* scheme. An alternative scheme for the total gas energy is 
\begin{multline}
\label{noestar_}
\frac{d}{d t}{E_{G,j k l}}
+ \mathcal{D}\left\{ \left( \mathcal{E}_\mathrm{G} + p \right) \mathbf{u} \right\}_{j k l}   \ = \mathcal{V} \left\{ \mathcal{E}_\mathrm{G} \right\}_{j k l} 
 -  \frac{s_{R, j k l}}{2 \Delta} \left( \Phi_{\mathrm{eff},j + 1 k l} - \Phi_{\mathrm{eff},j - 1 k l}\right) \\
 -  \frac{l_{z, j k l} - \rho_{j k l}  R_j^2 \Omega }{2 R_j^2 \Delta} \left( \Phi_{\mathrm{eff},j k + 1 l} - \Phi_{\mathrm{eff},j k - 1 l}\right) 
 -  \frac{s_{z, j k l}}{2 \Delta} \left( \Phi_{\mathrm{eff},j k l + 1} - \Phi_{\mathrm{eff},j k l - 1}\right).
\end{multline}
Equation (\ref{noestar_}) is obtained by applying equation (\ref{method}) to the total gas energy and adding first-order discrete derivatives to the RHS to account for the gravitational term. As shown above in \S \ref{singlesection}, application of equation (\ref{noestar_}) to stellar models results in an ever increasing total system energy. Over many dynamical timescales, the polytrope will dissipate.

Another way to handle the gas energy is to not evolve $E_G$ or $\tau$ at all. Instead we obtain the pressure using the polytropic relation of equation (\ref{polytropiceos}) and setting $\gamma := 1 + \frac{1}{n}$. As shown in \S \ref{singlesection}, the resulting model does \textit{not} appear to lose energy indefinitely.

\section{Center of Mass Correction}
\label{com_correct}
The center of mass correction applied to some of the simulations in \S \ref{singlesection} is formulated according to
\begin{equation}
\label{comcorrect}
	\mathbf{a}_{\mathrm{COM}} := -\omega^2 \left(\mathbf{X}_\mathrm{COM} - \mathbf{X}_{0,\mathrm{COM}}\right) - 
                             2 \omega \frac{d}{dt}\mathbf{X}_\mathrm{COM},
\end{equation}
where $\mathbf{a}_{\mathrm{COM}}$ is the spatially constant acceleration used to correct for center of mass motion, $\mathbf{X}_\mathrm{COM}$ is the current center of mass of the system, $\mathbf{X}_{0,\mathrm{COM}}$ is the center of mass at $t=0$, and $\omega$ is a suitably chosen constant. In the polytrope simulations discussed in \S \ref{singlesection}, $\omega := 1$, which is close to the inverse of the dynamical timescale. Equation (\ref{comcorrect}) is the equation for a critically damped harmonic oscillator. We chose this as our correction because we would like to counteract any displacements from the original center of mass without overshooting. At each time step, the center of mass of the system is computed. The velocity of the center of mass $\frac{d}{dt}\mathbf{X}_\mathrm{COM}$ is determined using $\mathbf{X}_\mathrm{COM}$ from the current and previous time step. The term 
\begin{equation}
\rho_{j k l} \left( a_{\mathrm{COM},x} \cos \phi_k - a_{\mathrm{COM},y} \sin \phi_k \right)
\end{equation}
is added to the RHS of radial momentum equation (\ref{discrete2}). The term
\begin{equation}
\rho_{j k l} R_j \left( a_{\mathrm{COM},x} \sin \phi_k + a_{\mathrm{COM},y} \cos \phi_k \right)
\end{equation}
is added to the RHS of the angular momentum equation (\ref{discrete3}). The total gas energy equation  (\ref{discrete5}) is modified by adding 
\begin{equation}
\rho_{j k l} \mathbf{u}_{j k l} \cdot \mathbf{a}_{\mathrm{COM}}
\end{equation}
to its RHS.

Note that this correction was not used in the $q=0.7a$ and $q=0.7b$ runs in \S \ref{binary_section}. We have demonstrated its success in \S \ref{singlesection} and intend it for future use.

\end{appendices}

\bibliography{paper}

\end{document}